\documentclass[smallextended]{svjour3}

\smartqed  
\usepackage{cite,lineno,hyperref}
\usepackage{color}

\modulolinenumbers[5]

\journalname{Journal of Mathematical Imaging and Vision}

\usepackage{setspace}
\usepackage{calc}
\usepackage{stix2}
\usepackage{amsmath,amsfonts,amssymb,bm,bbm,prettyref} 
\usepackage{mathtools}

\usepackage{tikz}
\usepackage[toc,page]{appendix}
\usepackage{makecell}
\usepackage{caption} 
\usepackage{subcaption} 
\usepackage{float}

\usepackage{tabularx}
\usepackage[ruled,vlined]{algorithm2e}
    
\usepackage{version}
\usepackage[english]{babel}
\excludeversion{hide}

\newcommand{\Jcal}{\mathcal{J}}

\renewcommand{\P}{{\mathcal P}}

\newcommand{\cond}{\,|\,}
\newcommand{\define}{\, := \,}
\newcommand{\diag}{\operatorname{diag}}
\newcommand{\Rf}{\mathbb{R}}

\newcommand{\Z}{\mathbb{Z}}
\newcommand{\N}{\mathbb{N}}

\newcommand{\ndim}{d}

\newcommand{\bphi}{\bm{\phi}}

\newcommand{\bpsi}{\bm{\psi}}
\newcommand{\Lrw}{{L_{\mathrm{rw}}}}
\newcommand{\Lsym}{{L_{\mathrm{sym}}}}
\newcommand{\jmax}{{j_{\mathrm{max}}}}

\newcommand{\vol}{\operatorname{vol}}

\newcommand{\cut}{\text{cut}}

\newcommand{\f}{\bm{f}}

\newcommand{\bx}{\bm{x}}

\newcommand{\bF}{\bm{F}}
\newcommand{\bI}{\bm{I}}

\newcommand{\transp}{{\scriptscriptstyle{\mathsf{T}}}}
\newcommand{\inner}[2]{\left\langle {#1}, {#2} \right\rangle}
\newcommand{\linner}[2]{\langle {#1}, {#2} \rangle}
\newcommand{\Span}{\operatorname{span}}

\graphicspath{{./figures/}}

\usepackage{graphicx}
\usepackage{caption} 
\usepackage{subcaption} 

\begin{document}

\title{eGHWT: The Extended Generalized Haar-Walsh Transform}

\author{Naoki Saito \and Yiqun Shao\thanks{Y.\ S.\ is now at Meta Platforms, Inc.}}
\institute{Department of Mathematics, University of California, Davis, CA, 95616
\\ \email{saito@math.ucdavis.edu ; yqshao@ucdavis.edu}}

\maketitle

\begin{abstract}
  Extending computational harmonic analysis tools from the classical setting of
  regular lattices to the more general setting of graphs and networks is very
  important and much research has been done recently.
  The Generalized Haar-Walsh Transform (GHWT) developed by Irion and Saito (2014)
  is a multiscale transform for signals on graphs, which is a generalization of
  the classical Haar and Walsh-Hadamard Transforms. We propose the
  \emph{extended} Generalized Haar-Walsh Transform (eGHWT), which is a
  generalization of the adapted time-frequency tilings of Thiele and Villemoes
  (1996). The eGHWT examines not only the efficiency of graph-domain partitions
  but also that of ``sequency-domain'' partitions \emph{simultaneously}.
  Consequently, the eGHWT and its associated best-basis selection algorithm for
  graph signals significantly improve the performance of the previous GHWT
  with the similar computational cost, $O(N \log N)$, where $N$ is the number of
  nodes of an input graph. While the GHWT best-basis algorithm seeks the most
  suitable orthonormal basis for a given task among more than $(1.5)^N$ possible
  orthonormal bases in $\Rf^N$, the eGHWT best-basis algorithm can find a better
  one by searching through more than $0.618\cdot(1.84)^N$ possible orthonormal
  bases in $\Rf^N$. This article describes the details of the eGHWT best-basis
  algorithm and demonstrates its superiority using several examples including
  genuine graph signals as well as conventional digital images viewed as graph
  signals. Furthermore, we also show how the eGHWT can be extended to 2D signals
  and matrix-form data by viewing them as a tensor product of graphs generated
  from their columns and rows and demonstrate its effectiveness on applications
  such as image approximation. 
\end{abstract}


\section{Introduction}
In recent years, research on graphs and networks is experiencing rapid growth
due to a confluence of several trends in science and technology:
the advent of new sensors, measurement technologies, and social network
infrastructure has provided huge opportunities to visualize complicated
interconnected network structures, record data of interest at various locations
in such networks, analyze such data, and make inferences and diagnostics.  
We can easily observe such network-based problems in truly diverse fields: 
biology and medicine (e.g., connectomes); computer science (e.g.,
social networks); electrical engineering (e.g., sensor networks);
hydrology and geology (e.g., ramified river networks); 
and civil engineering (e.g., road networks), to name just a few.
Consequently, there is an explosion of interest and demand to analyze data 
sampled on such graphs and networks, which are often called
``network data analysis'' or ``graph signal processing'';
see e.g., recent books~\cite{CHUNG-LU, EASLEY-KLEINBERG, LOVASZ-BOOK, NEWMAN2}
and survey articles~\cite{SHUMAN-ETAL, ORTEGA-ETAL},
to see the evidence of this trend.
This trend has forced the signal processing and applied mathematics communities to extend classical
techniques on regular domains to the setting of graphs.
Much efforts have been done to develop wavelet transforms for signals on
graphs (or the so-called graph signals)~\cite{SZLAM-HAAR-GRAPH,
  COIF-MAGG-DW, BREMER-COIF-MAGG-SZLAM, MURTAGH-Haar,
  LEE-NADLER-WASSERMAN, JANSEN-NASON-SILVERMAN, COIF-GAVISH,
  HAMMOND-VANDERGHEYNST-GRIBONVAL, RUSTAMOV, TREMBLAY-BORGNAT}.
Comprehensive reviews of transforms for signals on graphs have also been
written~\cite{SHUMAN-ETAL, ORTEGA-ETAL}. 

The Generalized Haar-Walsh Transform (GHWT)~\cite{IRION-SAITO-GHWT,
  IRION-SAITO-SPIE, IRION-SAITO-TSIPN}, developed by Irion and Saito, has
achieved superior results over other transforms in terms of both approximation
and denoising of signals on graphs (or graph signals for short).
It is a generalization of the classical Haar and Walsh-Hadamard Transforms.
In this article, we propose and develop the \emph{extended Generalized Haar-Walsh
Transform} (eGHWT). The eGHWT and its associated best-basis selection algorithm
for graph signals significantly improve the performance of the previous
GHWT with the similar computational cost, $O(N \log N)$ where $N$ is the number
of nodes of an input graph. While the previous GHWT best-basis algorithm seeks
the most suitable orthonormal basis (ONB) for a given task among more than
$(1.5)^N$ possible orthonormal bases in $\Rf^N$, the eGHWT best-basis
algorithm can find a better one by searching through more than
$0.618\cdot(1.84)^N$ possible orthonormal bases in $\Rf^N$.
It can be extended to 2D signals and matrix-form data in a more subtle way than
the GHWT. 
In this article, we describe the details of the eGHWT basis-basis algorithm and
demonstrate its superiority. Moreover, we showcase the versatility of the eGHWT
by applying it to genuine graph signals and classical digital images.
  
The organization of this article is as follows. In Sect.~\ref{sec:back},
we review background concepts, including graph signal processing and recursive
graph partitioning, which is a common strategy used by researches to develop
graph signal transforms.
In Sect.~\ref{sec:ghwt}, the GHWT and its best-basis algorithm are reviewed.
In Sect.~\ref{sec:eghwt}, we provide an overview of the eGHWT. We start by
reviewing the algorithm developed by~\cite{THIELE-VILLEMOES}. Then we illustrate
how that algorithm can be modified to construct the eGHWT. A simple example
illustrating the difference between the GHWT and the eGHWT is given. We finish
the section by explaining how the eGHWT can be extended to 2D signals and
matrix-form data.
In Sect.~\ref{sec:appl}, we demonstrate the superiority of the eGHWT over the
GHWT (including the graph Haar and Walsh bases) using real datasets.
Section~\ref{sec:disc} concludes this article and discusses potential future
projects.

We note that the most of the figures in this article are \emph{reproducible}.
The interested readers can find our scripts to generate the figures (written in
the \emph{Julia} programming language~\cite{JULIA}) at our software website:\\
\url{https://github.com/UCD4IDS/MultiscaleGraphSignalTransforms.jl}; \\
in particular, see its subfolder, \href{https://github.com/UCD4IDS/MultiscaleGraphSignalTransforms.jl/tree/master/test/paperscripts/eGHWT2021}{test/paperscripts/eGHWT2021}.

\section{Background}
\label{sec:back}
\subsection{Basics of Spectral Graph Theory and Notation}
In this section, we review some fundamentals of spectral graph theory and
introduce the notation that will be used throughout this article.

A \emph{graph} is a pair $G = (V, E)$, where
$V = V(G) = \{v_1, v_2, \ldots, v_N\}$ is the \emph{vertex} (or \emph{node}) set
of $G$, and $E = E(G) = \{e_1, e_2, \ldots, e_M\}$ is the \emph{edge} set,
where each edge connects two nodes $v_i, v_j$ for some
$1 \leq i \neq j \leq N$. We only deal with finite $N$ and $M$ in this article.
For simplicity, we often write $i$ instead of $v_i$.

An edge connecting a node $i$ and itself is called a \emph{loop}.
If there exists more than one edge connecting some $i, j$, then they are called
\emph{multiple edges}. A graph having loops or multiple edges is called a
\emph{multiple graph} (or multigraph); a graph with neither of these is called a
\emph{simple graph}. A \emph{directed graph} is a graph in which edges have
orientations while \emph{undirected graph} is a graph in which edges do not have
orientations. If each edge $e \in E$ has a weight (normally nonnegative), then
$G$ is called a \emph{weighted graph}.
A \emph{path} from $i$ to $j$ in a graph $G$ is a subgraph of $G$ 
consisting of a sequence of distinct nodes starting with $i$ and ending 
with $j$ such that consecutive nodes are adjacent.  
A path starting from $i$ that returns to $i$ (but is not a loop)
is called a \emph{cycle}.
For any two distinct nodes in $V$, if there is a path connecting them,
then such a graph is said to be \emph{connected}.
In this article, we mainly consider \emph{undirected weighted simple connected
graphs}. Our method can be easily adapted to other undirected graphs,
but we do not consider directed graphs here.

Sometimes, each node is associated with spatial coordinates in $\Rf^\ndim$.
For example, if we want to analyze a network of sensors and build a graph
whose nodes represent the sensors under consideration, then these nodes
have spatial coordinates in $\Rf^2$ or $\Rf^3$ indicating their current
locations.  In that case, we write $\bx[i] \in \Rf^\ndim$ for the location of
node $i$. Denote the functions supported on graph as
$\f = (f[1], \ldots, f[N])^\transp \in \Rf^N$. It is a data vector (often called
a \emph{graph signal}) where $f[i] \in \Rf$ is the value measured at the node
$i$ of the graph.  

We now discuss several matrices associated with undirected simple graphs.
The information in both $V$ and $E$ is captured by the \emph{edge weight matrix}
$W(G) \in \Rf^{N \times N}$, where $W_{ij} \geq 0$ is the edge weight between nodes
$i$ and $j$. In an unweighted graph, this is restricted to be either $0$ or $1$,
depending on whether nodes $i$ and $j$ are adjacent, and we may refer to $W(G)$
as an \emph{adjacency matrix}. In a weighted graph, $W_{ij}$ indicates the
\emph{affinity} between $i$ and $j$. In either case, since $G$ is undirected,
$W(G)$ is a symmetric matrix. We then define the \emph{degree matrix}
$$D(G) \define \diag(d_1, \ldots, d_N), \text{ where }d_i \define \sum_j W_{ij}.$$
With this in place, we are now able to define the
\emph{(unnormalized) Laplacian} matrix, \emph{random-walk normalized Laplacian}
matrix, and \emph{symmetric normalized Laplacian} matrix respectively, as
$$L(G) \define D(G)-W(G),$$
$$\Lrw(G) \define D(G)^{-1} L(G),$$
$$\Lsym(G) \define D(G)^{-1/2} L(G) D(G)^{-1/2}.$$
See \cite{VONLUX} for the details of the relationship between these
three matrices and their spectral properties.
We use $0=\lambda_0 < \lambda_1 \leq \ldots \leq \lambda_{N-1}$ to denote the
sorted Laplacian eigenvalues and $\bphi_0,\bphi_1,\ldots,\bphi_{N-1}$ to denote
the corresponding eigenvectors, where the specific Laplacian matrix to which
they refer will be clear from either context or subscripts.

Laplacian eigenvectors can then be used for graph partitioning. Spectral
clustering~\cite{VONLUX} performs $k$-means on the first few
eigenvector coordinates to partition the graph. This approach is justified from
the fact that it is an approximate minimizer of the graph-cut criterion called
\emph{Ratio Cut}~\cite{RATIOCUT} (or the \emph{Normalized Cut}~\cite{SHI-MALIK})
when $L$ (or $\Lrw$, respectively) is used. Suppose the nodes in $V(G)$ is
partitioned into two disjoint sets $A$ and $A^c$, then \emph{Ratio Cut} and
\emph{Normalized Cut} are defined by
\begin{align*}
\text{Ratio Cut}(A,A^c) &\define \frac{\cut(A,A^c)}{|A|} + \frac{\cut(A,A^c)}{|A^c|} \\
\text{Normalized Cut}(A,A^c) &\define \frac{\cut(A,A^c)}{\mathrm{vol}(A)} + \frac{\cut(A,A^c)}{\mathrm{vol}(A^c)},
\end{align*} 
where $\cut(A,A^c) \define \sum_{i \in A, j \in A^c} W_{ij}$ indicates the quality
of the cut (the smaller this quantity, the better the cut in general),
$\vol(A) \define \sum_{i\in A}d_i$ is the so-called
\emph{volume} of the set $A$, and $| A |$ is the cardinality of (i.e.,
the number of nodes in) $A$.

To reduce the computational complexity (as we did for the GHWT construction),
we only use the \emph{Fiedler vector}~\cite{FIEDLER}, i.e.,
the eigenvector $\bphi_1$ corresponding to the smallest nonzero eigenvalue
$\lambda_1$, to bipartition a given graph (or subgraph) in this article.
For a connected graph $G$, Fiedler showed that its Fiedler vector partitions
the vertices into two sets by letting
$$ V_1 = \{i \in V \cond \bphi_1[i] \geq 0 \}, \quad V_2 = V \setminus V_1,$$
such that the subgraphs induced on $V_1$ and $V_2$ by $G$ are both connected
graphs~\cite{FIEDLER}.
In this article, we use the Fiedler vector of $\Lrw$ of a given graph and
its subgraphs unless stated otherwise.
See e.g., \cite{VONLUX}, which suggests the use of the Fiedler vector of
$\Lrw$ for spectral clustering over that of the other Laplacian matrices.

\subsection{Recursive Partitioning of Graphs}
The foundation upon which the GHWT and the eGHWT are constructed is
a \emph{binary partition tree} (also known as a
\emph{hierarchical bipartition tree}) of an input graph $G(V,E)$:
a set of tree-structured subgraphs of $G$ constructed by recursively
bipartitioning $G$. This bipartitioning operation ideally splits each subgraph
into two smaller subgraphs that are roughly equal in size while keeping
tightly-connected nodes grouped together. 
More specifically, let $G^j_k$ denote the $k$th subgraph on level $j$ of
the binary partition tree of $G$ and $N^j_k \define | V(G^j_k) |$, where
$j, k \in \Z_{\geq 0}$.
Note $G^0_0 = G$, $N^0_0 = N$, i.e., level $j=0$ represents the root node of
this tree.
Then the two children of $G^j_k$ in the tree, $G^{j+1}_{k'}$ and $G^{j+1}_{k'+1}$,
are obtained through partitioning $G^j_k$ using the Fiedler vector of
$\Lrw(G^j_k)$. The graph partitioning is recursively performed until each
subgraph corresponding to the leaf contains only one node. Note that
$k' = 2k$ if the resulting binary partition tree is a perfect binary tree.

In general, other spectral clustering methods with different number of
eigenvectors or different Laplacian matrices are applicable as well.
However, we impose the following five conditions on the binary partition
tree:
\begin{enumerate}
\item The root of the tree is the entire graph, i.e., $G^0_0 = G$;
\item The leaves of the tree are single-node graphs, i.e.,
  $N^{\jmax}_k = 1$, where $\jmax$ is the height of the tree;
\item All regions (i.e., nodes in the subgraphs) on the same level are
  disjoint, i.e., $V(G^j_k) \cap V(G^j_{k'}) = \emptyset$ if $k \neq k'$;
\item Each subgraph with more than one node is partitioned into exactly
  two subgraphs;
\item (Optional) In practice, the size of the two children, $N^{j+1}_{k'}$
  and $N^{j+1}_{k'+1}$ should not be too far apart to reduce inefficiency.
\end{enumerate}
Even other (non-spectral) graph cut methods can be used to form the binary
partition tree, as long as those conditions are satisfied.
The flexibility of a choice of graph partitioning methods in the GHWT/eGHWT
construction is certainly advantageous.

\begin{figure}
  \centering\includegraphics[width = 0.5\textwidth]{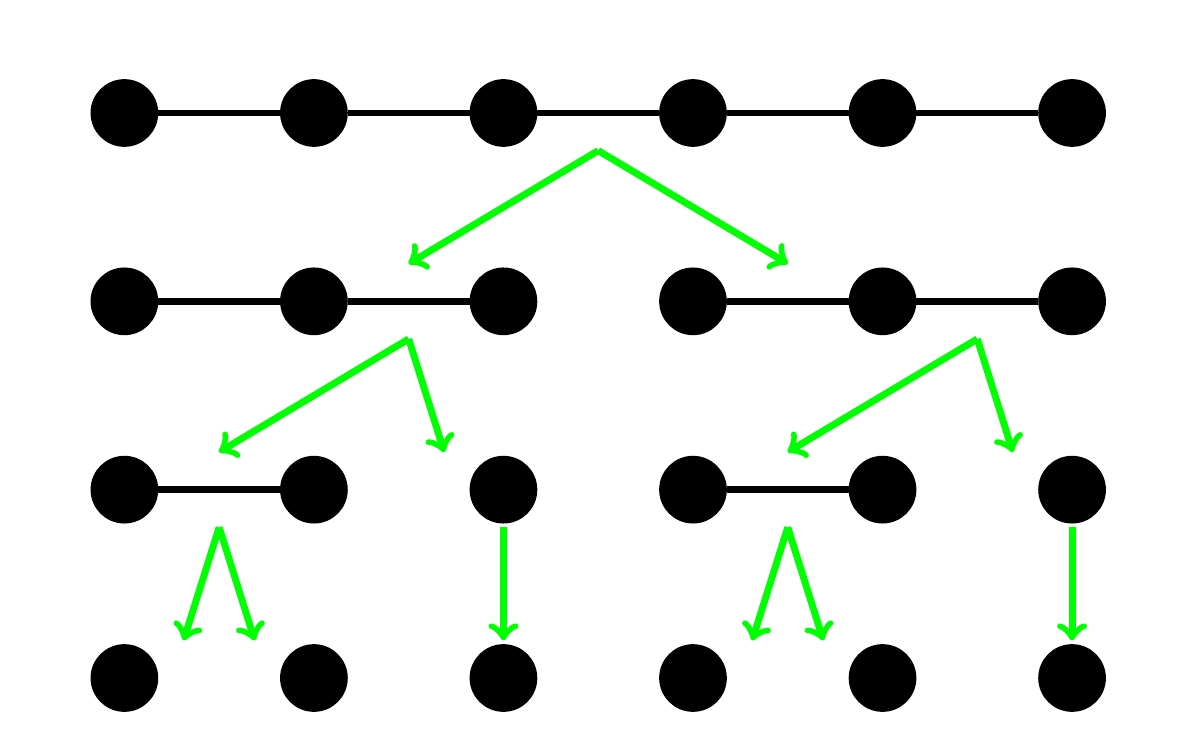}
  \caption{An example of a hierarchical bipartition tree for a path graph with
    $N = 6$ nodes, where the edge weights are equal. The root is the whole graph.
  }
  \label{fig:partition_tree_path}
\end{figure}

We demonstrate two examples illustrating the binary partition tree here.
The first one is a simple 6-node path graph, $P_6$. It has five edges with equal
weights connecting adjacent nodes. Figure~\ref{fig:partition_tree_path} is the
binary partition tree formed on the graph. In the first iteration, it is
bipartitioned into two subgraphs with $3$ nodes each. Then each of those $3$
node graphs is bipartitioned into a $2$-node graph and an $1$-node graph.
In the end, the subgraphs are all $1$-node graph.

\begin{figure}
  \begin{subfigure}{0.32\textwidth}
    \centering\includegraphics[width=\textwidth]{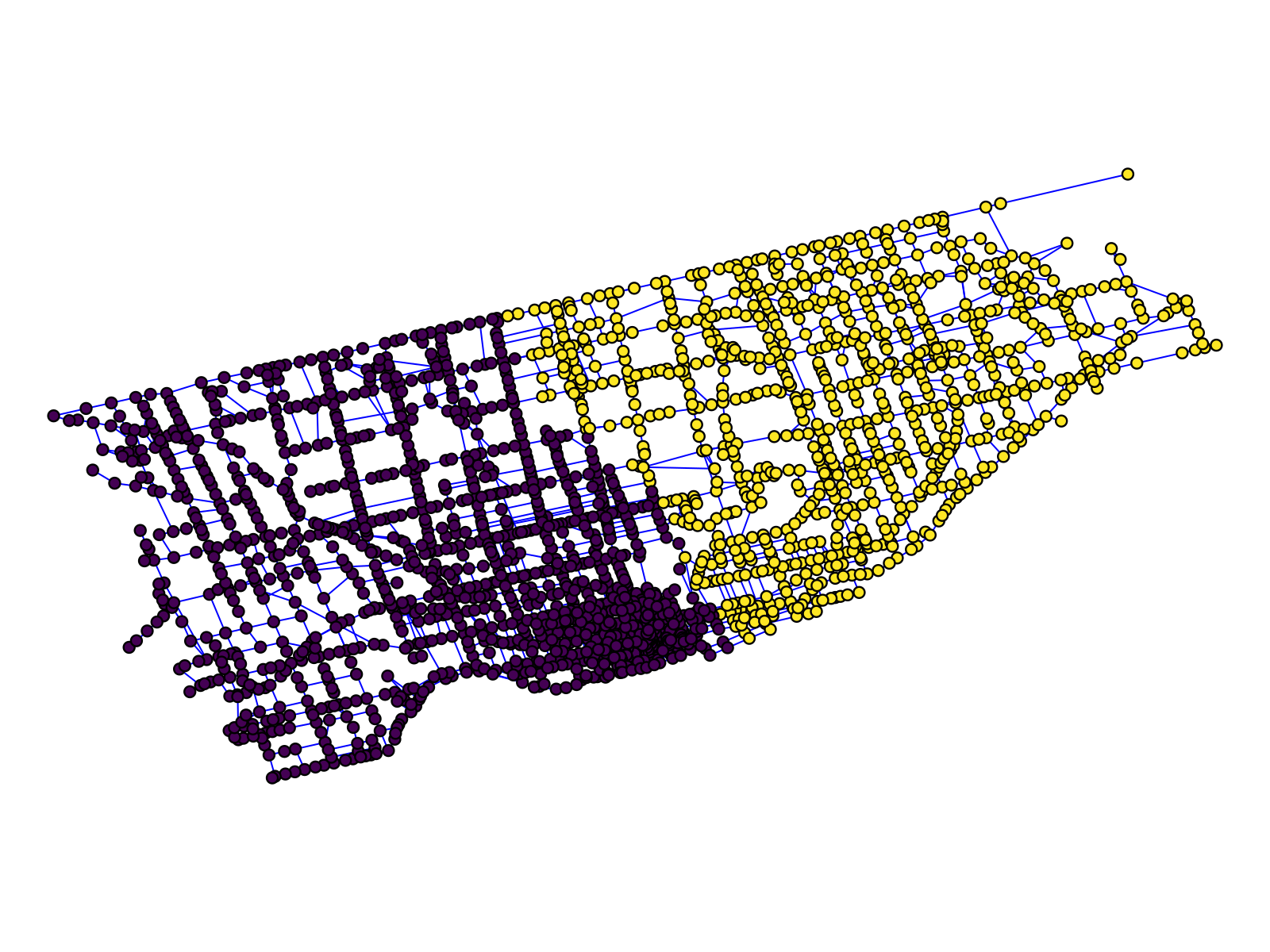}
    \caption{Level $j = 1$}
  \end{subfigure}
  \begin{subfigure}{0.32\textwidth}
    \centering
    \includegraphics[width=\textwidth]{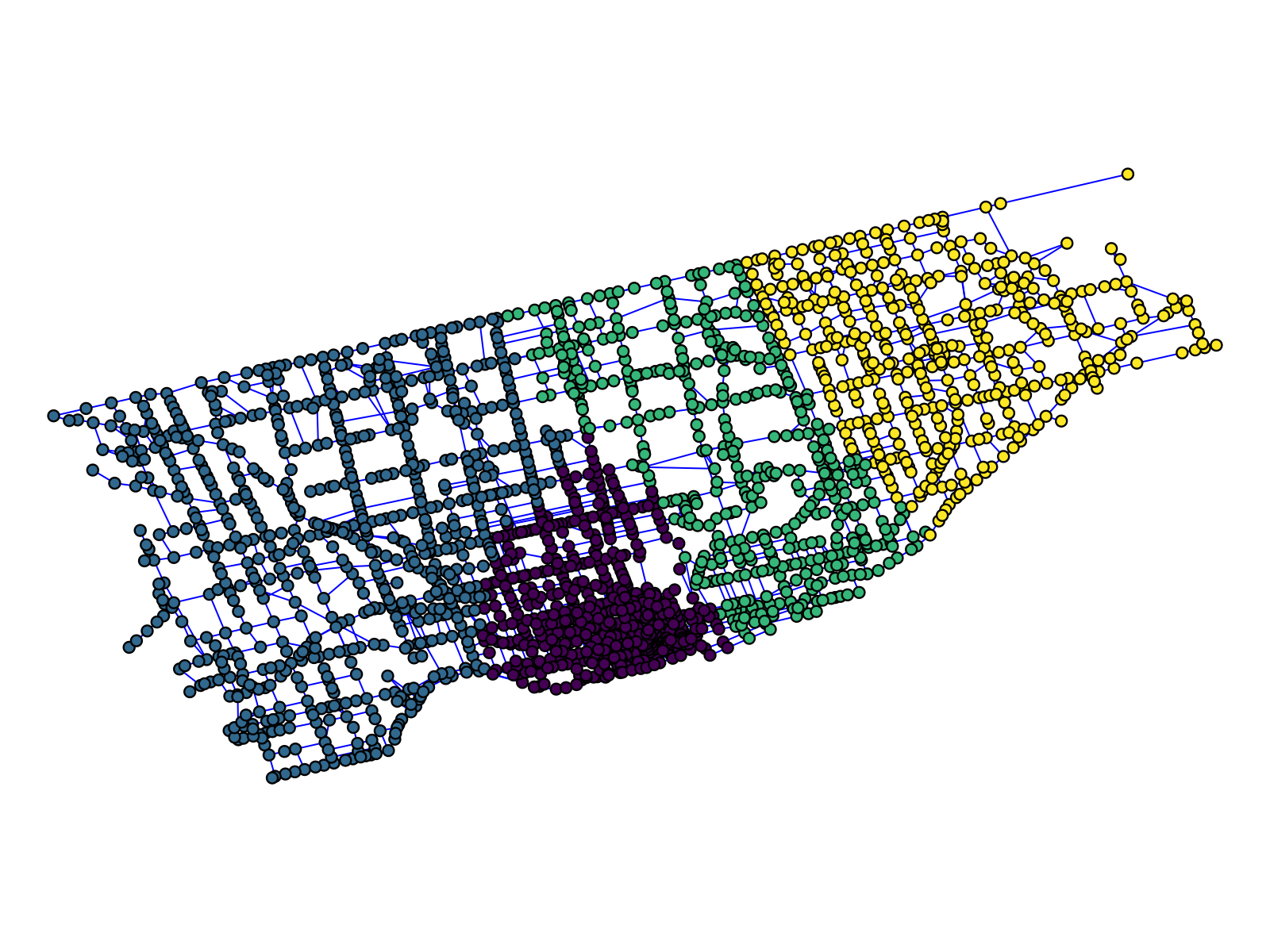}
    \caption{Level $j = 2$}
  \end{subfigure}
  \begin{subfigure}{0.32\textwidth}
    \centering
    \includegraphics[width=\textwidth]{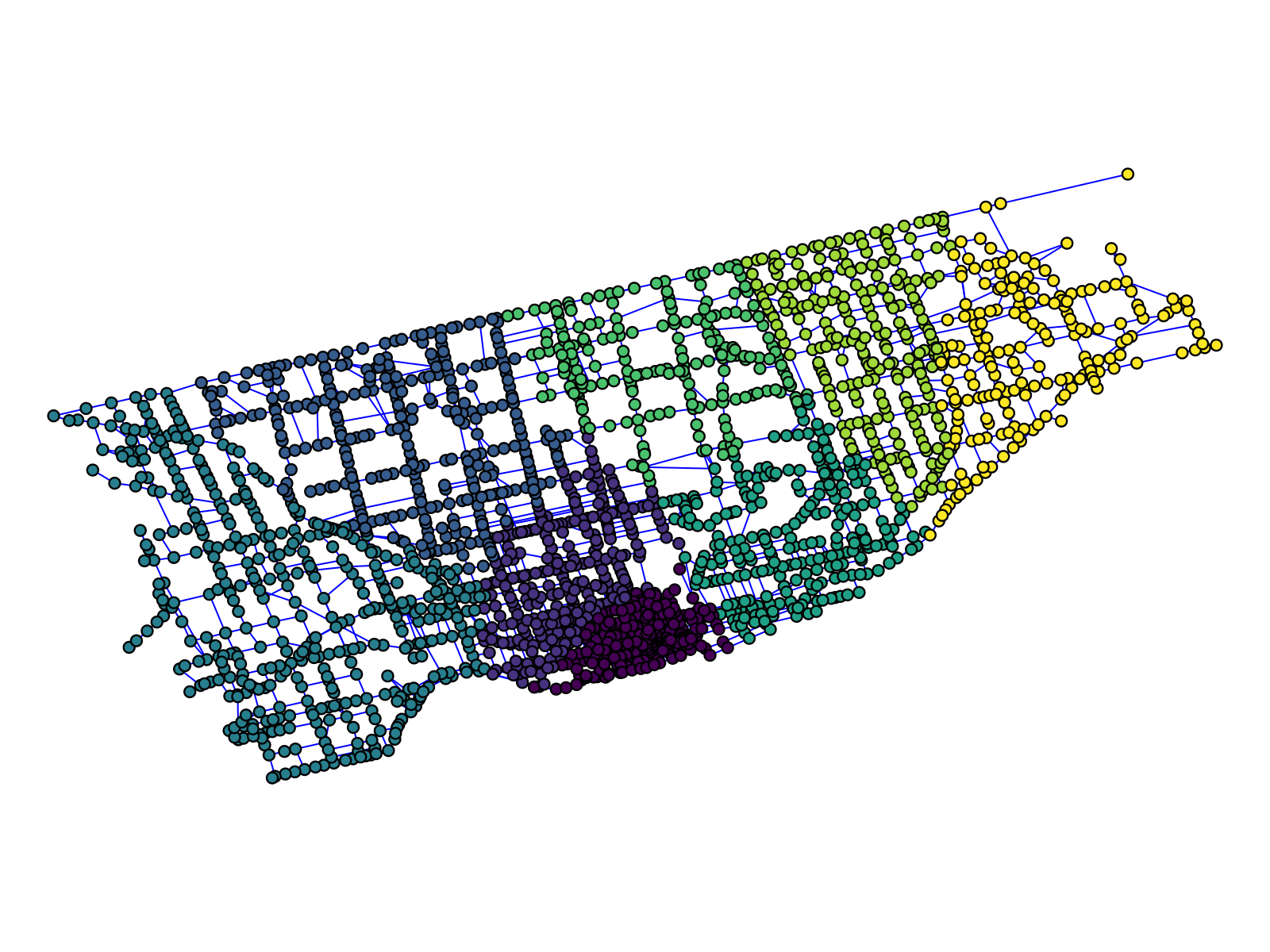}
    \caption{Level $j = 3$}
  \end{subfigure}
  \caption{A demonstration of hierarchical bipartition tree on the Toronto
    street network. On the same level, different colors correspond to different
    regions. The situation at level $j=0$ is not shown since there is no
    partition at $j=0$.
  }
  \label{fig:partition_tree_toronto} 
\end{figure}

The second example is the street network of the City of Toronto, which we
obtained from its open data portal\footnote{URL: \url{https://open.toronto.ca/dataset/traffic-signal-vehicle-and-pedestrian-volumes}}.
Using the street names and intersection coordinates included in the dataset,
we constructed the graph representing the street network there with $N = 2275$
nodes and $M = 3381$ edges. Each edge weight was set as the reciprocal of
the Euclidean distance between the endpoints of that edge.
Figure~\ref{fig:partition_tree_toronto} gives us a visualization of the first
three levels of the binary partition tree on this Toronto street network.

\section{The Generalized Haar-Walsh Transform (GHWT)}
\label{sec:ghwt}
In this section, we will review the
Generalized Haar-Walsh Transform (GHWT) \cite{IRION-SAITO-GHWT,
  IRION-SAITO-SPIE, IRION-SAITO-TSIPN}. It is a multiscale transform for graph
signals and a \emph{true} generalization of the classical Haar and Walsh-Hadamard
transforms: if an input graph is a simple path graph whose number of nodes is
dyadic, then the GHWT reduces to the classical counterpart \emph{exactly}.

\subsection{Overcomplete Dictionaries of Bases}
After the binary partition tree of the input graph $G$ with depth $\jmax$ is
generated, an overcomplete dictionary of basis vectors is composed.
Each basis vector is denoted as $\bpsi^j_{k,l}$, where $j \in [0, \jmax]$
denotes the \emph{level}, $k\in [0, K_j)$ denotes the \emph{region},
and $l$ denotes the \emph{tag}.
$K_j$ is the number of subgraphs of $G$ on level $j$. The tag $l$ assumes a
distinct integer value within the range $[0, 2^{\jmax - j})$.
The tag $l$, when expressed in binary, specifies the sequence of average and
difference operations by which $\bpsi^j_{k,l}$ was generated. For example,
$l = 6$ written in binary is $110$, which means that the basis vector/expansion
coefficient was produced by two differencing operations (two $1$s) followed by
an averaging operation (one $0$). Generally speaking, a larger value of $l$
indicates more oscillations in $\bpsi^j_{k, l}$,
with exceptions when imbalances occur in the recursive partitioning.
We refer to basis vectors with tag $l = 0$ as \emph{scaling vectors},
those with tag $l = 1$ as \emph{Haar vectors},
and those with tag $l \geq 2$ as \emph{Walsh vectors}.

The GHWT begins by defining an orthonormal basis on level $\jmax$ and obtaining
the corresponding expansion coefficients. The standard basis of $\Rf^N$ is used
here since each region at level $\jmax$ is a 1-node graph:
$\bpsi^{\jmax}_{k,0} \define \bm{1}_{V(G^{\jmax}_k)} \in \Rf^N$, where $k \in [0,N)$,
  $N^{\jmax}_k = 1$, and $\bm{1}_i$ is the indicator vector of node $i$,
  i.e., $\bm{1}_i[m] = 0$ if $m \neq i$ and $\bm{1}_i[i]=1$.
  The expansion coefficients $\{ d^{\jmax}_{k,0} \}_{k=0:N-1}$ are then simply
  the reordered input signal $\bm{f}$. From here the algorithm proceeds
  recursively, and the basis vectors and expansion coefficients on level $j-1$
  are computed from those on level $j$.
  The GHWT proceeds as in Algorithm~\ref{alg:ghwt}.

\begin{algorithm}
\SetAlgoLined
\DontPrintSemicolon 
\SetKwData{bvecq}{bvecq}
\SetKwData{true}{true}
\SetKwData{false}{false}
\KwIn{A binary partition tree $\{G^j_k\}$ of the graph $G$, $0 \leq j \leq \jmax$ and $0 \leq k < K_j$; Input graph signal $\bm{f}$ supported on $V(G)$;
  a flag \bvecq (default: \false) to compute the GHWT basis vectors}
\KwOut{The set of expansion coefficients $\{d^j_{k,l}\}$ of signal $\bm{f}$ on
  the overcomplete GHWT dictionary vectors $\{\bpsi^j_{k,l}\}$, which are
  also returned if \bvecq is set to \true}

\BlankLine

\tcp{The algorithm starts here.}

\For(\tcp*[h]{Basis vectors on level $\jmax$ are the standard unit vectors of $\Rf^N$}){$k = 0:N-1$}{$d_{k,0}^\jmax \leftarrow \inner{\bm{f}}{\bm{1}_{V(G_k^\jmax)}}$
  \tcp*[h]{$\inner{\cdot}{\cdot}$ is the standard inner product in $\Rf^N$}

  \lIf{\bvecq $\eqeq$ \true}{$\bpsi_{k,0}^\jmax \leftarrow \bm{1}_{V(G_k^\jmax)} \in \Rf^N$}
}

\For(\tcp*[h]{Compute coefficients/basis vectors on level $j-1$ from level $j$}){$j = \jmax:-1:1$}{
  \For{$k = 0:K_{j-1}-1$}{
    \tcp*[h]{Compute the scaling coeff's/vectors}
    $d^{j-1}_{k,0} \leftarrow \inner{\bm{f}}{\bm{1}_{V(G^{j-1}_k)}/\sqrt{N_k^{j-1}}}$

    \lIf{\bvecq $\eqeq$ \true}{$\bpsi^{j-1}_{k,0} \leftarrow \bm{1}_{V(G^{j-1}_k)}/\sqrt{N_k^{j-1}}$}

    \tcp*[h]{Below, we assume $V(G^{j-1}_k) = V(G^j_{k'}) \sqcup V(G^j_{k'+1})$}

    \If(\tcp*[h]{Compute the Haar coeff's/vectors}){$N_k^{j-1} > 1$}{
      $d^{j-1}_{k,1} \leftarrow \frac{N^j_{k'+1}\sqrt{N^j_{k'}} d^j_{k',0} - N^j_{k'}\sqrt{N^j_{k'+1}}d^j_{k'+1,0}}{\sqrt{N^j_{k'}(N^j_{k'+1})^2 + N^j_{k'+1}(N^j_{k'})^2}}$

      \lIf{\bvecq $\eqeq$ \true}{
        $\bpsi^{j-1}_{k,1} \leftarrow \frac{N^j_{k'+1}\sqrt{N^j_{k'}} \bpsi^j_{k',0} - N^j_{k'}\sqrt{N^j_{k'+1}}\bpsi^j_{k'+1,0}}{\sqrt{N^j_{k'}(N^j_{k'+1})^2 + N^j_{k'+1}(N^j_{k'})^2}}$}
          
    }
    
    \If(\tcp*[h]{Compute the Walsh coeff's/vectors}){$N^{j-1}_k > 2$}{
      \For{$l = 1:2^{\jmax-j} - 1$}{
        \uIf{both subregions $k'$ and $k'+1$ have a basis vector with tag $l$}{
          $d^{j-1}_{k,2l} \leftarrow (d^j_{k',l} + d^j_{k'+1,l})/\sqrt{2}$\;
          $d^{j-1}_{k,2l+1} \leftarrow (d^j_{k',l} - d^j_{k'+1,l})/\sqrt{2}$
          
          \If{\bvecq $\eqeq$ \true}{
            $\bpsi^{j-1}_{k,2l} \leftarrow (\bpsi^j_{k',l} + \bpsi^j_{k'+1,l})/\sqrt{2}$\;
            $\bpsi^{j-1}_{k,2l+1} \leftarrow (\bpsi^j_{k',l} - \bpsi^j_{k'+1,l})/\sqrt{2}$}
          }
        
        \uElseIf{(without loss of generality) only subregion $k'$ has a basis vector with tag $l$}{
          $d^{j-1}_{k,2l} \leftarrow d^j_{k',l}$
        
          \lIf{\bvecq $\eqeq$ \true}{$\bpsi^{j-1}_{k,2l} \leftarrow \bpsi^j_{k',l}$}
          }
        
        \Else{Do nothing \tcp*[h]{Neither of subregions has a basis vector with tag $l$}}
        
        }}}}
\caption{Generating expansion coefficients relative to the basis vectors in the
GHWT basis dictionary~\cite{IRION-SAITO-GHWT, IRION-SAITO-SPIE, IRION-SAITO-TSIPN}}
\label{alg:ghwt}
\end{algorithm}
Note that when analyzing the input signal $\bm{f}$, we only need to compute
the expansion coefficients without generating the basis vectors in
Algorithm \ref{alg:ghwt} in general.

\begin{figure}
  \begin{subfigure}{0.475\textwidth}
    \centering\includegraphics[width=\textwidth]{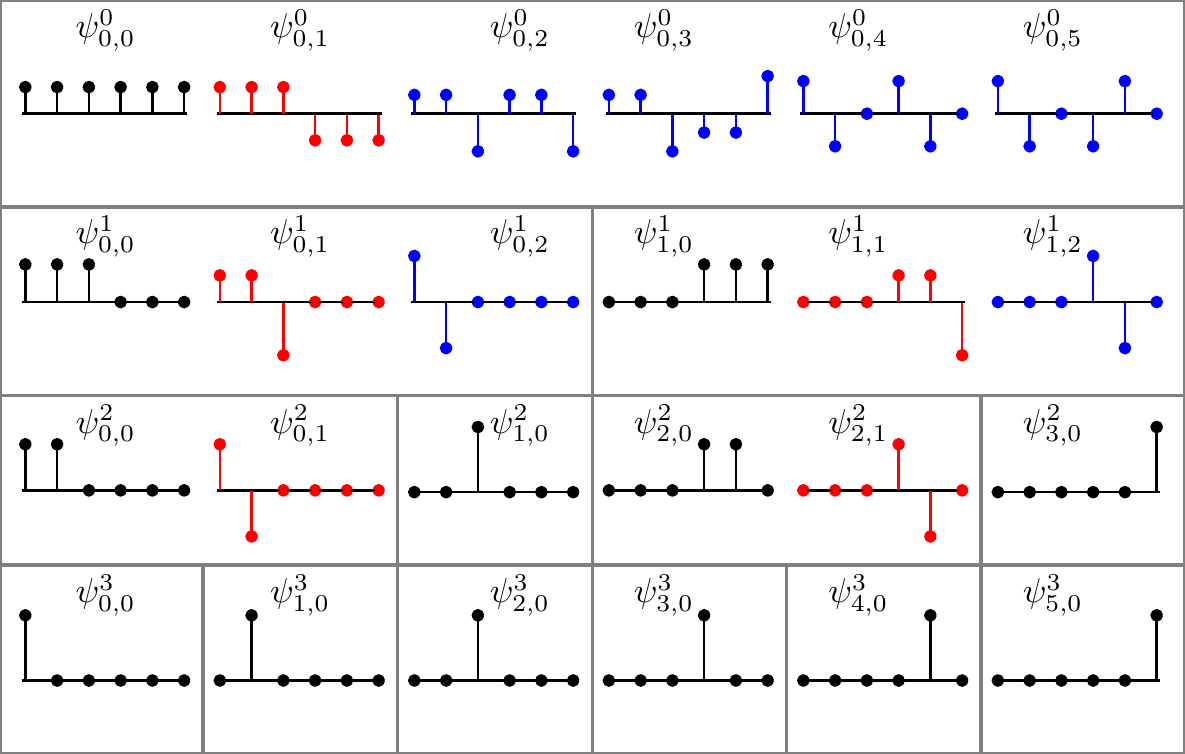}
    \caption{The GHWT c2f dictionary}
  \end{subfigure}
  \hspace{1em}
  \begin{subfigure}{0.475\textwidth}
    \centering\includegraphics[width=\textwidth]{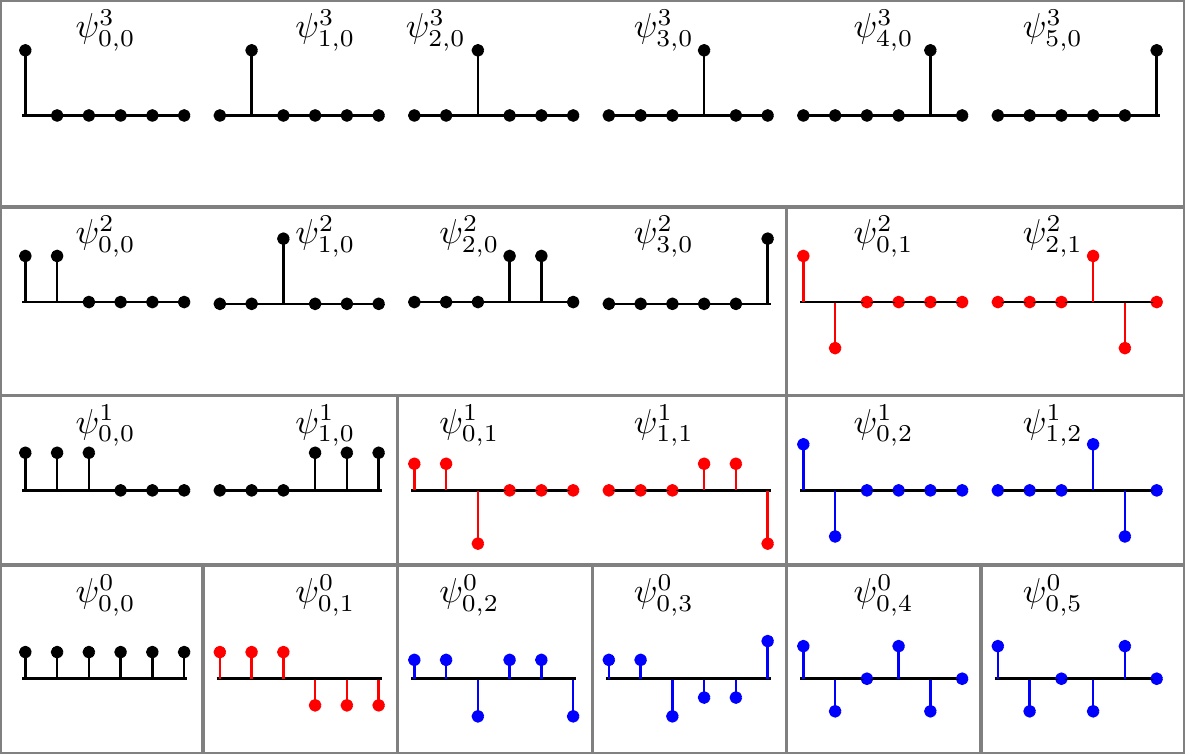}
  \caption{The GHWT f2c dictionary}
  \end{subfigure}
  \caption{The GHWT c2f and f2c dictionaries on $P_6$.
    Stem plots with black, red, blue colors correspond to the scaling,
    Haar, and Walsh vectors, respectively.}
\label{alg:ghwt_dict}
\end{figure}

\begin{figure}
  \begin{subfigure}{0.32\textwidth}
    \centering
    \includegraphics[width=\textwidth]{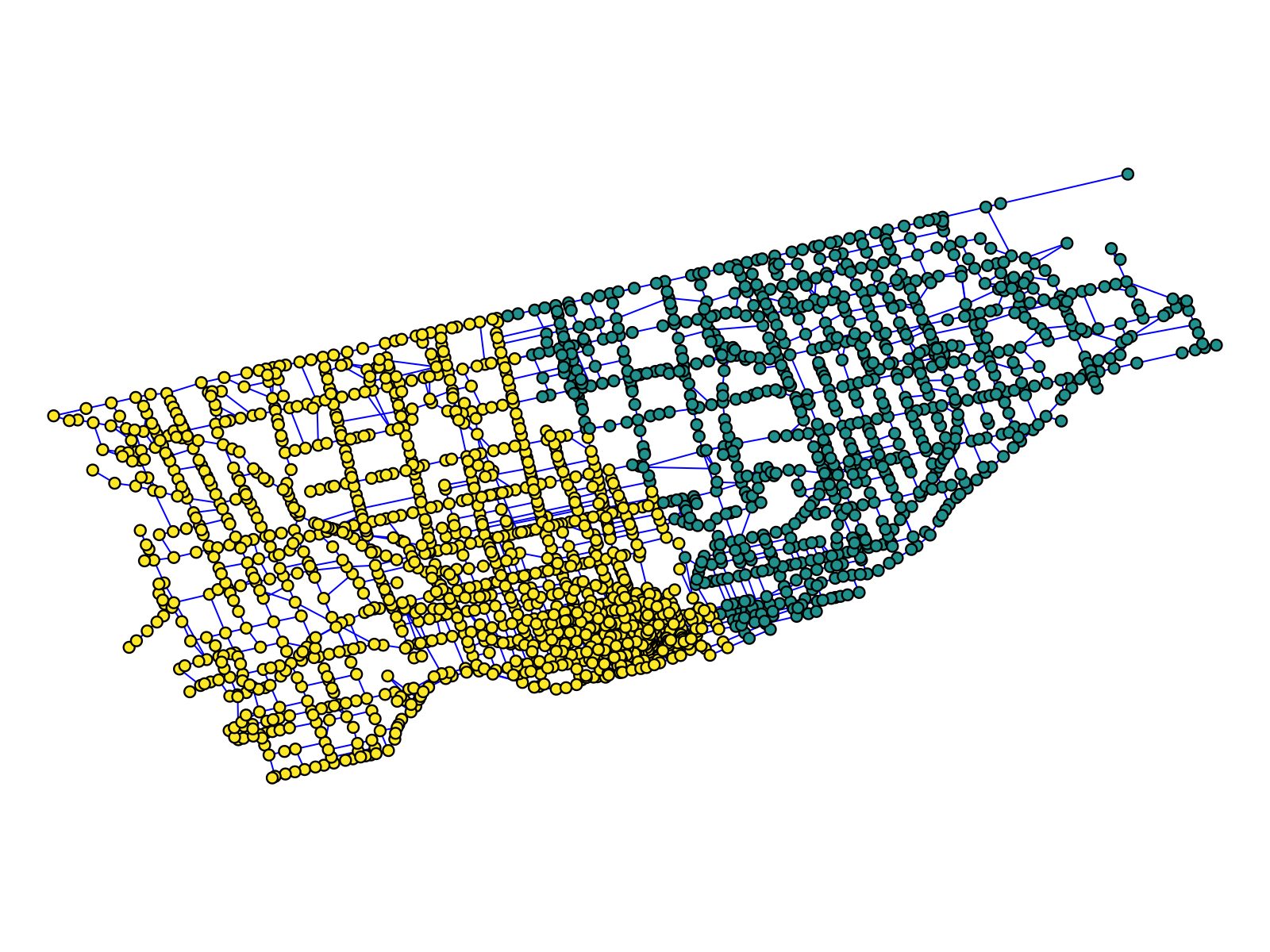}
    \caption{Scaling vector $\bpsi^1_{0,0}$}
  \end{subfigure}
  \begin{subfigure}{0.32\textwidth}
    \centering
    \includegraphics[width=\textwidth]{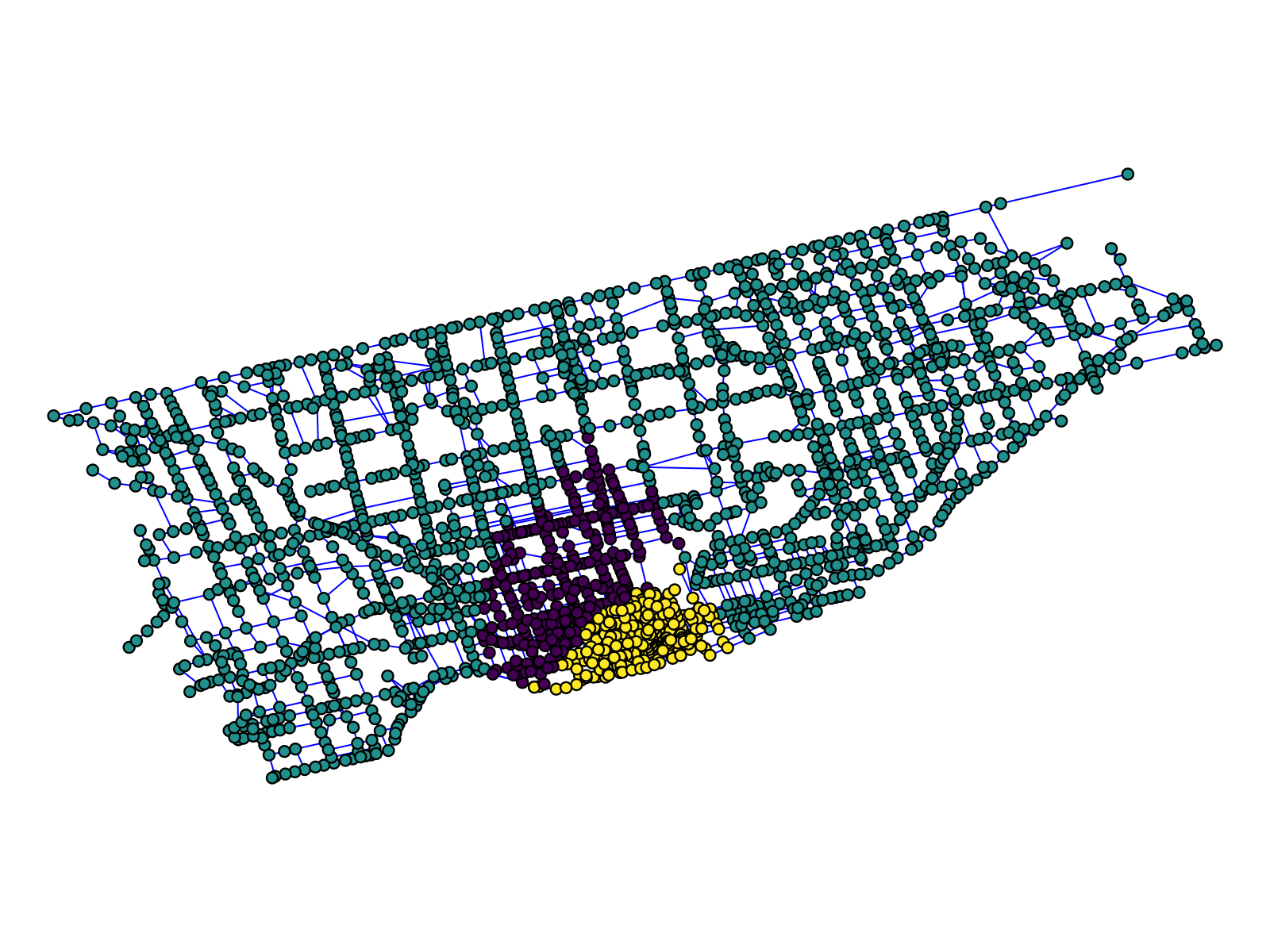}
    \caption{Haar vector $\bpsi^2_{0,1}$}
  \end{subfigure}
  \begin{subfigure}{0.32\textwidth}
    \centering
    \includegraphics[width=\textwidth]{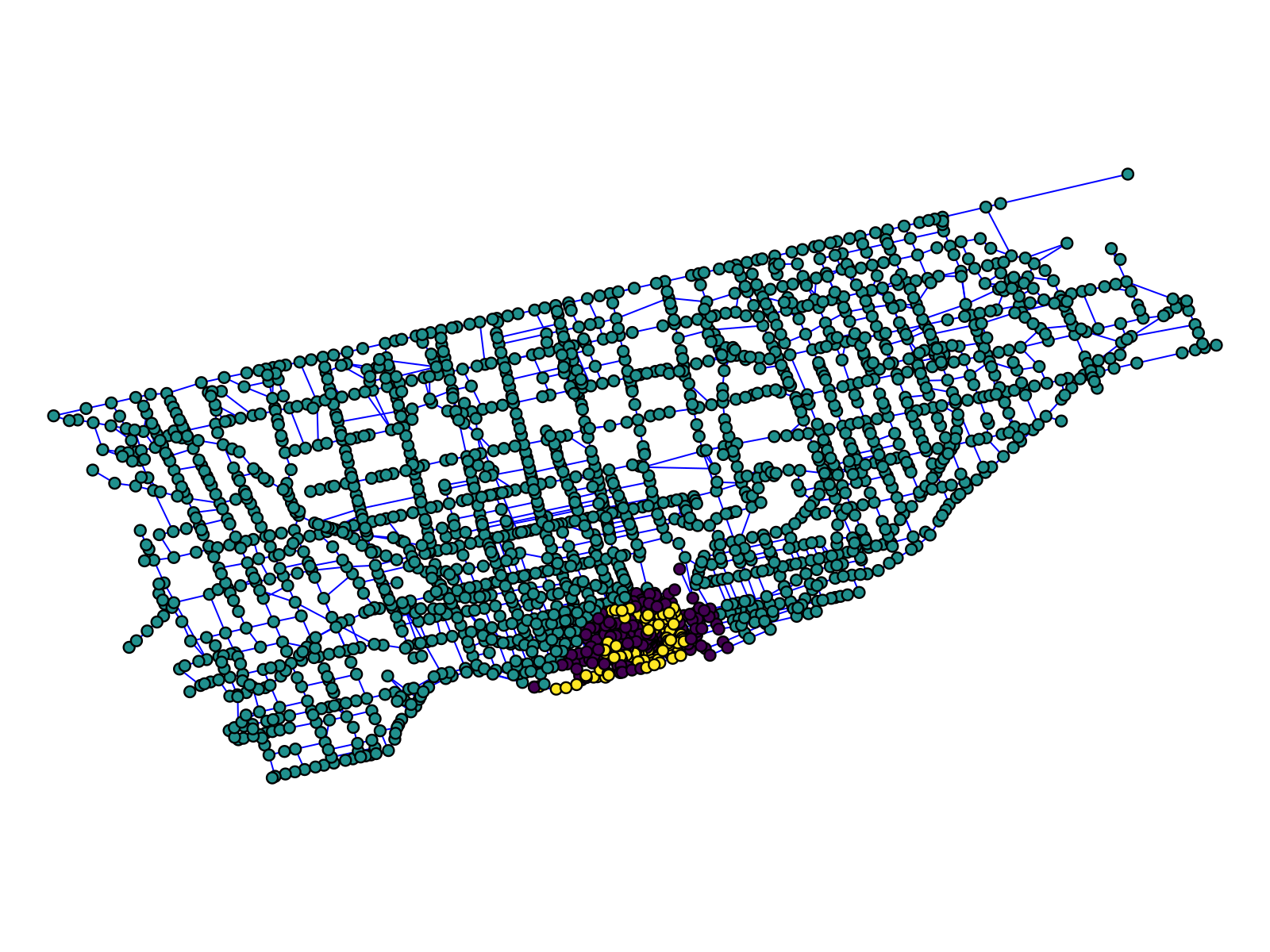}
    \caption{Walsh vector $\bpsi^3_{0,9}$}
  \end{subfigure}
  \caption{Examples of basis vectors from the GHWT dictionary computed on
    Toronto street network. The color scheme called \emph{viridis} \cite{VIRIDIS}
    is used to represent the amplitude of these vectors ranging from deep violet
    (negative) to dark green (zero) to yellow (positive).}
  \label{fig:example_vector_toronto}
\end{figure}

For the dictionary of basis vectors, several remarks are in order.
\begin{enumerate}
\item The basis vectors on each level are localized. In other words,
  $\bpsi^j_{k,l}$ is supported on $V(G^j_k)$.
  If $V(G^j_k) \cap V(G^{j'}_{k'}) = \emptyset$, then the basis vectors
  $\{ \bpsi^j_{k,l} \}_l$ and $\{ \bpsi^{j'}_{k',l'}\}_{l'}$ are mutually orthogonal.
\item The basis vectors on $V(G^j_k)$ span the same subspace as the union of
  those on $V(G^{j+1}_{k'})$ and $V(G^{j+1}_{k'+1})$, where
  $V(G^j_k) = V(G^{j+1}_{k'}) \sqcup V(G^{j+1}_{k'+1})$.
\item The depth of the dictionary is the same as the binary partition tree,
which is approximately $O(\log N)$ if the tree is nearly balanced.
There are $N$ vectors on each level, so the total number of basis vectors
is approximately $O(N \log N)$. 
\end{enumerate}
Note that Algorithm~\ref{alg:ghwt} groups the GHWT basis vectors by region
(i.e., the index $k$) and arranges them from the coarse scale to the fine scale,
which we call the \emph{coarse-to-fine} (c2f) dictionary.
Alternatively, we can group them by tag (i.e., the index $l$) and reverse
the order of the levels (i.e., scales), which we call the \emph{fine-to-coarse}
(f2c) dictionary~\cite{IRION-SAITO-GHWT, IRION-SAITO-SPIE, IRION-SAITO-TSIPN}.
The c2f dictionary corresponds to a collection of basis vectors by
recursively partitioning the ``time'' domain information of the input graph
signal while the f2c dictionary corresponds to those by recursively
partitioning the ``frequency'' (or ``sequency'') domain information of the
input graph signal.
Each dictionary contains more than $(1.5)^N$ choosable ONBs; see, e.g.,
Thiele and Villemoes~\cite{THIELE-VILLEMOES} for the details on
how to compute or estimate this number.
Note, however, that exceptions can occur when the recursive partitioning
generates a highly imbalanced tree.
Figure~\ref{alg:ghwt_dict} shows these dictionaries for $P_6$.
Figure~\ref{fig:example_vector_toronto} shows some basis vectors from the
GHWT dictionary computed on the Toronto street network.

\subsection{The Best-Basis Algorithm in the GHWT}
To select an ONB from a dictionary of wavelet packets that is ``best'' for
approximation/compression, we typically use the so-called
\emph{best-basis algorithm}. The one used for the GHWT in
\cite{IRION-SAITO-GHWT, IRION-SAITO-SPIE, IRION-SAITO-TSIPN} was
a straightforward generalization of the Coifman-Wickerhauser
algorithm~\cite{COIF-WICK}, which was developed for non-graph signals
of dyadic length. This algorithm requires a real-valued cost function $\Jcal$
measuring the approximation/compression inefficiency of the subspaces in the
dictionary, and aims to find an ONB whose coefficients minimize $\Jcal$,
(i.e., the most efficient ONB for approximating/compressing an input signal),
which we refer to as the ``best basis'' chosen from the GHWT dictionary.
The algorithm initiates the best basis as the whole set of vectors at the bottom
level of the dictionary. Then it proceeds upwards, comparing the cost of the
expansion coefficients corresponding to two children subgraphs to the cost
of those of their parent subgraph. The best basis is updated if the cost of
the parent subgraph is smaller than that of its children subgraphs.
The algorithm continues until it reaches the top (i.e., the root) of the binary
partition tree (i.e., the dictionary).

The best-basis algorithm works as long as $\Jcal$ is nonnegative and
additive~\footnote{The additivity property can be dropped in principle by
following the work of Saito and Coifman on the local regression basis~\cite{SAITO-COIF-SONIC}} of the form $\Jcal(\bm{d}) \define \sum_i g(d_i)$ with
$g: \Rf \to \Rf_{\geq 0}$, where $\bm{d}$ is the expansion coefficients of an
input graph signal on a region.
For example, if one wants to promote sparsity in graph signal representation or
approximation, $\Jcal(\bm{d})$ function can be chosen as: either the $p$th
power of $\ell^p$-(quasi)norm $\sum_i |d_i|^p$ for $0 < p < 2$ or the
$\ell^0$-pseudonorm $| \{ d_i \cond d_i \neq 0 \} |$. Note that the smaller the
value of $p$ is, the more emphasis in sparsity is placed.

Note that one can search the c2f and f2c dictionaries separately to obtain two
sets of the best bases, among which the one with smaller cost is chosen
as the final best basis of the GHWT dictionaries.
We also note here that the graph Haar basis is selectable \emph{only in the GHWT
f2c dictionary} while the graph Walsh basis is selectable in either dictionary.

\section{The Extended GHWT (eGHWT)}
\label{sec:eghwt}
In this section, we describe the \emph{extended} GHWT (eGHWT):
our new best-basis algorithm on the GHWT dictionaries, which
\emph{simultaneously} considers the ``time'' domain split \emph{and}
``frequency'' (or ``sequency'') domain split of an input graph signal.
This transform will allow us to deploy the modified best-basis algorithm that
can select the best ONB for one's task (e.g., efficient approximation, denoising,
etc.) among a much larger set ($> 0.618 \cdot (1.84)^N$) of ONBs than the GHWT
c2f/f2c dictionaries could provide ($> (1.5)^N$). The previous best-basis
algorithm only searches through the c2f dictionary and f2c dictionary separately,
but this new method makes use of those two dictionaries \emph{simultaneously}.
In fact, the performance of the eGHWT, by its construction, is
always superior to that of the GHWT, which is clearly demonstrated
in our numerical experiments in Section~\ref{sec:appl}.

Our eGHWT is the graph version of the Thiele-Villemoes algorithm~\cite{THIELE-VILLEMOES} that finds the best basis among the ONBs of $\Rf^N$ consisting of
discretized and rescaled Walsh functions for an 1D non-graph signal (i.e.,
a signal discretized on a 1D regular grid) of length $N$, where $N$ must be
a dyadic integer. Their algorithm operates in the time-frequency plane
and constructs its tiling with minimal cost among all possible tilings
with dyadic rectangles of area one, which clearly depends on the input signal.
Here we adapt their method to our graph setting that does not require dyadic $N$.
In addition, the generalization of the Thiele-Villemoes algorithm for 2D signals
developed by Lindberg and Villemoes~\cite{LINDBERG-VILLEMOES} can be generalized
to the 2D eGHWT, as we will discuss more in Sections~\ref{sec:2D-eGHWT} and
\ref{sec:2D-eGHWT-example}.

\subsection{Fast Adaptive Time-Frequency Tilings}
\label{sec:intro_tf}
In this subsection, we briefly review the Thiele-Villemoes
algorithm~\cite{THIELE-VILLEMOES}. First of all, let us define the
so-called \emph{Walsh system}, which forms an ONB for $L^2[0,1)$.
Let $W_0(t) = 1$ for $0\leq t <1$ and zero elsewhere, and define
$W_1, W_2, \ldots$ recursively by
\begin{equation}
\begin{split}
W_{2l}(t) &= W_l(2t) + (-1)^l W_l(2t-1),\\
W_{2l+1}(t) &= W_l(2t) - (-1)^l W_l(2t-1).
\end{split}
\end{equation}
Then $\{W_l\}^{\infty}_{l=0}$ is an ONB for $L^2[0,1)$ and is referred to as
\emph{the Walsh system in sequency order}.
Each basis function, $W_l(t)$, is piecewise equal to either $1$ or $-1$ on
$[0, 1)$. Note that the scaling and Haar vectors at the global scale are included
in this Walsh system.

Viewing $S \define [0,1) \times [0,\infty)$ as a time-frequency plane, the
\emph{tile} corresponding to the \emph{rescaled} and \emph{translated}
Walsh function (which we also refer to as the \emph{Haar-Walsh function}),
$$w_p(t) \define 2^{j/2} W_l(2^j t-k),$$ 
is defined as
\begin{equation}
 p = p(j,k,l) \define [ 2^{-j}k, 2^{-j}(k+1) ) \times [ 2^{j}l, 2^{j}(l+1) ).
\end{equation}
Note that the area of $p$ is $1$.
Thiele and Villemoes showed that the functions $w_p$ and $w_q$ are orthogonal
if and only if the tiles $p$ and $q$ are disjoint. Moreover, if a collection of
tiles form a disjoint covering of a given dyadic rectangle (i.e., a rectangle
with dyadic sides) in the time-frequency plane, then the Haar-Walsh functions of
those tiles form an ONB of the subspace corresponding to that dyadic rectangle.

Now the 1D discrete signal space $\Rf^N\ (N = 2^n)$ can be identified with the
subset $S_n \define [0,1) \times [0, N)$ of $S$ in the time-frequency plane.
Given the overcomplete dictionary of Haar-Walsh functions on $\Rf^N$,
the best-basis algorithm now is equivalent to finding a set of basis vectors
for a given input signal with minimal cost that also generates a disjoint tiling
of $S_n$.
That tiling is called the \emph{minimizing tiling}.
\begin{lemma}[Thiele \& Villemoes~\cite{THIELE-VILLEMOES}]
Let $T \subset S$ be a rectangle of area greater or equal to two, with left
half $L$, right half $R$, lower half $D$, and upper half $U$.
Assume each tile $p \subset T$ has the cost $c(p)$. Define
$$ m_T \define \min\left\{\sum_{p\in\mathcal{B}}c(p) \, \Bigg| \, \mathcal{B} \text{ is a disjoint covering of } T \right\},$$
and similarly $m_L$,$m_R$,$m_D$,$m_U$. Then
$$m_T = \min\{m_L + m_R, m_D + m_U\}. $$
\end{lemma}
This lemma tells us that the minimizing tiling of $T$ can be split either in
the time-direction into two minimizing tilings of $L$ and $R$, or in the
frequency-direction into those of $D$ and $U$. It enables a dynamic programming
algorithm to find the minimizing tiling of $S_n$; see Algorithm~3.3 of
\cite{THIELE-VILLEMOES}.

\subsection{Relabeling Region Indices}
If the input graph is a simple path graph $P_N$ with dyadic $N$ and
the partition tree is a balanced complete binary tree, then the GHWT dictionary
is the same as the classical Haar-Walsh wavelet packet dictionary for
1D regular signals, on which the Thiele-Villemoes algorithm\cite{THIELE-VILLEMOES}
can be applied in a straightforward manner. To adapt the algorithm to a graph
signal of an arbitrary length or an imperfect binary partition tree
of an input graph, we need to modify the GHWT dictionary first. 

Specifically, the region index $k$ of $G^j_k$ and $\bpsi^j_{k,l}$ needs to be
\emph{relabeled}.
Previously, on level $j$, the region index $k$ takes all the integer values in
$[0, K_j)$ where $K_j \leq 2^j$ is the total number of subgraphs (or regions) on
level $j$. After relabeling, $k$ takes an integer value in $[0, 2^j)$ according
to its location in the binary tree. Algorithm~\ref{alg:ghwt_modif} precisely
describes the whole procedure.
\begin{algorithm}
  \KwIn{A binary partition tree denoted by $\{G^j_k\}$, $0 \leq j \leq \jmax$, $0 \leq k < K_j$}
  \KwOut{A perfect binary partition tree $\{\widetilde{G}^j_k\}$ containing $\{G^j_k\}$ as its subset}
\SetAlgoLined

\BlankLine

\tcp{The algorithm starts here.}

$\widetilde{G}^0_0 \leftarrow G^0_0$ 
\tcp*[h]{On level $0$, there is only one region $G^0_0$, so no relabeling is required.}

\For{$j = 1:\jmax$}{
\For{$k = 0:K_{j-1}-1$}{
  \eIf{$G^{j-1}_k$ is split into $G^j_{k'}$ and $G^j_{k'+1}$}
      {$\widetilde{G}^j_{2k} \leftarrow G^j_{k'}; \quad
        \widetilde{G}^j_{2k+1} \leftarrow G^j_{k'+1}$}(\tcp*[h]{$G^{j-1}_k$ is kept as $G^j_{k'}$}){$\widetilde{G}^j_{2k} \leftarrow G^j_{k'}; \quad
        \widetilde{G}^j_{2k+1} \leftarrow \emptyset$}
}
}
\caption{Relabeling the GHWT Dictionary}
\label{alg:ghwt_modif}
\end{algorithm}
A couple of remarks are in order.
First, the region indices of the basis vectors $\{\bpsi^j_{k,l}\}$,
$0 \leq k < K_j$ are also relabeled accordingly as $\{\widetilde{\bpsi}^j_{k',l}\}$ supported on the subgraph $\widetilde{G}^j_{k'}$, where $0 \leq k' < 2^j$.
Note that some of the basis vectors in $\{\widetilde{\bpsi}^j_{k',l}\}$ that do
not have the corresponding basis vectors in the original GHWT dictionary
$\{\bpsi^j_{k,l}\}$ are ``fictitious'' (or ``non-existent'') ones and can be set
as the zero vectors. In practice, however, we even do not need to store them
as the zero vectors; we simply do not compute the cost corresponds to such
fictitious basis vectors. 
Second, to simplify our notation, we just assume the $\{\bpsi^j_{k,l}\}$ and
$\{G^j_k\}$ are those already relabeled by Algorithm~\ref{alg:ghwt_modif}
in the rest of our article.

\begin{figure}
  \begin{subfigure}{0.475\textwidth}
    \centering\includegraphics[width=\textwidth]{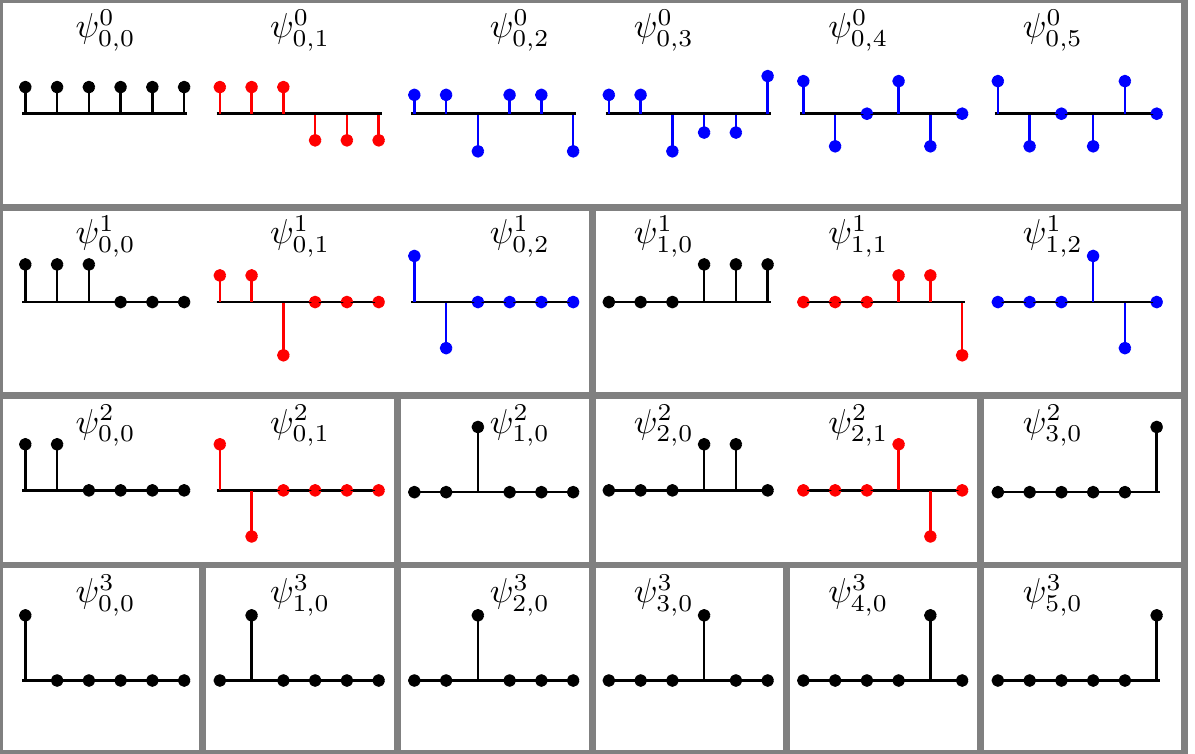}
    \caption{The GHWT c2f dictionary}
    \label{fig:c2f-orig}
  \end{subfigure} 
\hspace{1em}  
  \begin{subfigure}{0.475\textwidth}
    \centering\includegraphics[width=\textwidth]{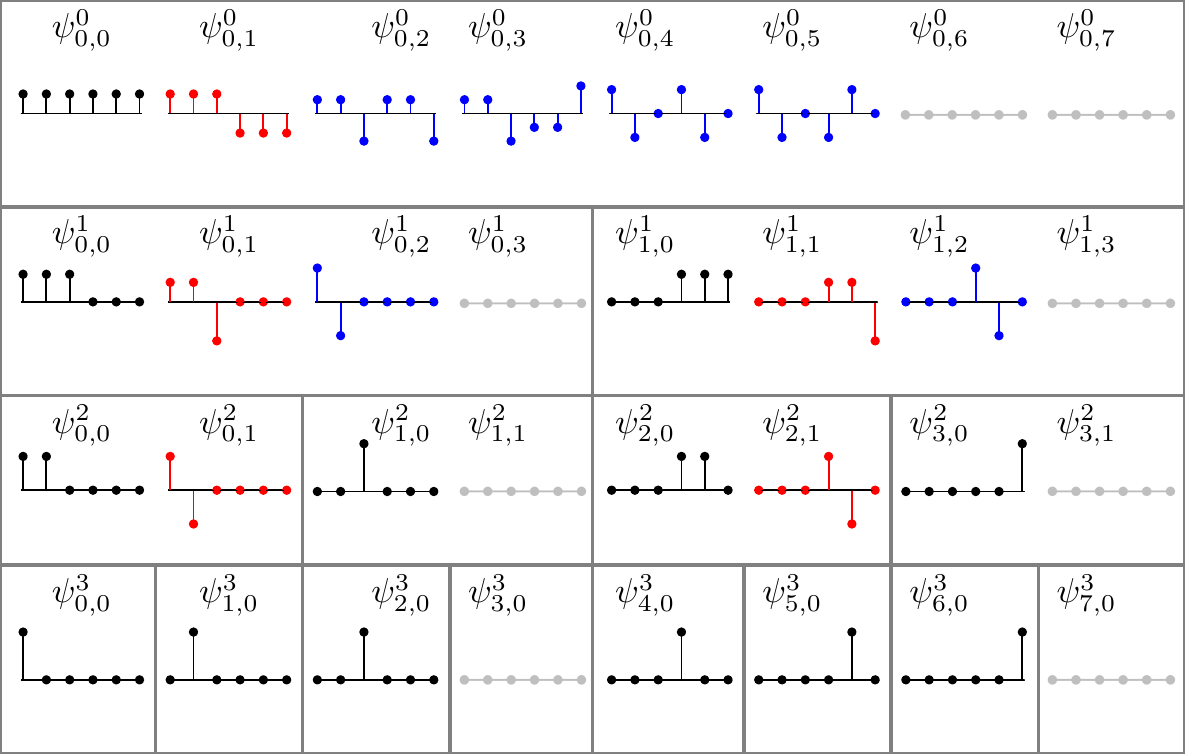}
  \caption{The relabeled GHWT c2f dictionary}
  \label{fig:c2f-relabeled}
  \end{subfigure}
  \caption{(a) The GHWT c2f dictionary on $P_6$.
    Stem plots with black, red, and blue colors correspond to the scaling,
    Haar, and Walsh vectors, respectively. (b) The relabeled GHWT c2f
    dictionary by Algorithm~\ref{alg:ghwt_modif} applied to the GHWT c2f
    dictionary shown in (a). The gray stem plots indicate the ``fictitious'' (or
    ``non-existent'') basis vectors newly generated by
    Algorithm~\ref{alg:ghwt_modif}.}
\end{figure}

Figure~\ref{fig:c2f-relabeled} shows the result of Algorithm~\ref{alg:ghwt_modif}
applied to the GHWT c2f dictionary shown in Fig.~\ref{fig:c2f-orig} on
$P_6$. Before the relabeling, the dictionary forms a complete but imperfect
binary tree. As one can see, after the relabeling,
the initial GHWT c2f dictionary is a subset of a perfect binary tree
shown in Fig.~\ref{fig:c2f-relabeled}.

\subsection{The New Best-Basis (eGHWT) Algorithm}
We can now apply Algorithm~\ref{alg:eghwt} to search for the best basis in the
relabeled GHWT dictionary that have a perfect binary partition tree,
similarly to the Thiele-Villemoes algorithm.
For simplicity, we refer to this whole procedure as the \emph{eGHWT}.
In order to understand this algorithm, let us first remark that
we use the so-called \emph{associative array}: an abstract data type
composed of a collection of (key, value) pairs such that each possible key
appears at most once in the collection.
The reason why we use the associative arrays instead of the regular arrays is
to save storage space while keeping the algorithm flexible and efficient
without losing the convenience of manipulating arrays.
This point is important since many basis vectors $\bpsi^j_{k,l}$ after relabeling
via Algorithm~\ref{alg:ghwt_modif} may be fictitious, which we need to neither
store nor compute: using regular matrices to store them will be wasteful.
For example, Algorithm~\ref{alg:eghwt} has a statement like
$A_0[j,k,l] \leftarrow g(\linner{\f}{\bpsi^j_{k,l}})$, where
$A_0$ is an associative array. 
This actually means that $((j,k,l), g(\linner{\f}{\bpsi^j_{k,l}}))$ is a pair of
(key, value) of the associative array $A_0$. Here we write $(j,k,l) \in A_0$ to
denote that $(j,k,l)$ is a valid key of $A_0$. Therefore, $(j,k,l) \in A_0$ if
and only if non-fictitious $\bpsi^j_{k,l}$ exists. Since we relabeled
$\bpsi^j_{k,l}$, there are fictitious $\bpsi^j_{k,l}$'s (i.e., zero vectors)
for some triple $(j,k,l)$. In that case, we write $(j,k,l) \notin A_0$.

\begin{algorithm}
\SetAlgoLined
\KwIn{The relabeled GHWT basis vectors $\{\bpsi^j_{k,l}\}$ generated
  by Algorithm~\ref{alg:ghwt_modif};
  a nonnegative function $g: \Rf \to \Rf_{\geq 0}$ to evaluate the cost
  of each expansion coefficient; an input graph signal $\f \in \Rf^N$.}
\KwOut{The eGHWT best basis $B$.}

\BlankLine

\tcp{Initialize two associative arrays $A_0$ and $I_0$
  that store costs and indices for the best basis search,
  respectively}

$A_0[j,k,l] \leftarrow g(\langle \f, \bpsi^j_{k,l} \rangle)$
if $\bpsi^j_{k,l}$ is \emph{not} fictitious;

$I_0[j,k,l] \leftarrow 1$ if $(j,k,l) \in A_0$\;

\For(\tcp*[h]{Compute $(A_{m+1}, I_{m+1})$ from $(A_m, I_m)$}){$m = 0:\jmax - 1$}{
Initialize $A_{m+1}$ and $I_{m+1}$ as empty associative arrays\;

\For{each $(j,k,l) \in A_m$ with $j < \jmax-m$ and $l\equiv 0 \pmod{2}$}{
    \eIf{$(A_m[j,k,l] + A_m[j,k,l+1]) \leq (A_m[j+1,2k,l/2] + A_m[j+1,2k+1,l/2])$}
    {$I_{m+1}[j,k,l/2]\leftarrow 0$\;
    $A_{m+1}[j,k,l/2]\leftarrow (A_m[j,k,l] + A_m[j,k,l+1])$\;}
    {$I_{m+1}[j,k,l/2]\leftarrow 1$\;
    $A_{m+1}[j,k,l/2]\leftarrow (A_m[j+1,2k,l/2] + A_m[j+1,2k+1,l/2])$\;}
    \tcp{Note that any of $A_m[j,k,l+1]$, $A_m[j+1,2k,l/2]$, $A_m[j+1,2k+1,l/2]$
      will be replaced by $0$ in the above steps if it corresponds to
      the fictitious basis vector.}
}
}
$I \leftarrow I_\jmax$\;
\For(\tcp*[f]{Recover the best basis from $\{I_m\}$}){$m = \jmax - 1:-1:0$}{
Initialize $I_{\mathrm{temp}}$ as an empty associative array\;
\For{each $(j,k,l) \in I$}{
\eIf{$I[j,k,l] \eqeq 0$}
{$I_{\mathrm{temp}}[j,k,2l] \leftarrow I_m[j,k,2l]$ if $(j,k,2l) \in I_m$\;$I_{\mathrm{temp}}[j,k,2l+1] \leftarrow I_m[j,k,2l+1]$ if $(j,k,2l+1) \in I_m$\;}
(\tcp*[h]{i.e., $I[j,k,l] \eqeq 1$})
{$I_{\mathrm{temp}}[j+1,2k,l] \leftarrow I_m[j+1,2k,l]$ if $(j+1,2k,l) \in I_m$\;$I_{\mathrm{temp}}[j+1,2k+1,l] \leftarrow I_m[j+1,2k+1,l]$ if $(j+1,2k+1,l) \in I_m$\;}
}
$I \leftarrow I_{\mathrm{temp}}$\;
}
$B \leftarrow \emptyset$; \tcp*[h]{Initialize $B$ as an empty set}\\
\For{each $(j,k,l) \in I$}{$B \leftarrow B \cup \{ \bpsi^j_{k,l} \}$}

\tcp{$B$ is the best basis!}

\caption{The New Best-Basis (eGHWT) Algorithm}
\label{alg:eghwt}
\end{algorithm}

\begin{hide}
\footnotetext[1]{An \emph{associative array} is an abstract data type
  composed of a collection of (key, value) pairs such that each possible key
  appears at most once in the collection.}
\footnotetext[2]{$((j,k,l), g(\langle \bpsi^j_{k,l}, \f \rangle))$ is a pair of
  (key, value) of the associative array $A_0$. Here we use $(j,k,l) \in A_0$ to
  denote that $(j,k,l)$ is a valid key of $A_0$. Therefore, $(j,k,l) \in A_0$ if
  and only if $\bpsi^j_{k,l}$ exists. Since we relabeled $\bpsi^j_{k,l}$, there is
  no corresponding $\bpsi^j_{k,l}$ for some triple $(j,k,l)$. In that case,
  $(j,k,l) \notin A_0$.}
\end{hide}

Several remarks on this algorithm are in order:
\begin{itemize}
\item The associative array, $A_m$, holds the minimum cost of ONBs
  in a subspace. The value of $A_m[j,k,l]$ is set to the smaller value
  between $A_{m-1}[j,k,2l] + A_{m-1}[j,k,2l+1]$ and
  $A_{m-1}[j+1,2k,l] + A_{m-1}[j+1,2k+1,l]$.
  The subspace corresponding to $A_m[j,k,l]$ is the direct sum of
  the two subspaces corresponding to $A_{m-1}[j,k,2l]$ and
  $A_{m-1}[j,k,2l+1]$, which is the same as the direct sum of those corresponding
  to $A_{m-1}[j+1,2k,l]$ and $A_{m-1}[j+1,2k+1,l]$.
  In other words, when we compute $A_{m}$ from $A_{m-1}$, we concatenate the
  subspaces. This process is similar to finding the best tilings
  for dyadic rectangles from those with half size in Thiele-Villemoes algorithm
  \cite{THIELE-VILLEMOES} as described in Sect.~\ref{sec:intro_tf}.

\item If the input graph $G$ is a simple 1D path $P_N$ with dyadic $N$, and
  if we view an input graph signal $\f$ as a discrete-time signal, then
  $(A_{m-1}[j,k,2l], A_{m-1}[j,k,2l+1])$ corresponds to splitting the subspace of
  $A_m[j,k,l]$ in the frequency domain in the time-frequency plane
  while $(A_{m-1}[j+1,2k,l], A_{m-1}[j+1,2k+1,l])$ does the split in the time
  domain.

\item The subspace of each entry in $A_0$ is one dimensional since it is
  spanned by a single basis vector. In other words, $A_0[j,k,l]$ corresponds to
  $\Span\{\bpsi^j_{k,l}\}$: the one-dimensional subspace spanned by
  $\bpsi^j_{k,l}$.

\item $A_\jmax$ has only one entry $A_\jmax[0,0,0]$, which corresponds to the
  whole $\Rf^N$. Its value is the minimum cost among all the choosable ONBs,
  i.e., the cost of the best basis.

\item If an input graph signal is of dyadic length, Algorithm~\ref{alg:eghwt}
  selects the best basis among a much larger set ($> 0.618 \cdot (1.84)^N$)
  of ONBs than what each of the GHWT c2f and f2c dictionaries would provide
  ($> (1.5)^N$)~\cite{THIELE-VILLEMOES}. 
  The numbers are similar even for non-dyadic cases as long as the partition
  trees are essentially balanced.
  The essence of this algorithm is that at each step of the recursive evaluation
  of the costs of subspaces, it compares the cost of the parent subspace with
  not only its two children subspaces partitioned in the ``frequency'' domain
  (as the GHWT f2c does), but also its two children subspaces partitioned in
  the ``time'' domain (as the GHWT c2f does).

\item 
  If the underlying graph of an input graph signal is a simple
  unweighted path graph of dyadic length, i.e., $P_{2^n}$, $n \in \N$, 
  then Algorithm~\ref{alg:eghwt} reduces to the Thiele-Villemoes algorithm
  \emph{exactly}. Note that in such a case, neither computing the Fiedler
  vectors of subgraphs nor relabeling the subgraphs via
  Algorithm~\ref{alg:ghwt_modif}, is necessary;
  it is more efficient to force the midpoint partition at each level explicitly
  in that case.
\end{itemize}

\subsection{The eGHWT illustrated by a simple graph signal on $P_6$}
The easiest way to appreciate and understand the eGHWT algorithm is to use
a simple example. Let $\f = [2,-2,1,3,-1,-2]^\transp \in \Rf^6$ be an example
graph signal on $G=P_6$. The $\ell^1$-norm is chosen as the cost function.

\begin{figure}
  \begin{subfigure}{0.475\textwidth}
    \centering\includegraphics[width=\textwidth]{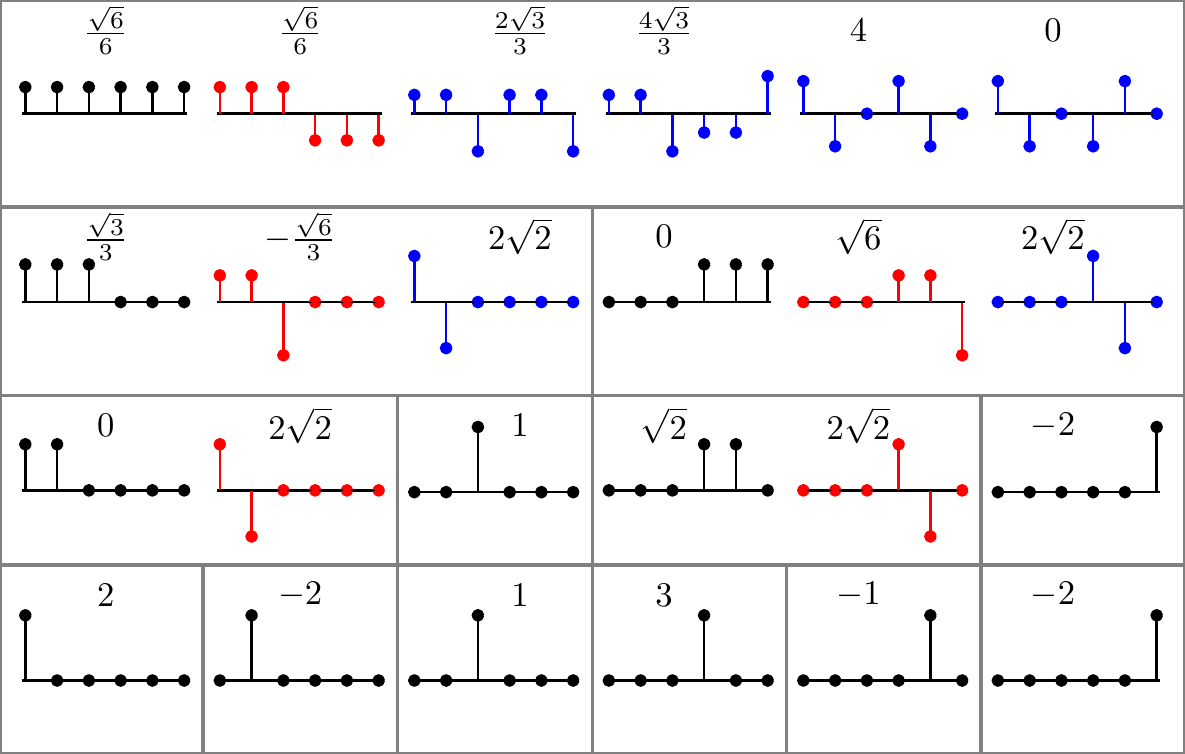}
    \caption{The GHWT c2f dictionary}
    \label{fig:p6_c2fcoef}
  \end{subfigure}
  \hspace{1em}
  \begin{subfigure}{0.475\textwidth}
    \centering\includegraphics[width=\textwidth]{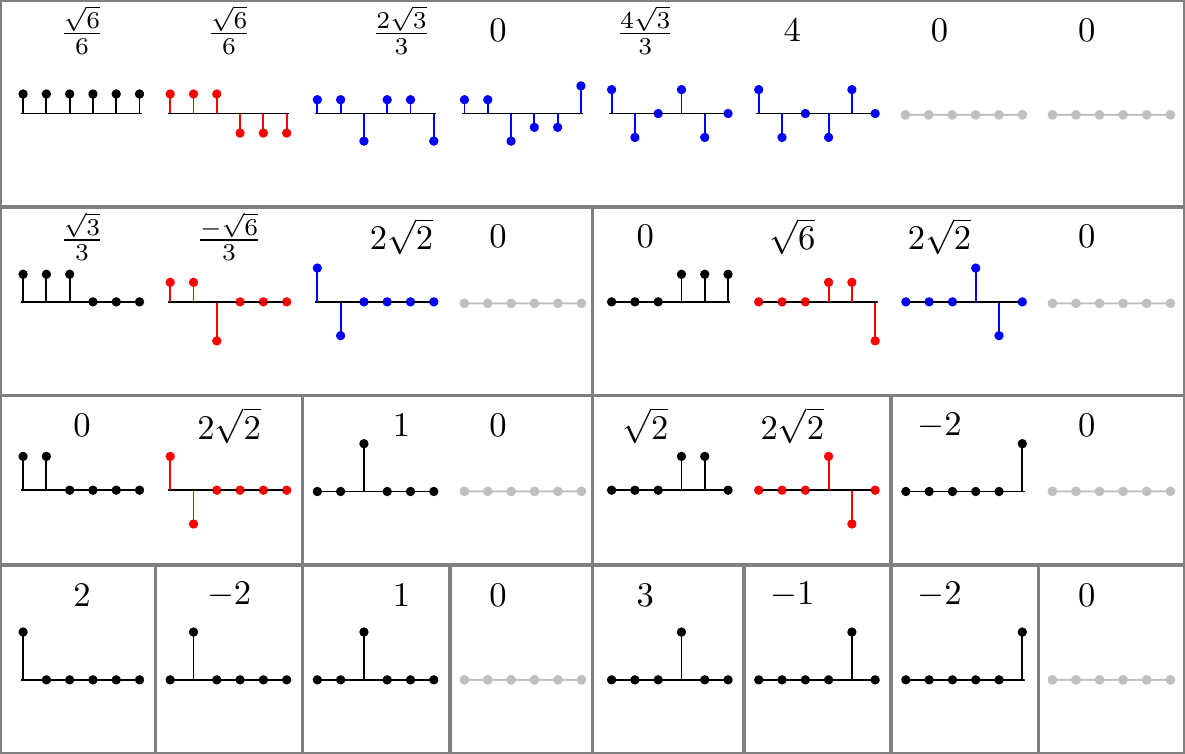}
    \caption{The GHWT c2f dictionary after relabeling}
    \label{fig:p6_c2fcoef_relabeled}
  \end{subfigure}
  \caption{(a) The expansion coefficients and the basis vectors of
    the simple graph signal $\f = [2,-2,1,3,-1,-2]^\transp$ on $P_6$
    relative to the GHWT c2f dictionary;
    (b) those correspond to the GHWT c2f dictionary after relabeling}
  \label{fig:p6_coef}
\end{figure}

\begin{figure}
  \centering
  \includegraphics[width = \textwidth]{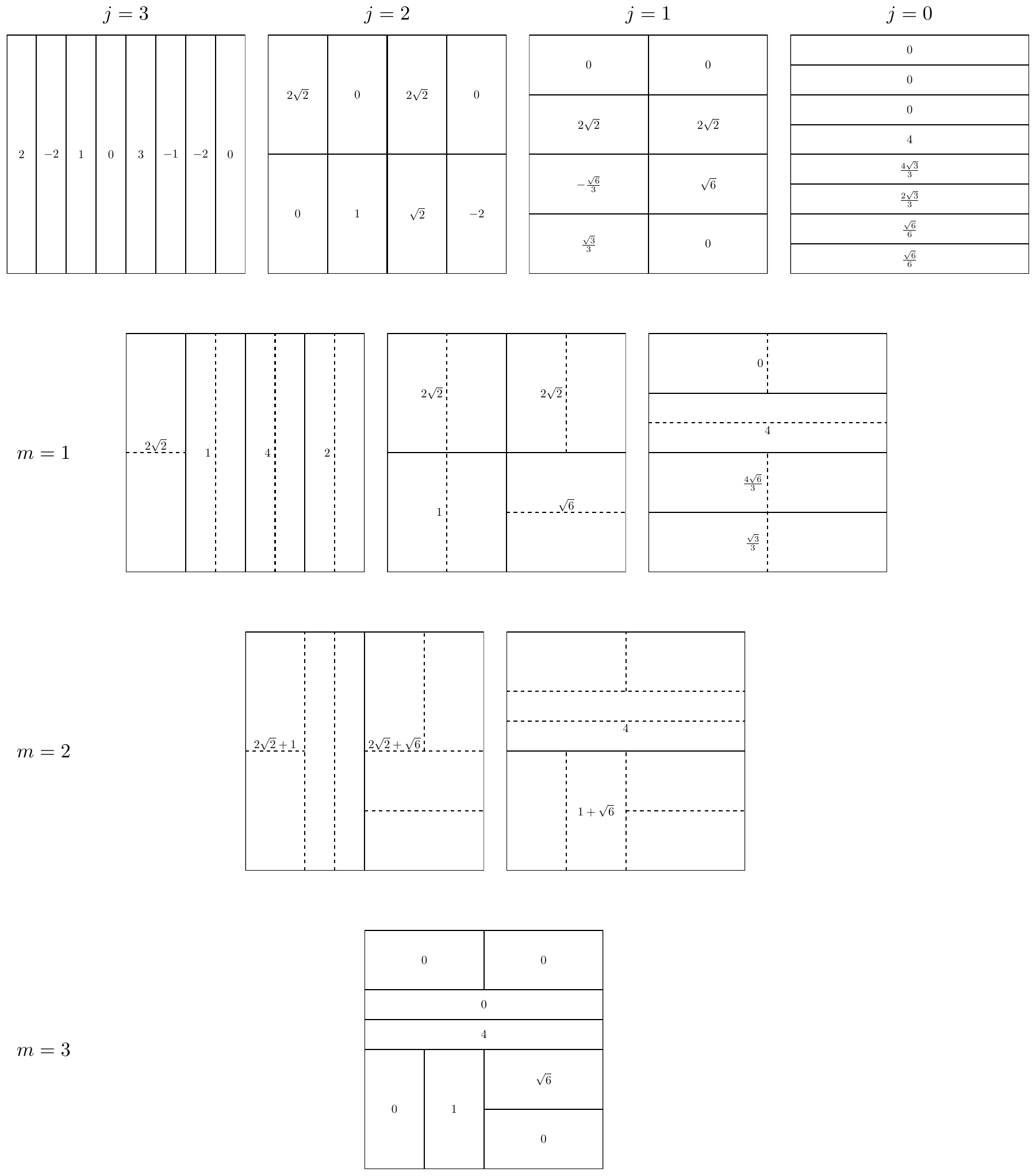}
  \caption{Graphical illustration of Algorithm~\ref{alg:eghwt} for the simple signal
    $\f = [2,-2,1,3,-1,-2]^\transp \in \Rf^6$. The cost function is the
    $\ell^1$-norm of the expansion coefficients. The top row contains all
    possible tiles and coefficients. The bottom row represents the
    eGHWT best basis.}
  \label{fig:tf_p8}
\end{figure}

Figure~\ref{fig:p6_c2fcoef} shows the coefficients of that signal on the GHWT c2f
dictionary and Fig.~\ref{fig:p6_c2fcoef_relabeled} corresponds to the relabeled
GHWT c2f dictionary by Algorithm~\ref{alg:ghwt_modif}.
After the GHWT c2f for $P_6$ is relabeled via Algorithm~\ref{alg:ghwt_modif},
the dictionary tree has the same structure as that of $P_8$.
Figure~\ref{fig:tf_p8} illustrates the progression of Algorithm~\ref{alg:eghwt}
on this simple graph signal. The time-frequency plane in this case is
$S_3 = [0,1) \times [0,8)$ and the frequency axis is scaled in this figure
so that $S_3$ is a square for visualization purpose.
The collection of all 32 tiles and the corresponding expansion coefficients
are placed on the four copies of $S_3$ in the top row of Fig.~\ref{fig:tf_p8},
which are ordered from the finest time/coarsest frequency partition ($j=3$)
to the coarsest time/finest frequency partition ($j=0$).
Note that each tile has the unit area.

In the first iteration ($m=1$) in Fig.~\ref{fig:tf_p8}, the minimizing tilings
for all the dyadic rectangles of area $2$ are computed from those with unit area.
For example, the left most dyadic rectangle (showing its cost $2\sqrt{2}$)
at $m = 1$, can be composed by either a pair of tiles
$(\bpsi^3_{0,0}, \bpsi^3_{1,0})$ or $(\bpsi^2_{0,0}, \bpsi^2_{0,1})$.
The corresponding costs are $|2| + |-2| = 4$ or $|2\sqrt{2}| + |0| = 2\sqrt{2}$,
and the minimum of which is $2\sqrt{2}$. The minimizing tiling for that dyadic
rectangle is hence $\bpsi^2_{0,0}$ and $\bpsi^2_{0,1}$ with its cost $2\sqrt{2}$.
In the second iteration ($m=2$), the algorithm finds the minimizing tilings for
all dyadic rectangles with area $4$. In the third iteration ($m=3$),
the algorithm finds the minimizing tiling for the whole $S_3$.
After the best basis is found in the relabeled GHWT c2f dictionary,
we remove those ``fictitious'' zero basis vectors and undo the relabeling
done by Algorithm~\ref{alg:ghwt_modif} to get back to the original $(j,k,l)$ indices.

\begin{figure}
  \begin{subfigure}{0.475\textwidth}
    \centering\includegraphics[width=\textwidth]{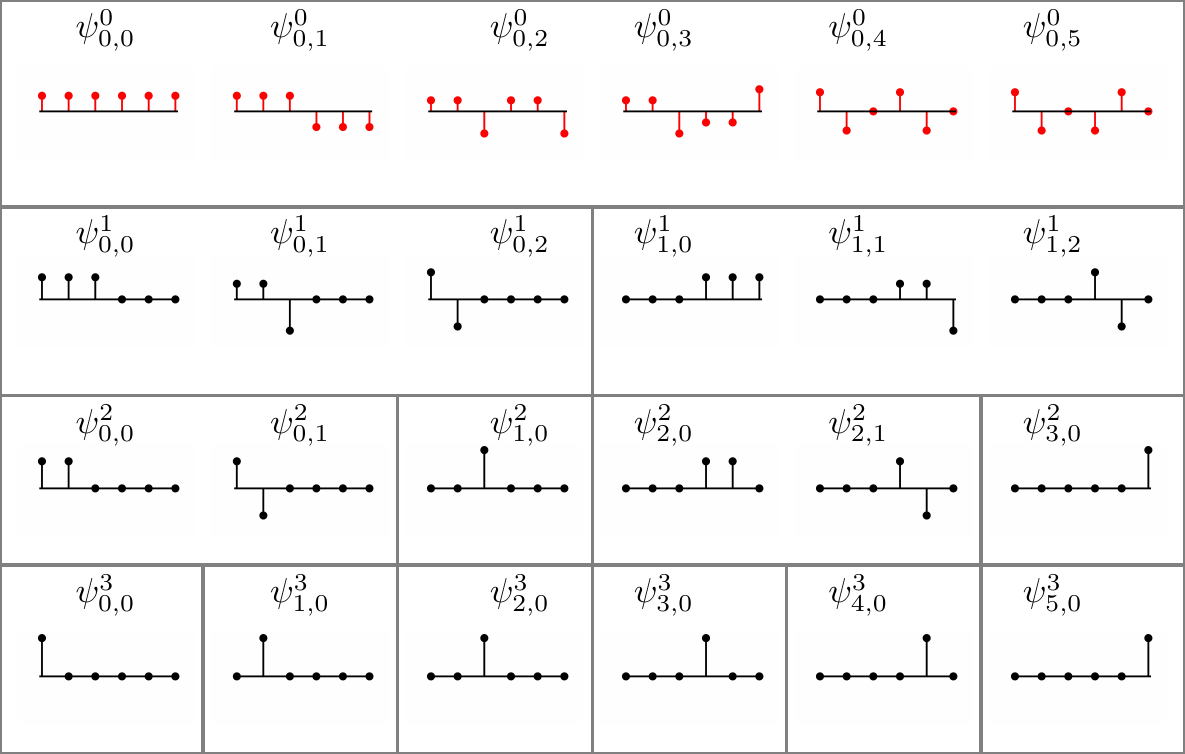}
    \caption{The GHWT c2f best basis}
  \end{subfigure}
  \hspace{1em}
  \begin{subfigure}{0.475\textwidth}
    \centering\includegraphics[width=\textwidth]{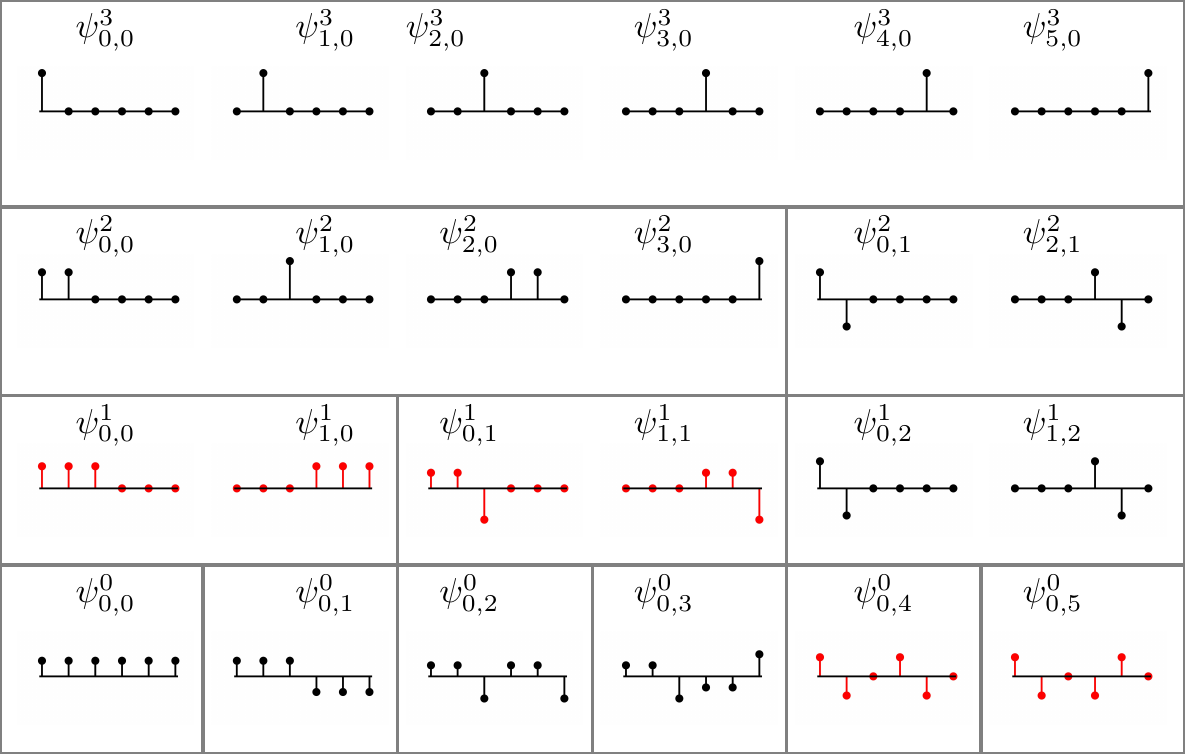}
    \caption{The GHWT f2c best basis}
  \end{subfigure}
  \caption{(a) The GHWT c2f best-basis vectors for the input graph signal
    $\f = [2,-2,1,3,-1,-2]^\transp$ (indicated by red); (b) the GHWT f2c
    best-basis vectors for the same signal.}
\label{fig:p6_fg_old}
\end{figure}

\begin{figure}
  \begin{subfigure}{0.475\textwidth}
    \centering\includegraphics[width=\textwidth]{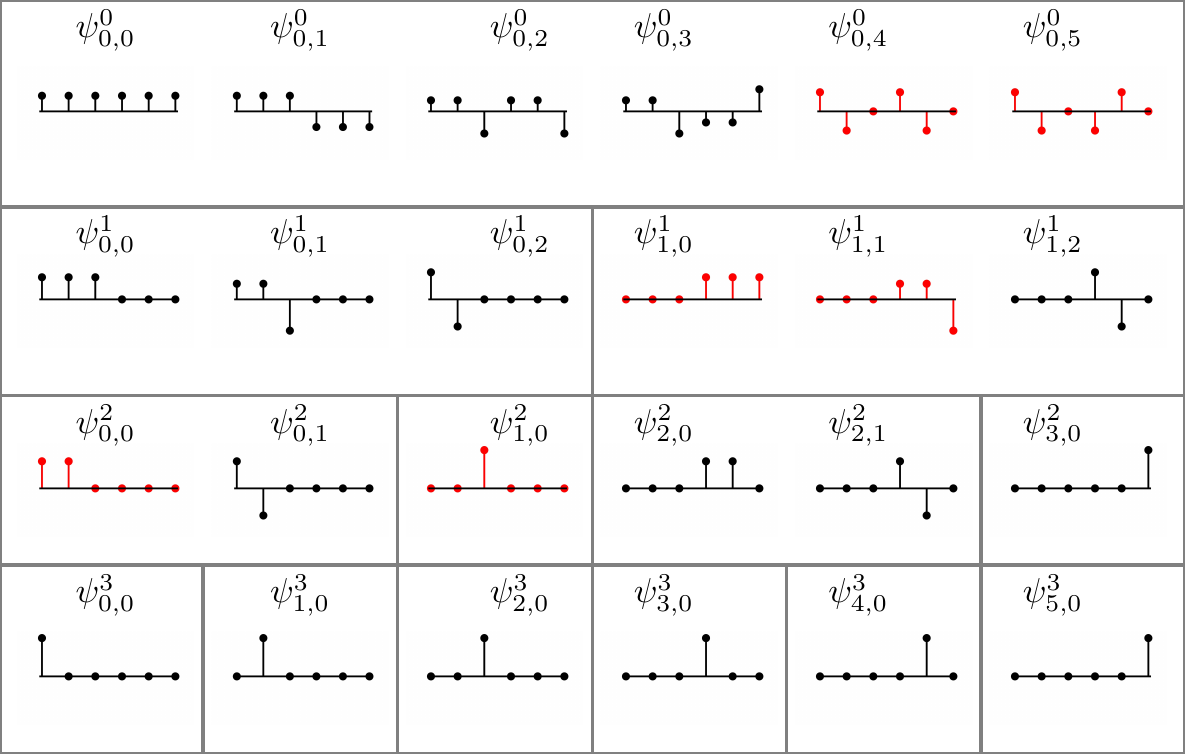}
    \caption{The eGHWT best basis shown in the GHWT c2f dictionary}
    \label{fig:p6_fg_new_c2f}
  \end{subfigure}
  \hspace{1em}
  \begin{subfigure}{0.475\textwidth}
    \centering\includegraphics[width=\textwidth]{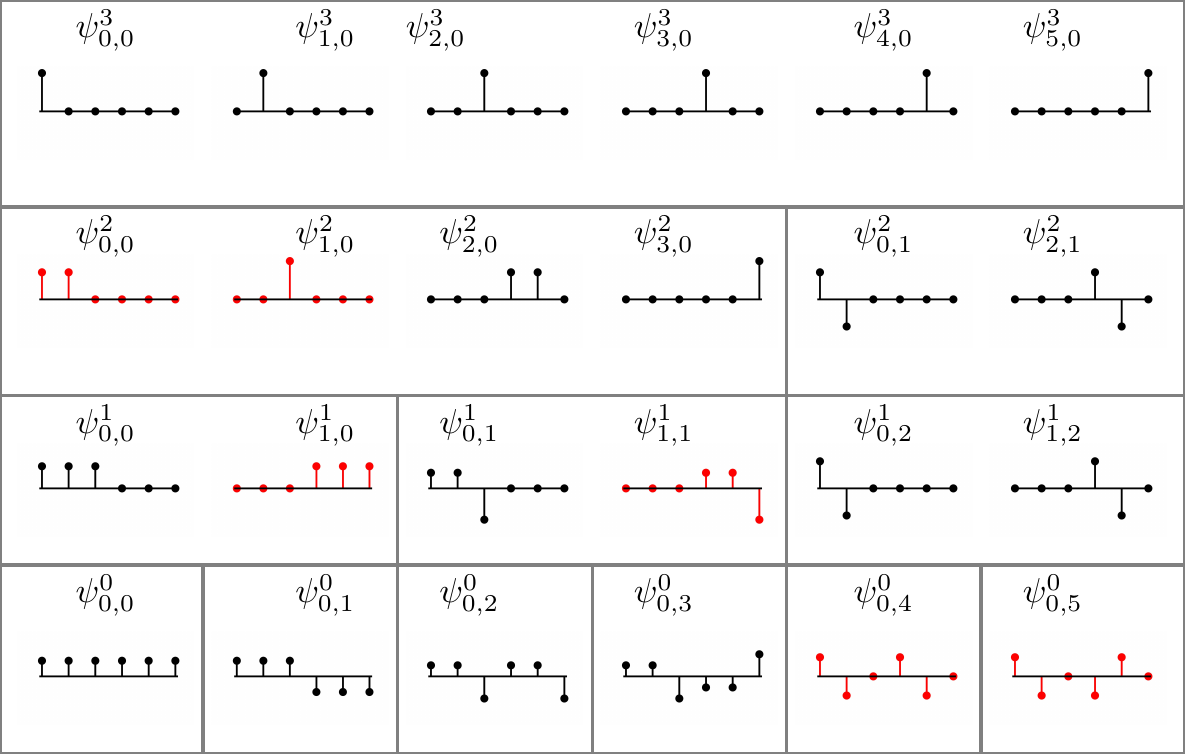}
    \caption{The eGHWT best basis shown in the GHWT f2c dictionary}
    \label{fig:p6_fg_new_f2c}
  \end{subfigure}
  \caption{The eGHWT best-basis vectors for the signal
    $\f = [2,-2,1,3,-1,-2]^\transp$ selected by Algorithm~\ref{alg:eghwt}
    (indicated by red) that are displayed within the GHWT c2f dictionary (a) and
    the GHWT f2c dictionary (b). Note the orthogonality of these vectors,
    and compare them with those shown in Fig.~\ref{fig:p6_fg_old}.}
  \label{fig:p6_fg_new}
\end{figure}

Figure~\ref{fig:p6_fg_old} shows that the GHWT c2f best basis for this simple
graph signal is actually the Walsh basis, and its representation is
$\frac{\sqrt{6}}{6}\bpsi_{0,0}^0+\frac{\sqrt{6}}{6}\bpsi_{0,1}^0+\frac{2\sqrt{3}}{3}\bpsi_{0,2}^0+\frac{4\sqrt{3}}{3}\bpsi_{0,3}^0+4\bpsi_{0,4}^0+0\bpsi_{0,5}^0$
with its cost $\approx 8.28$ and the f2c-GHWT best basis representation is
$\frac{\sqrt{3}}{3}\bpsi_{0,0}^1+0\bpsi_{1,0}^1+\frac{\sqrt{6}}{3}\bpsi_{0,1}^1+\sqrt{6}\bpsi_{1,1}^1+4\bpsi_{0,4}^0+0\bpsi_{0,5}^0$ with its cost $\approx 7.84$
while Fig.~\ref{fig:p6_fg_new} demonstrates that the best basis representation
chosen by the eGHWT algorithm is $0\bpsi_{0,0}^2+1\bpsi_{1,0}^2+0\bpsi_{1,0}^1+\sqrt{6}\bpsi_{1,1}^1+4\bpsi_{0,4}^0+0\bpsi_{0,5}^0$ with its cost $\approx 7.45$,
which is the smallest among these three best basis representations.
The indices used here are those of the original ones.

Figure~\ref{fig:p6_fg_new} clearly demonstrates that the eGHWT best basis
cannot be obtained by simply applying the previous GHWT best basis algorithm
on the c2f and f2c dictionaries.
More specifically, let us consider the vectors $\bpsi^1_{1,0}$ and $\bpsi^0_{0,4}$
in the eGHWT best basis. From Fig.~\ref{fig:p6_fg_new_c2f}, we can see that
$\bpsi^1_{1,0}$ is supported on the child graph $G^1_1$ that was generated by
bipartitioning the input graph $G^0_0$ where $\bpsi^0_{0,4}$ is supported.
Therefore, they cannot be selected in the GHWT c2f best basis simultaneously.
A similar argument applies to $\bpsi^2_{1,0}$ and $\bpsi^1_{1,0}$ in the eGHWT best
basis as shown in Fig.~\ref{fig:p6_fg_new_f2c}: they cannot be selected in the
GHWT f2c best basis simultaneously.

\subsection{Generalization to 2D Signals/Matrix Data}
\label{sec:2D-eGHWT}
The Thiele-Villemoes algorithm~\cite{THIELE-VILLEMOES} has been extended to 2D
signals by Lindberg and Villemoes~\cite{LINDBERG-VILLEMOES}.
Similarly to the former, the Lindberg-Villemoes algorithm works for only 2D
signals of dyadic sides.
We want to generalize the Lindberg-Villemoes algorithm for a more general 2D
signal or matrix data whose sides are not necessarily dyadic, as we did
for the previous GHWT dictionaries in \cite{IRION-SAITO-MLSP16}.
Before describing our 2D generalization of the eGHWT algorithm, let us briefly
review the Lindberg-Villemoes algorithm.
As we described in Sect.~\ref{sec:intro_tf}, the best tiling in each step in the
Thiele-Villemoes algorithm chooses a bipartition with a smaller cost between that
of the time domain and that of the frequency domain.
For 2D signals, the time-frequency domain has four axes instead of two.
It has time and frequency axes on each of the $(x,y)$ components (i.e.,
columns and rows of an input image).
The best tiling comes from the split in the time or frequency directions in
either the $x$ or $y$ component. This forces them to choose the best tiling/split
among four options instead of two for each split.
Similarly to the 1D signal case, dynamic programming is used to find the
minimizing tiling for a given 2D signal. 
We note that our 2D generalized
version of the eGHWT algorithm \emph{exactly} reduces to the Lindberg-Villemoes
algorithm if the input graph is a 2D regular lattice of dyadic sizes and
the midpoint (or midline to be more precise) partition is used throughout.

For a more general 2D signal or matrix data, we can compose the affinity
matrices on the row and column directions separately (see, e.g.,
\cite{IRION-SAITO-MLSP16}),
thus define graphs on which the rows and columns are supported.
In this way, the input 2D signal can be viewed as a tensor product of two graphs.
Then the eGHWT can be extended to 2D signal from 1D in a similar way
as how Lindberg and Villemoes \cite{LINDBERG-VILLEMOES} extended the
Thiele-Villemoes algorithm~\cite{THIELE-VILLEMOES}. Examples will be given in
Sect.~\ref{sec:2D-eGHWT-example}.

\section{Applications}
\label{sec:appl}
In this section, we will demonstrate the usefulness and efficiency of the eGHWT
using several real datasets, and compare its performance with that of
the classical Haar transform, graph Haar basis, graph Walsh basis, and
the GHWT c2f/f2c best bases.
We note that our performance comparison is to emphasize the difference
between the eGHWT and its close relatives. Hence we will not compare the
performance of the eGHWT with those graph wavelets and wavelet packets of
different nature; see, e.g., \cite{IRION-SAITO-TSIPN, CLONINGER-LI-SAITO} for
further information.

\subsection{Efficient Approximation of a Graph Signal}
\label{sec:toronto}
\begin{figure}
  \begin{subfigure}{0.475\textwidth}
    \centering\includegraphics[width=\textwidth]{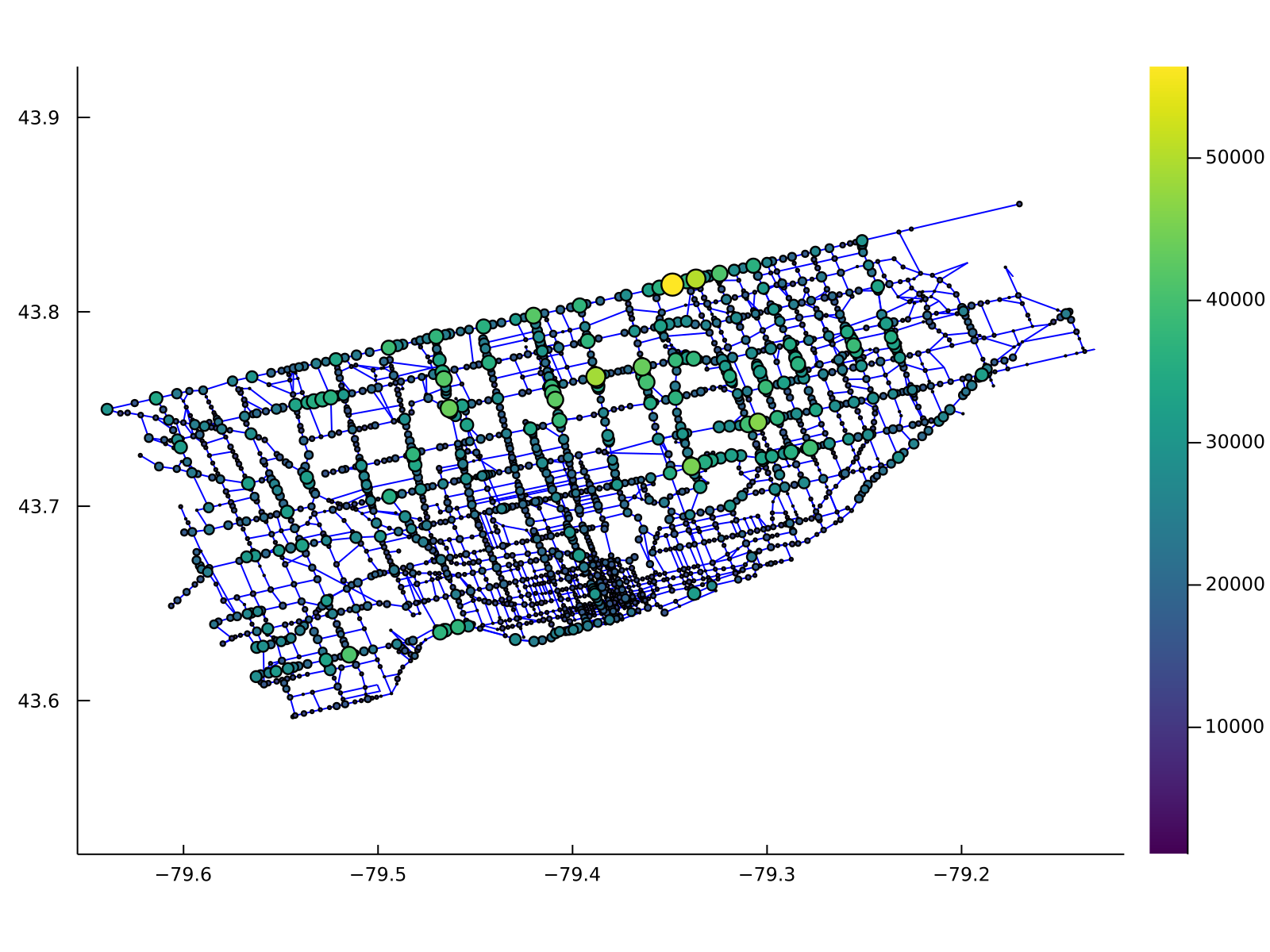}
    \caption{Vehicular volume counts}
    \label{fig:toronto_original}
  \end{subfigure}
  \quad
  \begin{subfigure}{0.475\textwidth}
    \centering\includegraphics[width=\textwidth]{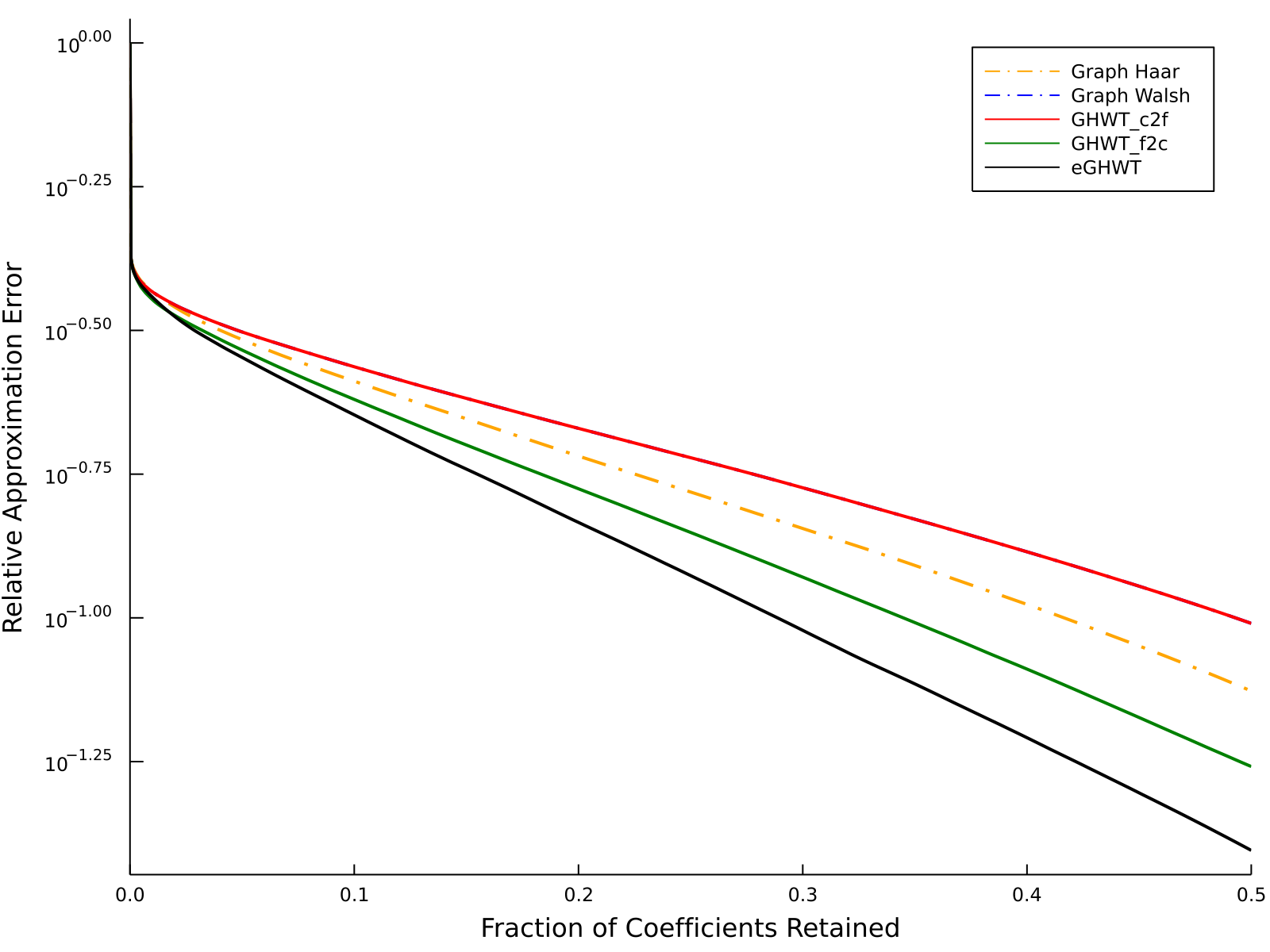}
    \caption{Relative $\ell^2$ approximation error}
    \label{fig:toronto_l2}
  \end{subfigure}
  \caption{(a) Vehicular volume data in the city of Toronto
    (the node diameter is proportional to its volume counts
    for visualization purpose); 
    (b) Relative $\ell^2$ approximation errors by various graph bases}
  \label{fig:toronto_vv}
\end{figure}

Here we analyze the eight peak-hour vehicle volume counts on the Toronto street
network, which is shown in Fig.~\ref{fig:toronto_vv}.
We have already described this street network in Sect.~\ref{sec:back} and
in Fig.~\ref{fig:partition_tree_toronto}.
The data was typically collected at the street intersections equipped with
traffic lights between the hours of 7:30 am and 6:00 pm, over the period of
03/22/2004--02/28/2018. As one can see, the vehicular volume are spread
in various parts of this street network with the concentrated northeastern
region.

\begin{figure}
  \begin{subfigure}{0.475\textwidth}
    \centering\includegraphics[width=\textwidth]{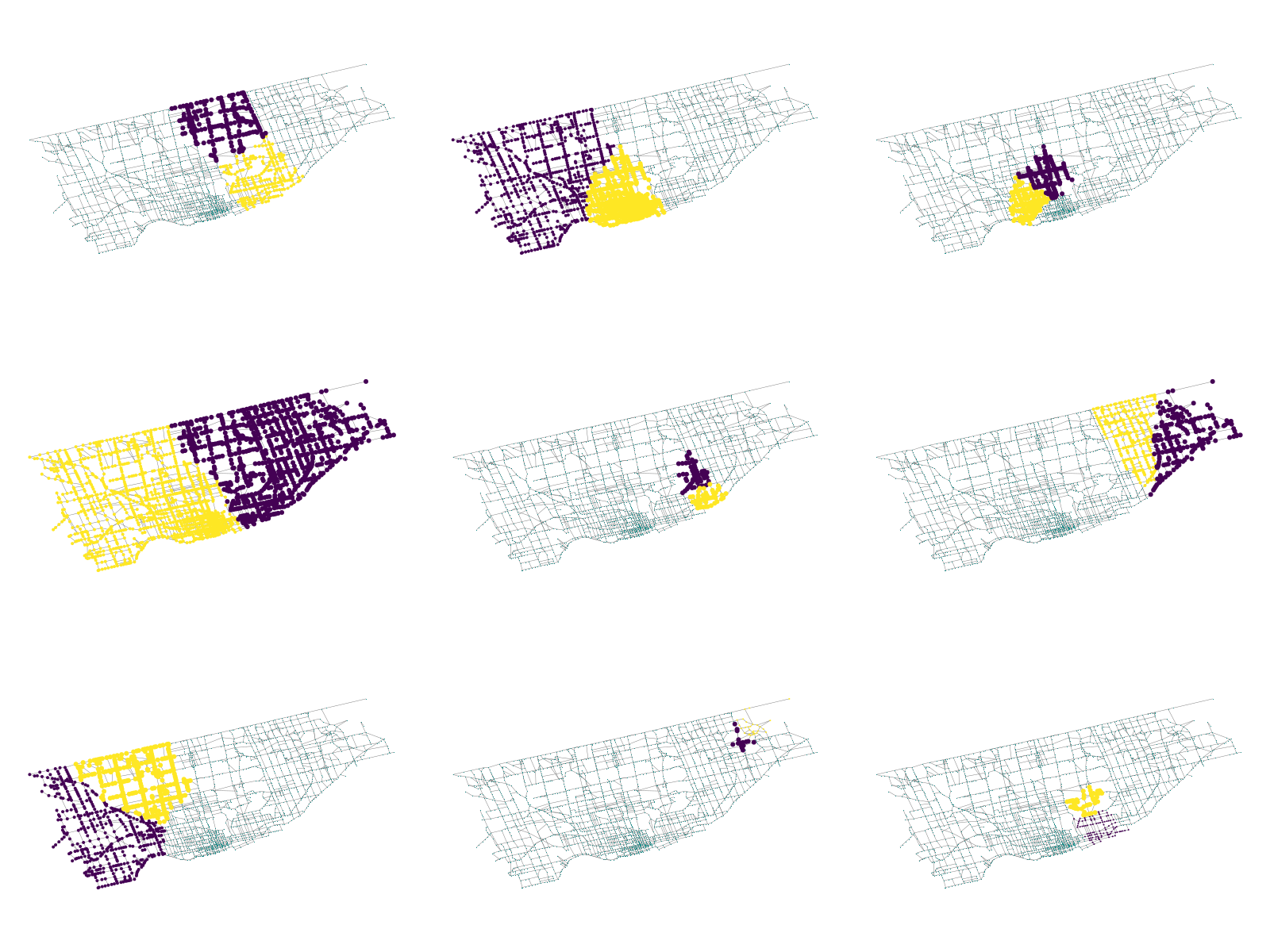}
    \captionsetup{skip=1pt}
      \caption{Haar}
  \end{subfigure}
    \hspace{1em}
    \begin{subfigure}{0.475\textwidth}
      \centering\includegraphics[width=\textwidth]{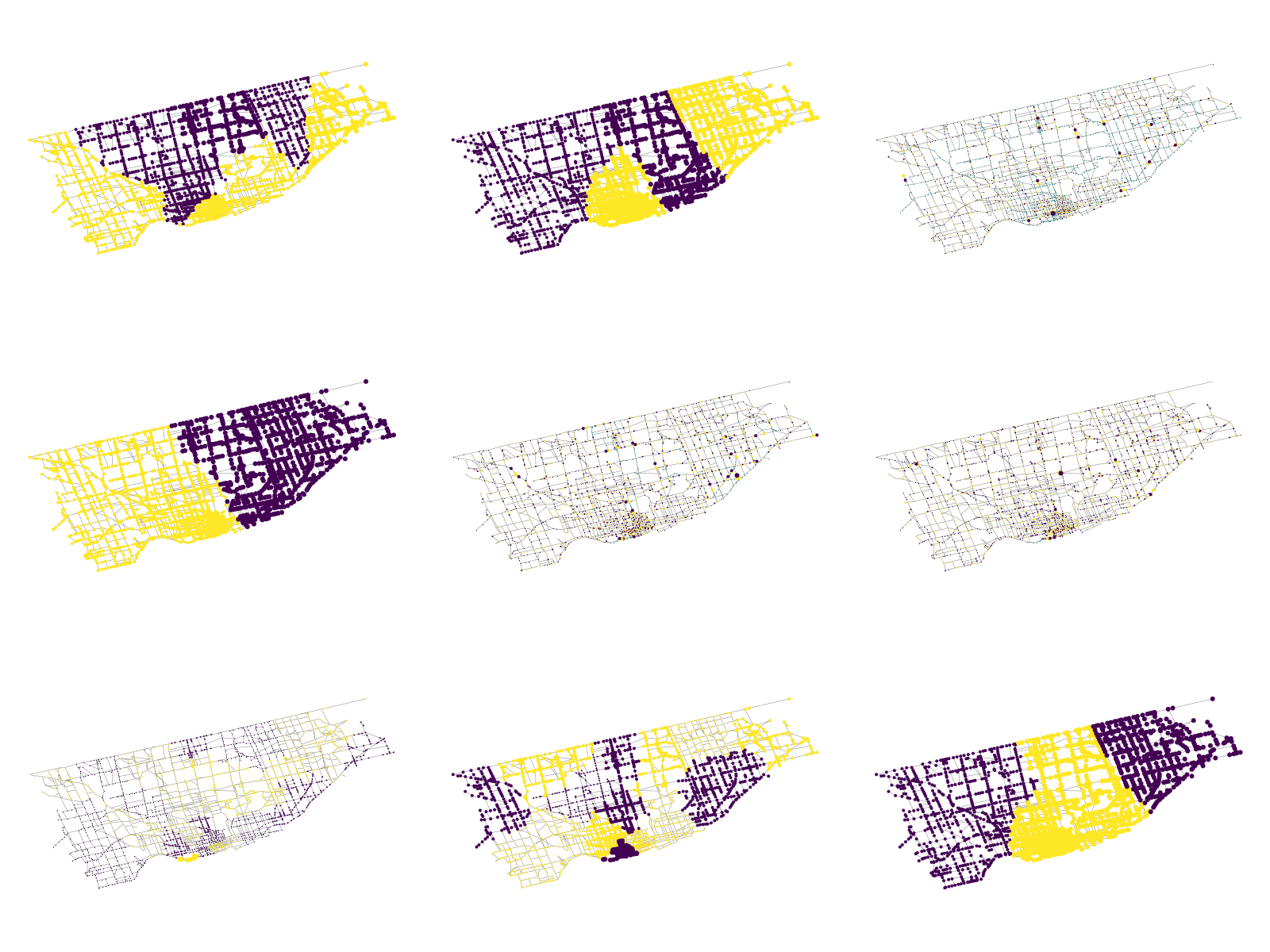}
      \captionsetup{skip=1pt}
      \caption{GHWT c2f = Walsh}
    \end{subfigure}
    \\[1em]
    \begin{subfigure}{0.475\textwidth}
      \centering\includegraphics[width=\textwidth]{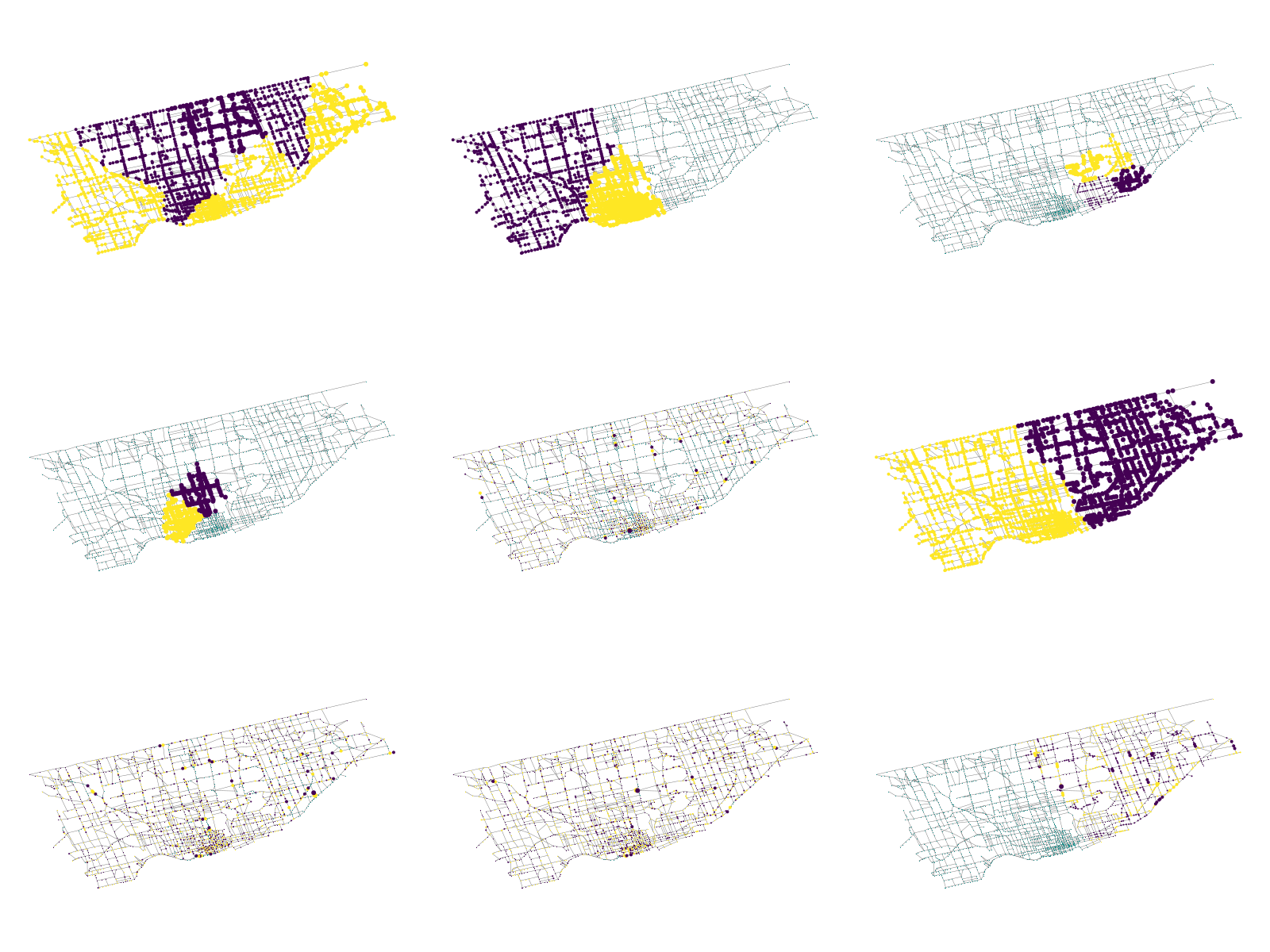}
      \captionsetup{skip=1pt}
      \caption{GHWT f2c}
    \end{subfigure}
    \hspace{1em}
    \begin{subfigure}{0.475\textwidth}
      \centering\includegraphics[width=\textwidth]{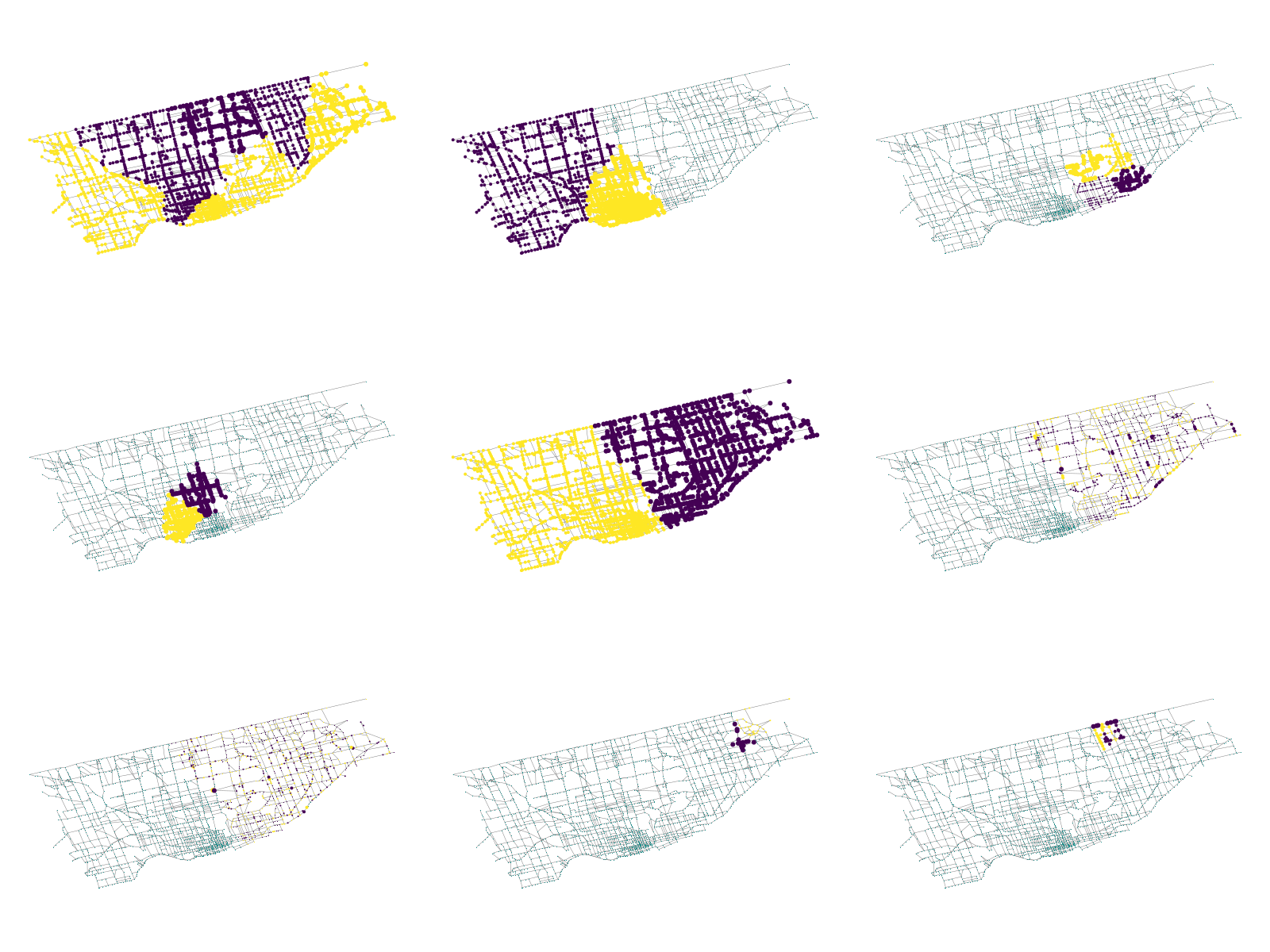}
      \captionsetup{skip=1pt}
      \caption{eGHWT}
    \end{subfigure}
    \caption{Nine most significant basis vectors of the graph Haar, the GHWT c2f
      best basis (= the graph Walsh in this case), the GHWT f2c best basis, and
      the eGHWT best basis. The diameters of the nodes are proportional to the
      log of the absolute values of the basis vector amplitudes.
      The basis vector amplitudes within $(-0.001, 0.001)$ are mapped to the
      viridis colormap to emphasize the difference between positive (yellow)
      and negative (deep violet) components.}
    \label{fig:toronto_top9}
  \end{figure}

In addition to the eGHWT best basis, the graph Haar basis, the graph Walsh basis,
the GHWT c2f/f2c best bases are used to compare the performance.
Figure~\ref{fig:toronto_l2} shows the performance comparison. The $y$-axis denotes
the relative approximation error $\| \f - \P_n\f \|_2 / \| \f \|_2$,
where $\P_n\f$ denotes the approximation of $\f$ with the basis vectors having
the $n$ largest coefficients in magnitude.
The $x$-axis denotes $n/N$, i.e., the fraction of coefficients retained.
We can see that the error of the eGHWT decays fastest, followed by the
GHWT f2c best basis, the graph Haar basis, and the GHWT c2f best basis that
chose the graph Walsh basis, in that order. 
In Fig.~\ref{fig:toronto_top9}, we display the nine most significant basis
vectors (except for the DC vector) for the graph Haar, the GHWT c2f best basis
(which selected the graph Walsh basis), the GHWT f2c best basis, and
the eGHWT best basis to approximate this traffic volume data.

These figures clearly demonstrate the superiority of the eGHWT best basis over
the graph Haar basis, the graph Walsh basis, and the regular GHWT best bases.
The top 9 basis vectors for these bases displayed in Fig.~\ref{fig:toronto_top9}
show the characteristics of each basis under consideration.
The graph Haar basis vectors are non-oscillatory blocky vectors with
positive and negative components at various locations and scales.
The graph Walsh basis vectors (= the GHWT c2f best-basis vectors) are all global
oscillatory piecewise-constant vectors.
The GHWT f2c best-basis vectors and the eGHWT best-basis vectors are similar
and can capture the multiscale and multi-frequency features of the input
graph signal. Yet, the performance of the eGHWT best basis exceeds that of
the GHWT f2c best basis simply because the former can search the best basis
among a much larger collection of ONBs than the latter can.

\subsection{Viewing a General Matrix Signal as a Tensor Product of Graphs}
\label{sec:2D-eGHWT-example}
To analyze and process a single graph signal, we can use the eGHWT to produce a
suitable ONB.
Then for a collection of signals in a matrix form (including regular digital
images), we can also compose the affinity matrix of the rows and that of
the columns separately, thus define graphs on which the rows and columns are
supported as was done previously~\cite{COIF-GAVISH, IRION-SAITO-MLSP16}.
Those affinity matrices can be either computed from the similarity of rows or
columns directly or can be composed from information outside the original matrix
signal. For example, Kalofolias et al.~\cite{KALOFOLIAS-ETAL} used row and
column graphs to analyze recommender systems.

After the affinity graphs on rows and columns are obtained, we can use the eGHWT
to produce ONBs on rows and columns separately. Then the matrix signal can be
analyzed or compressed by the tensor product of those two ONBs.
In addition, as mentioned in Sect.~\ref{sec:2D-eGHWT}, we have also extended
the eGHWT to the tensor product of row and column affinity graphs and search for
best 2D ONB on the matrix signal directly.
Note that we can also specify or compute the binary partition trees in a
non-adaptive manner (e.g., recursively splitting at the middle of each region),
typically for signals supported on a regular lattice.

\subsubsection{Approximation of the Barbara Image}
\begin{figure}
  \begin{subfigure}{0.475\textwidth}
    \centering\includegraphics[width=\textwidth]{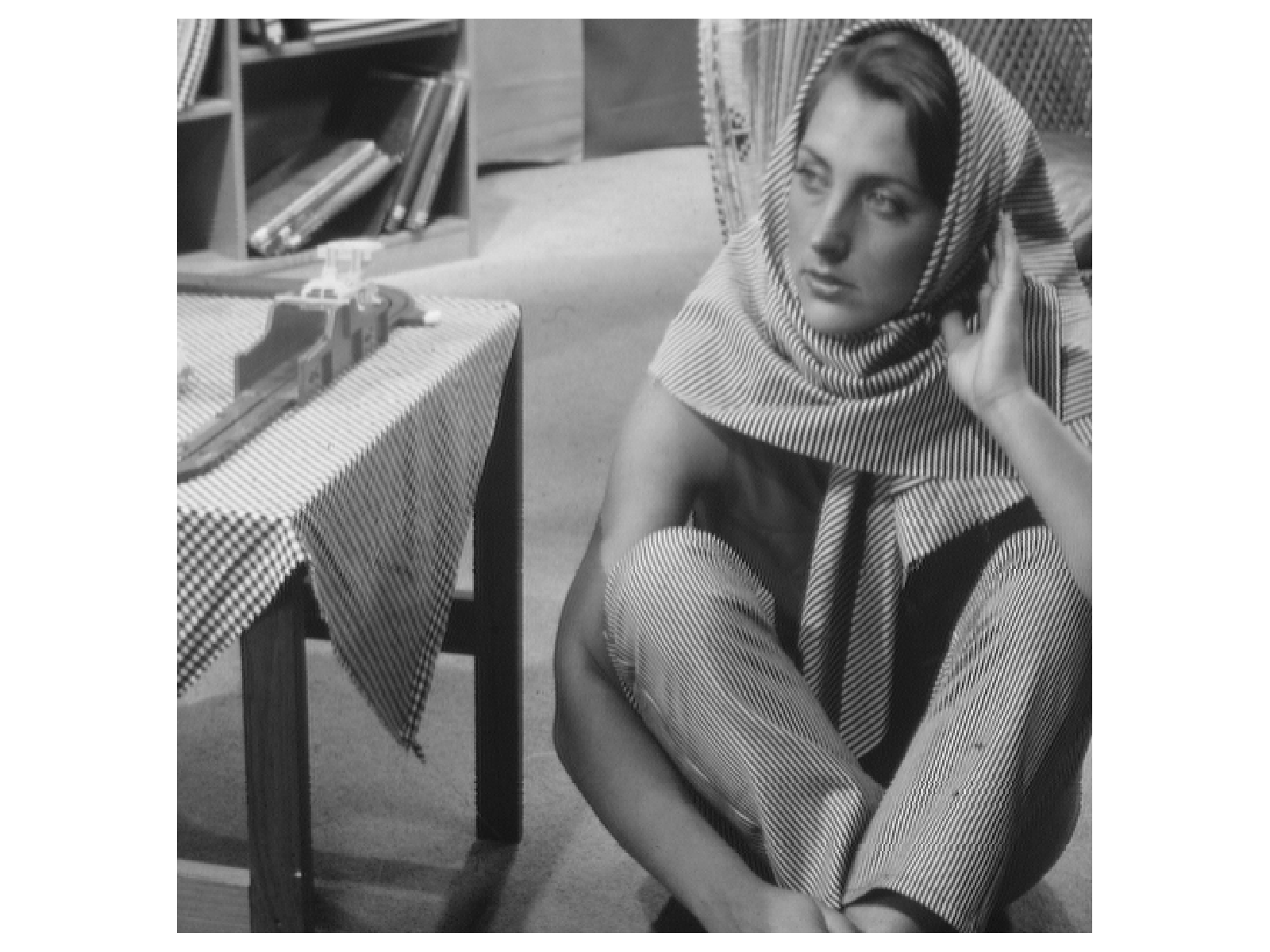}
    \caption{Barbara}
    \label{fig:barbara_original}
  \end{subfigure}
  \quad
  \begin{subfigure}{0.475\textwidth}
    \centering\includegraphics[width=\textwidth]{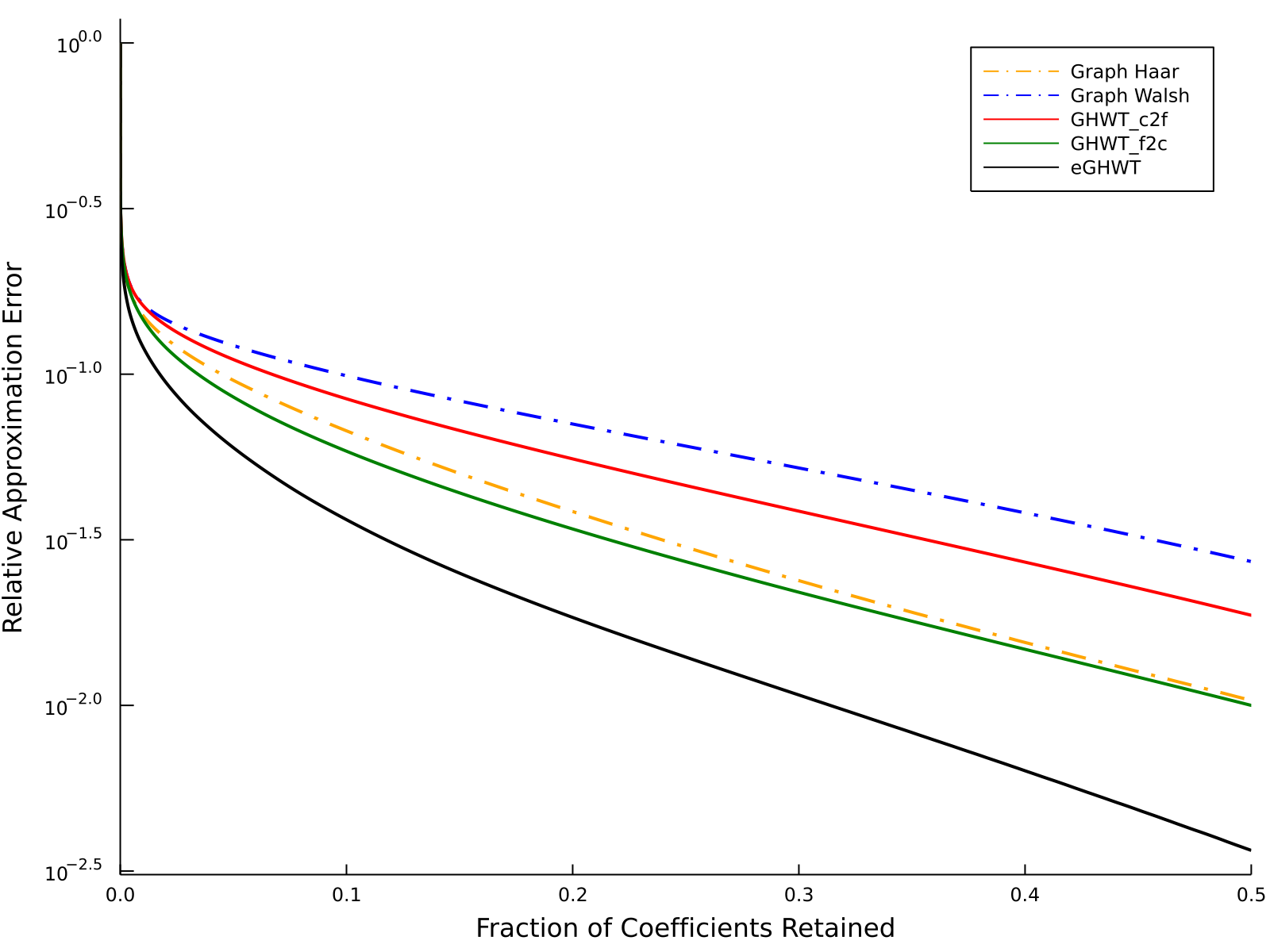}
    \caption{Relative $\ell^2$ approximation error}
    \label{fig:barbara_l2}
  \end{subfigure}
  \caption{(a) The original Barbara image of size $512 \times 512$ pixels;
    (b) Relative $\ell^2$ approximation errors by various graph bases}
  \label{fig:barbara}
\end{figure}

In this section, we compare the performance of various bases in approximating
the famous Barbara image shown in Fig.~\ref{fig:barbara_original}, and
demonstrate our eGHWT can be applied to a conventional digital image in a
straightforward manner, and outperforms the regular GHWT best bases.
This image consists of $512 \times 512$ pixels, and has been normalized to
have its pixel values in $(0, 1)$. The partition trees on the rows and columns
  are \emph{specified explicitly:} every bipartition is forced at the middle of
  each region. 
Therefore, those two trees are perfect binary trees with depth equal to
$\log_2(512) + 1 = 10$.

Figure~\ref{fig:barbara_l2} displays the relative $\ell^2$ errors of the
approximations by the graph Haar basis, the graph Walsh basis, the GHWT cf2/f2c
best bases, and the eGHWT best basis as a function of the fraction of
the coefficients retained.

\begin{figure}
  \begin{subfigure}{0.475\textwidth}
    \centering\includegraphics[width=\textwidth]{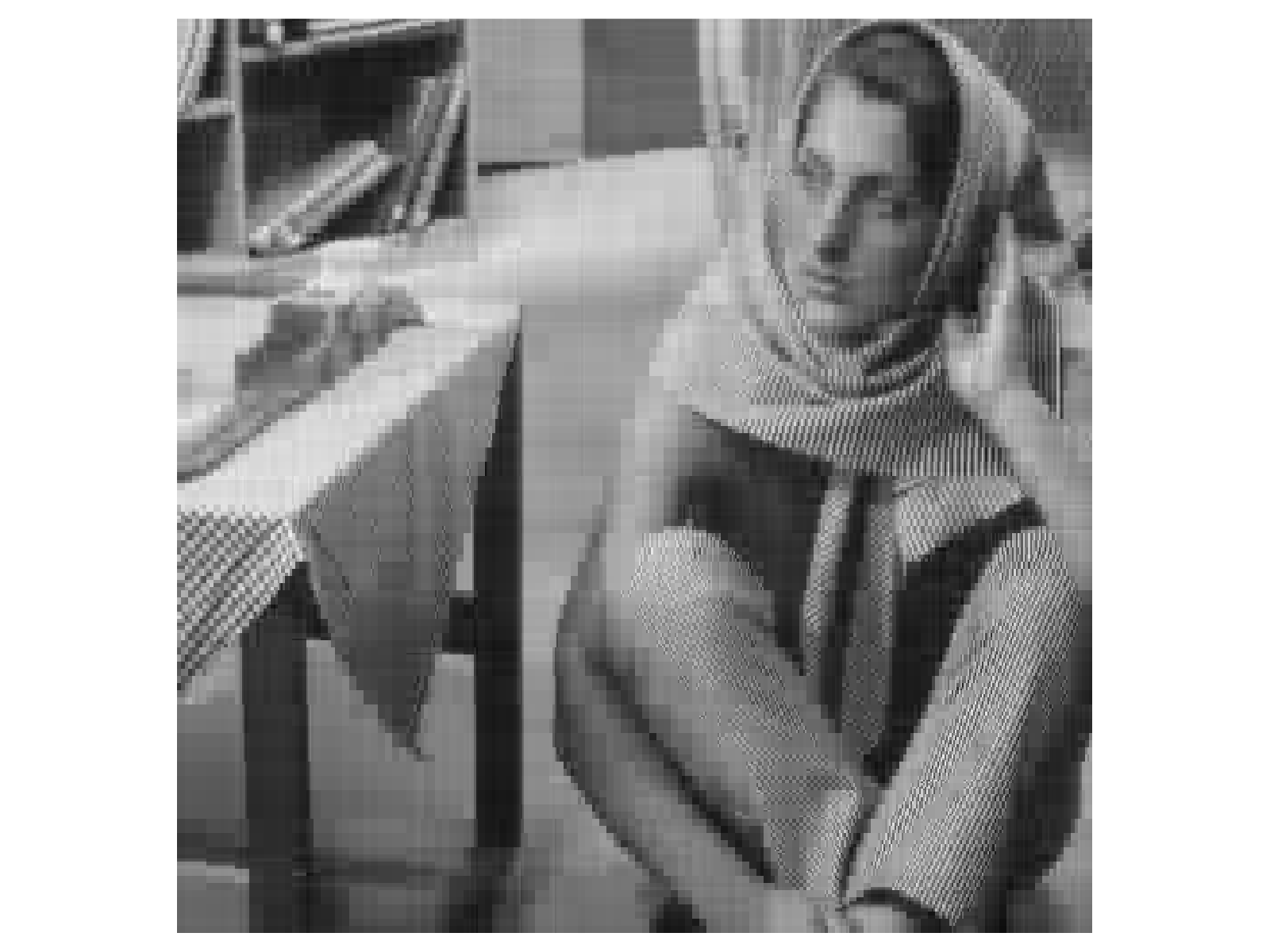}
    \captionsetup{skip=1pt}
    \caption{Haar, PSNR = 24.50dB}
  \end{subfigure}
  \begin{subfigure}{0.475\textwidth}
    \centering\includegraphics[width=\textwidth]{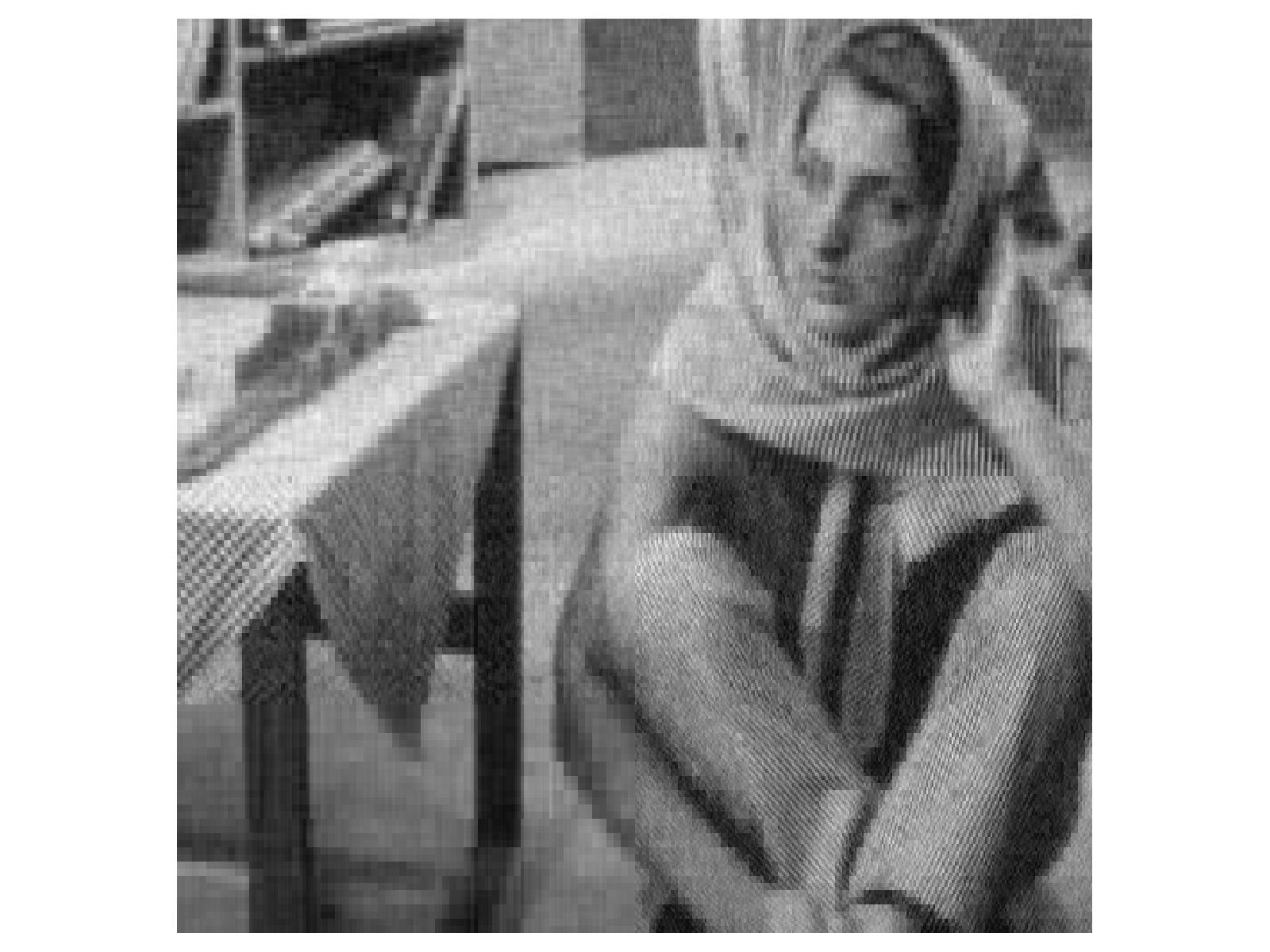}
    \captionsetup{skip=1pt}
    \caption{GHWT c2f, PSNR = 23.51dB}
  \end{subfigure}
    \\[1em]
    \begin{subfigure}{0.475\textwidth}
      \centering\includegraphics[width=\textwidth]{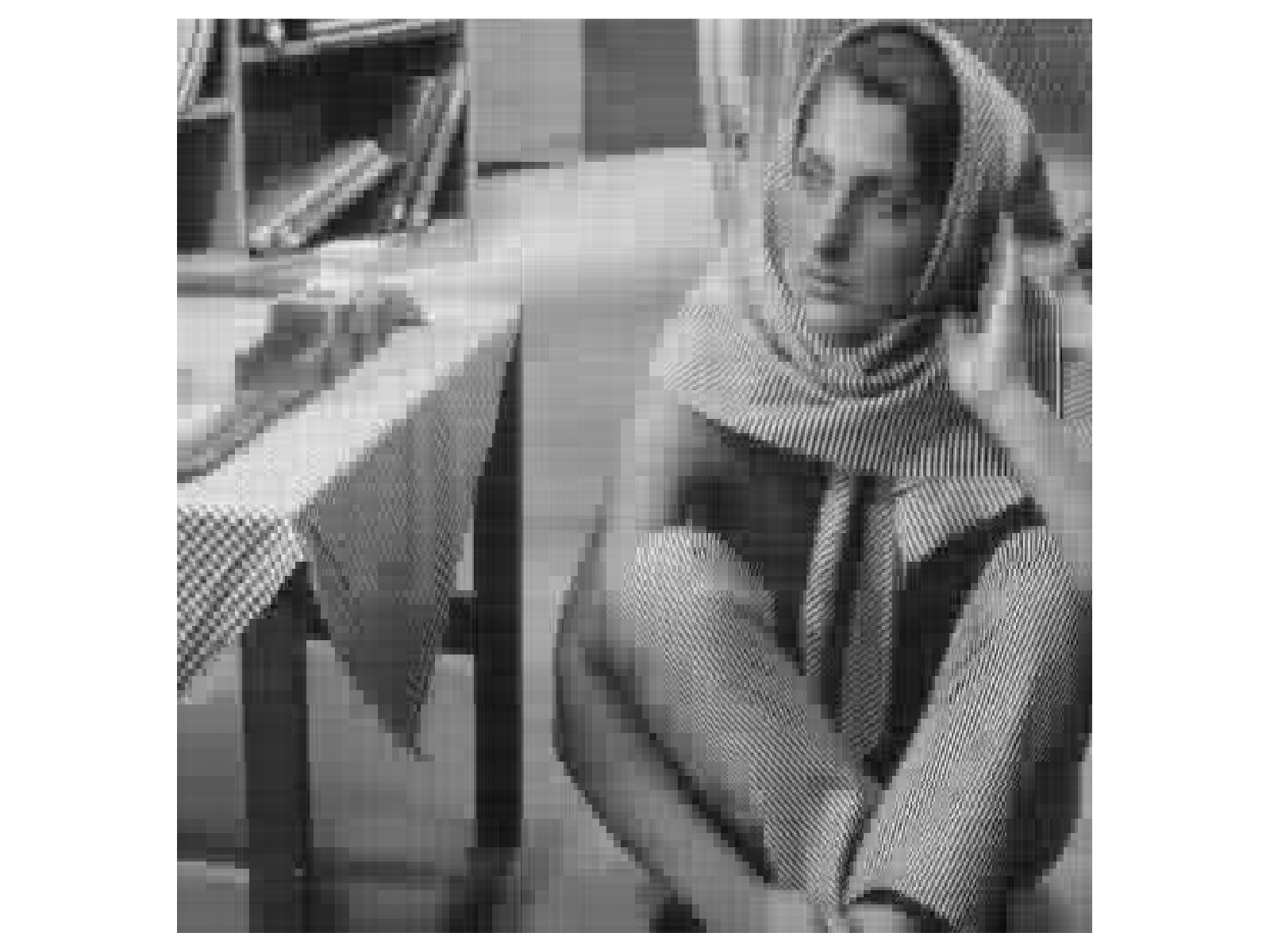}
      \captionsetup{skip=1pt}
      \caption{GHWT f2c, PSNR = 25.27dB}
    \end{subfigure}
    \begin{subfigure}{0.475\textwidth}
      \centering\includegraphics[width=\textwidth]{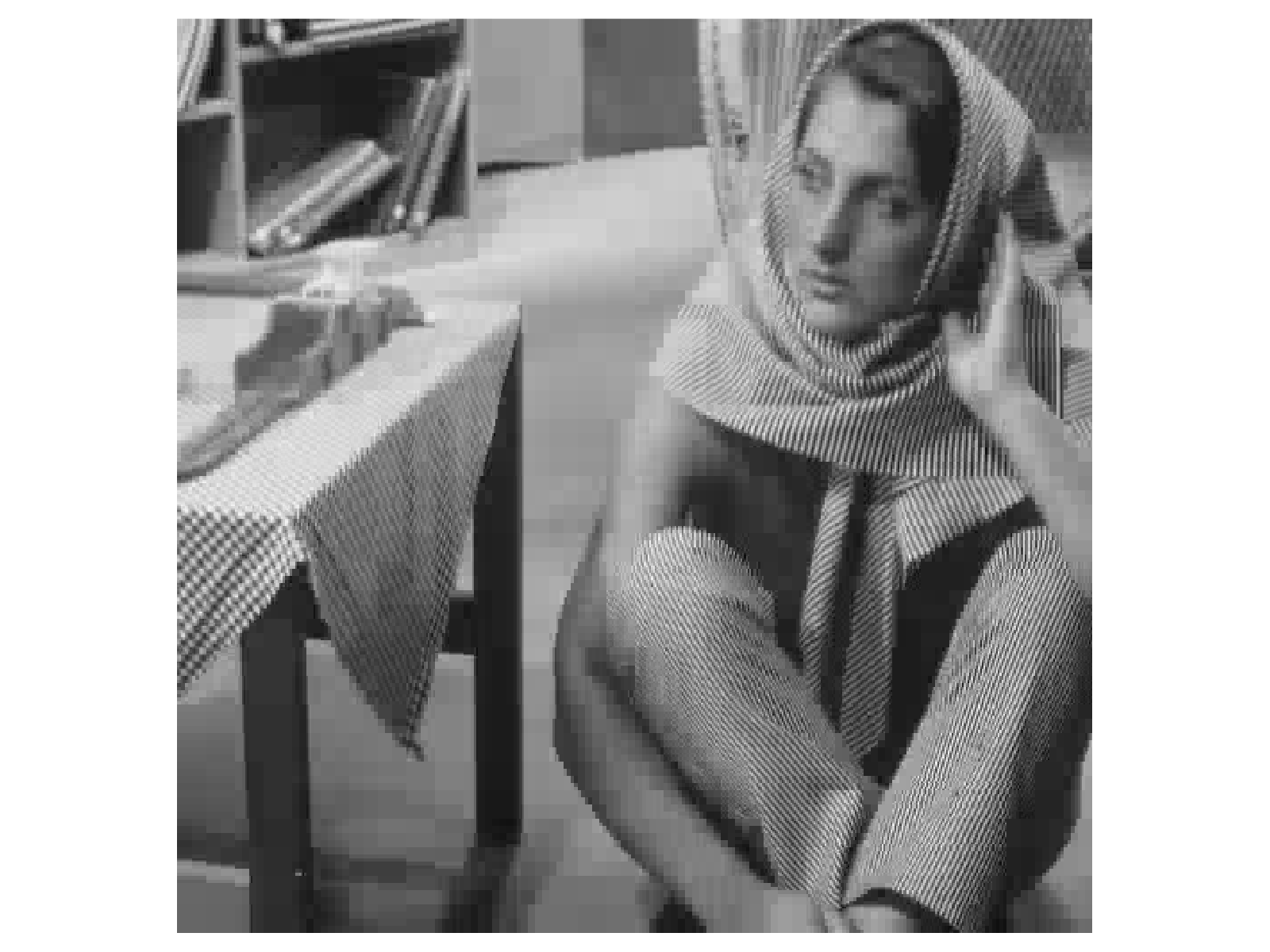}
      \captionsetup{skip=1pt}
      \caption{eGHWT, PSNR = 27.78dB}
    \end{subfigure}
    \caption{Approximations of the Barbara image using various bases using only 3.125\% of coefficients (online viewing is recommended for the details)}
    \label{fig:barbara_full}
  \end{figure}

\begin{figure}
  \begin{subfigure}{.475\textwidth}
    \centering\includegraphics[width=.475\textwidth]{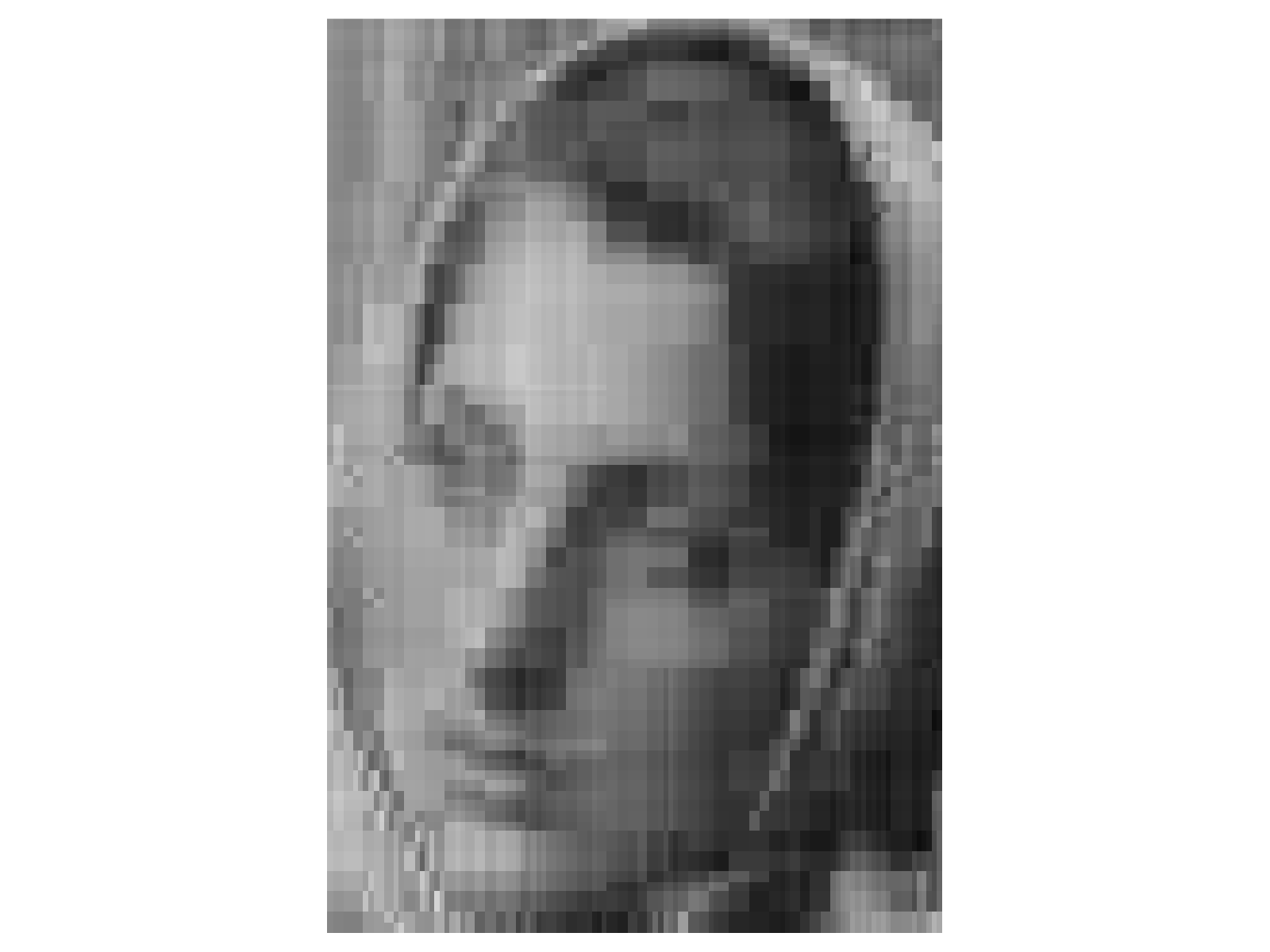}
    \centering\includegraphics[width=.475\textwidth]{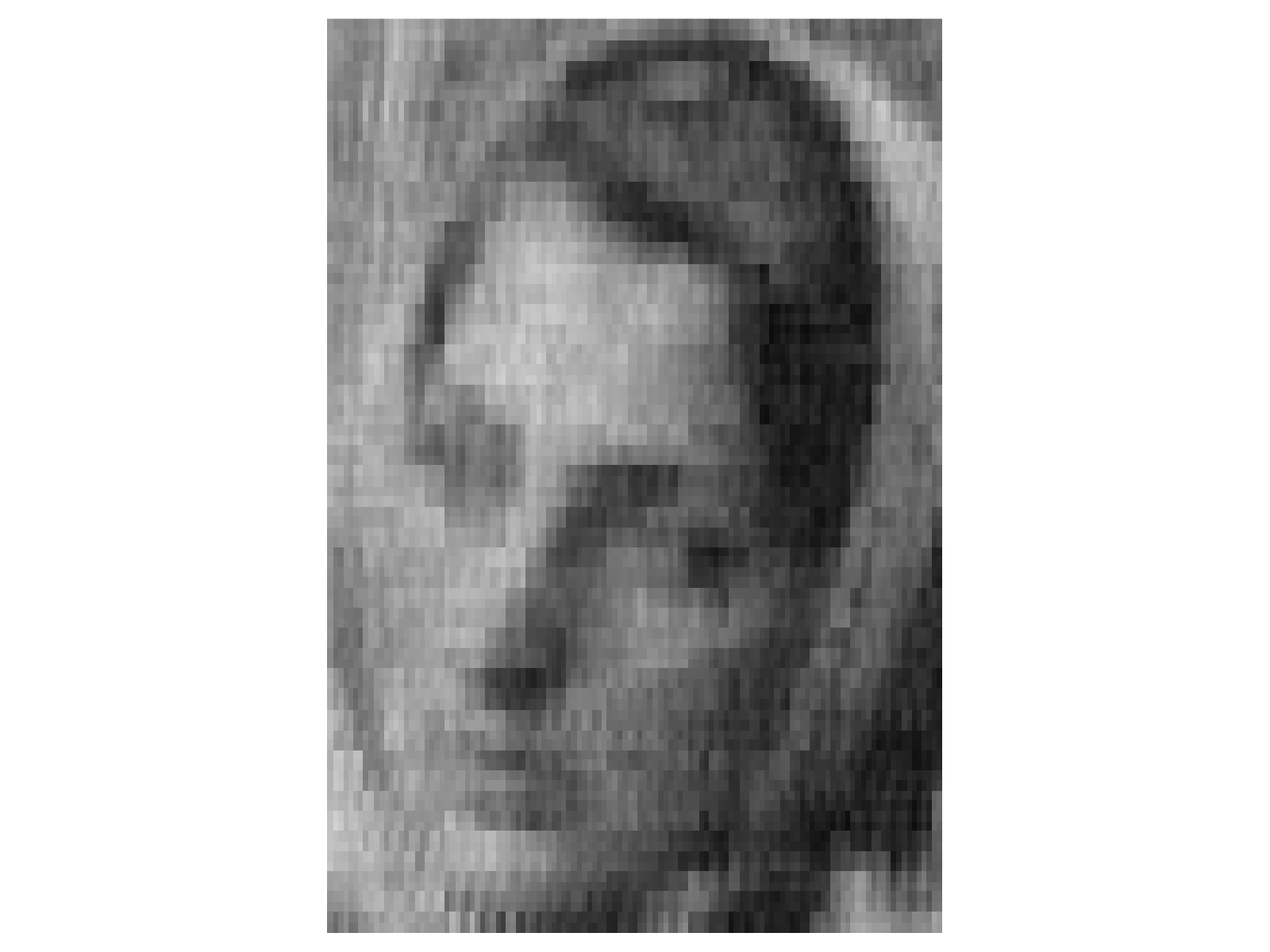}\\
    \centering\includegraphics[width=.475\textwidth]{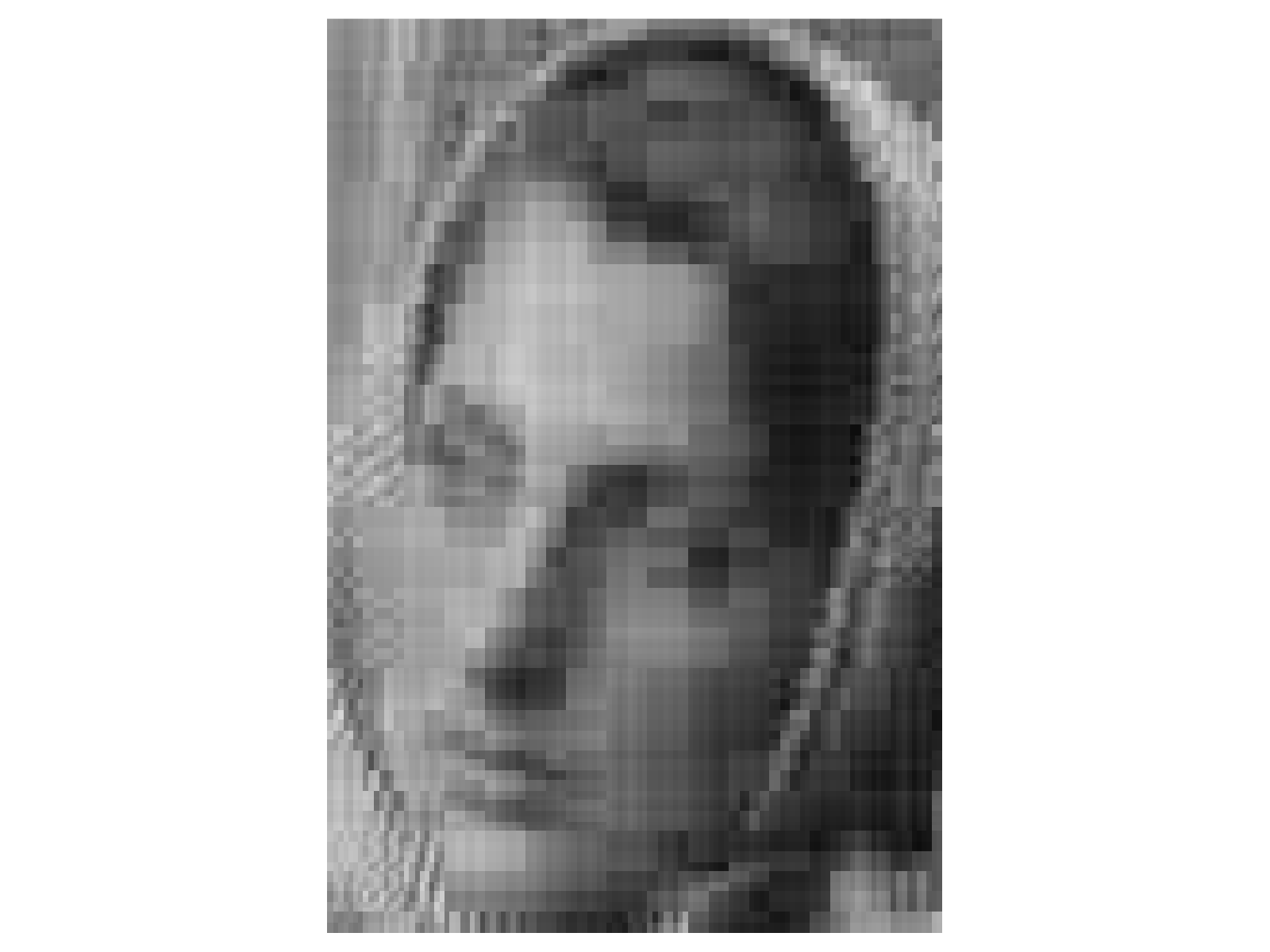}
    \centering\includegraphics[width=.475\textwidth]{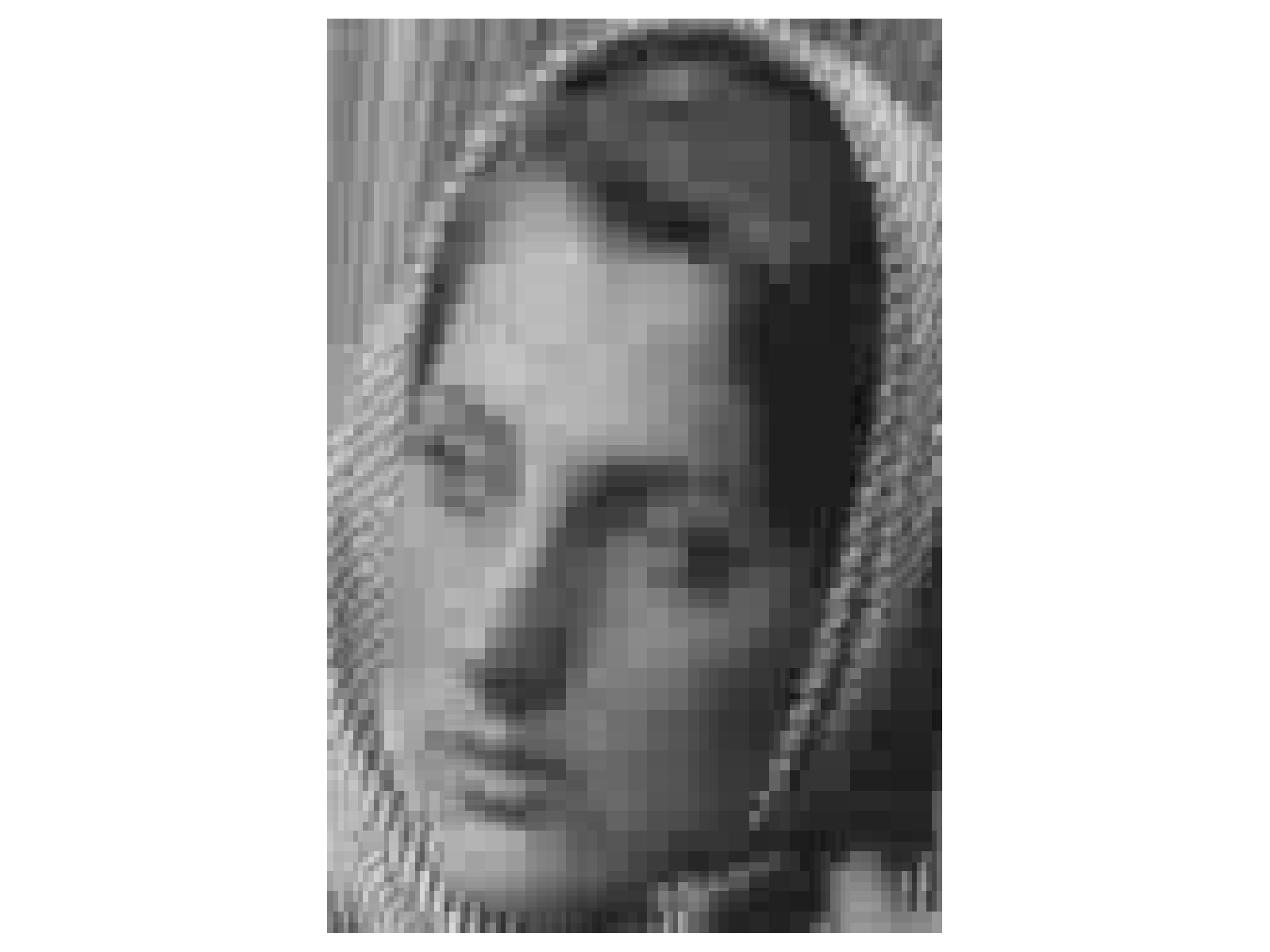}
    \caption{face}
  \end{subfigure}
  \quad
  \begin{subfigure}{.475\textwidth}
    \centering\includegraphics[width=.475\textwidth]{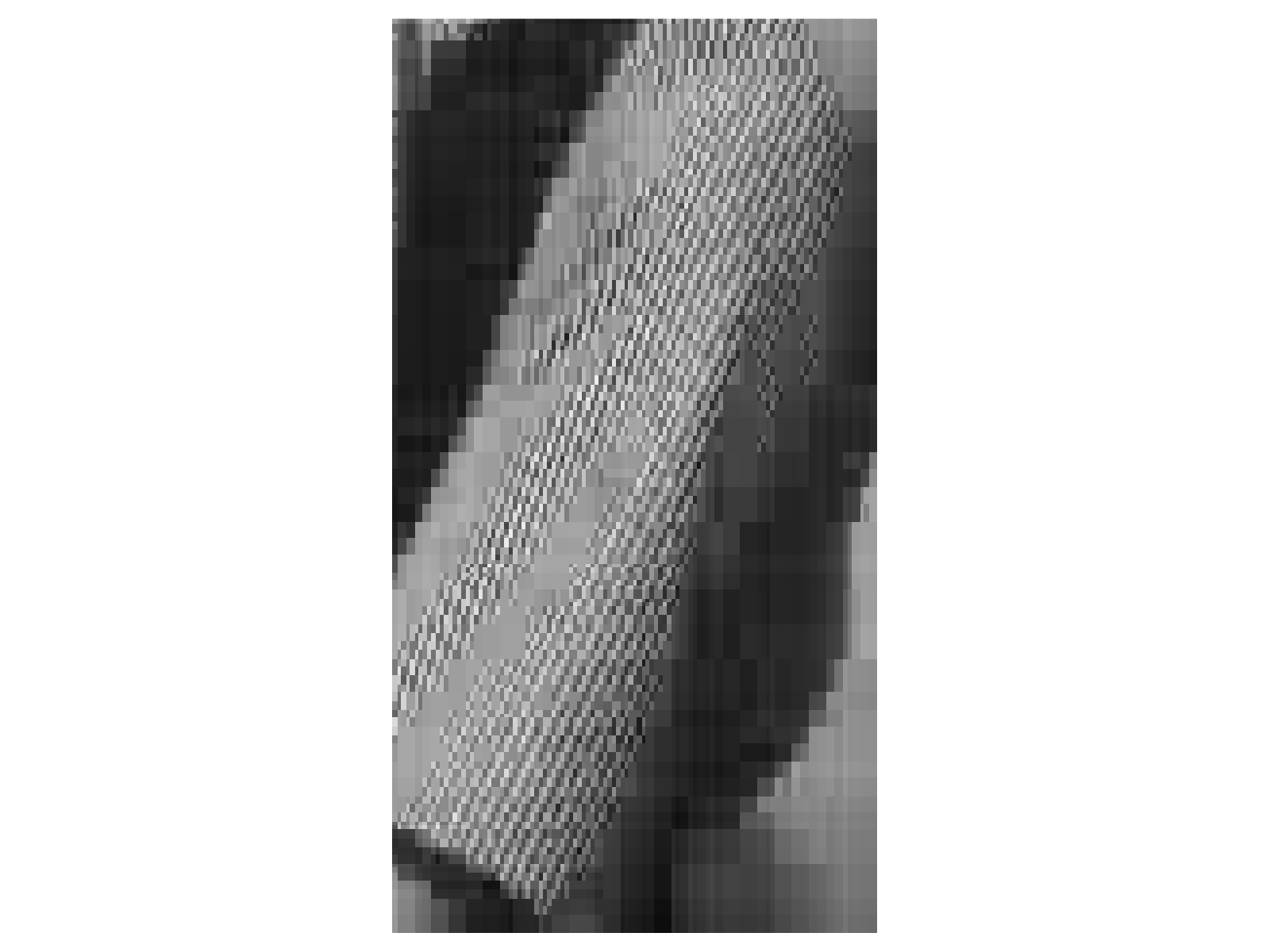}
    \centering\includegraphics[width=.475\textwidth]{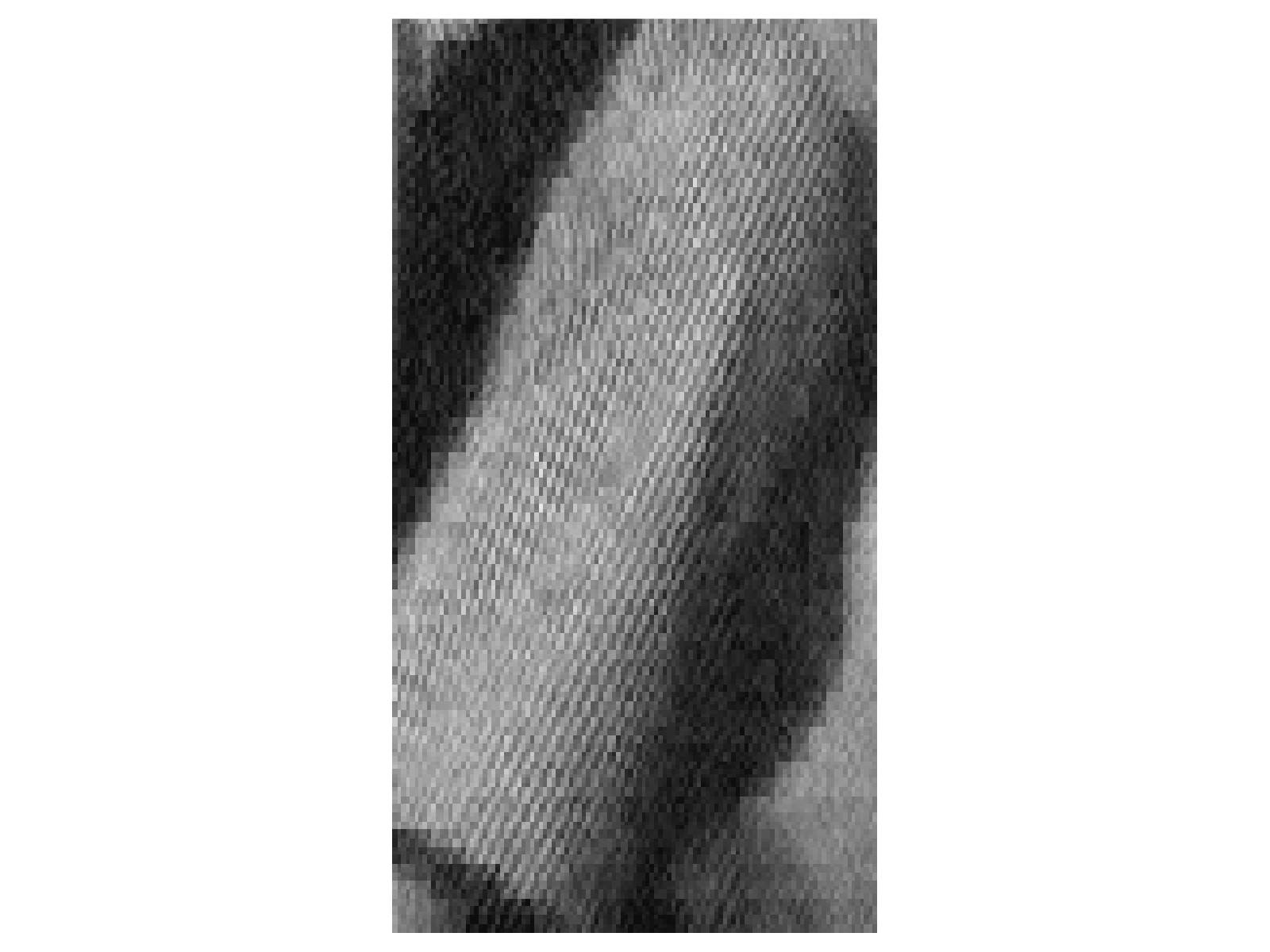}\\
    \centering\includegraphics[width=.475\textwidth]{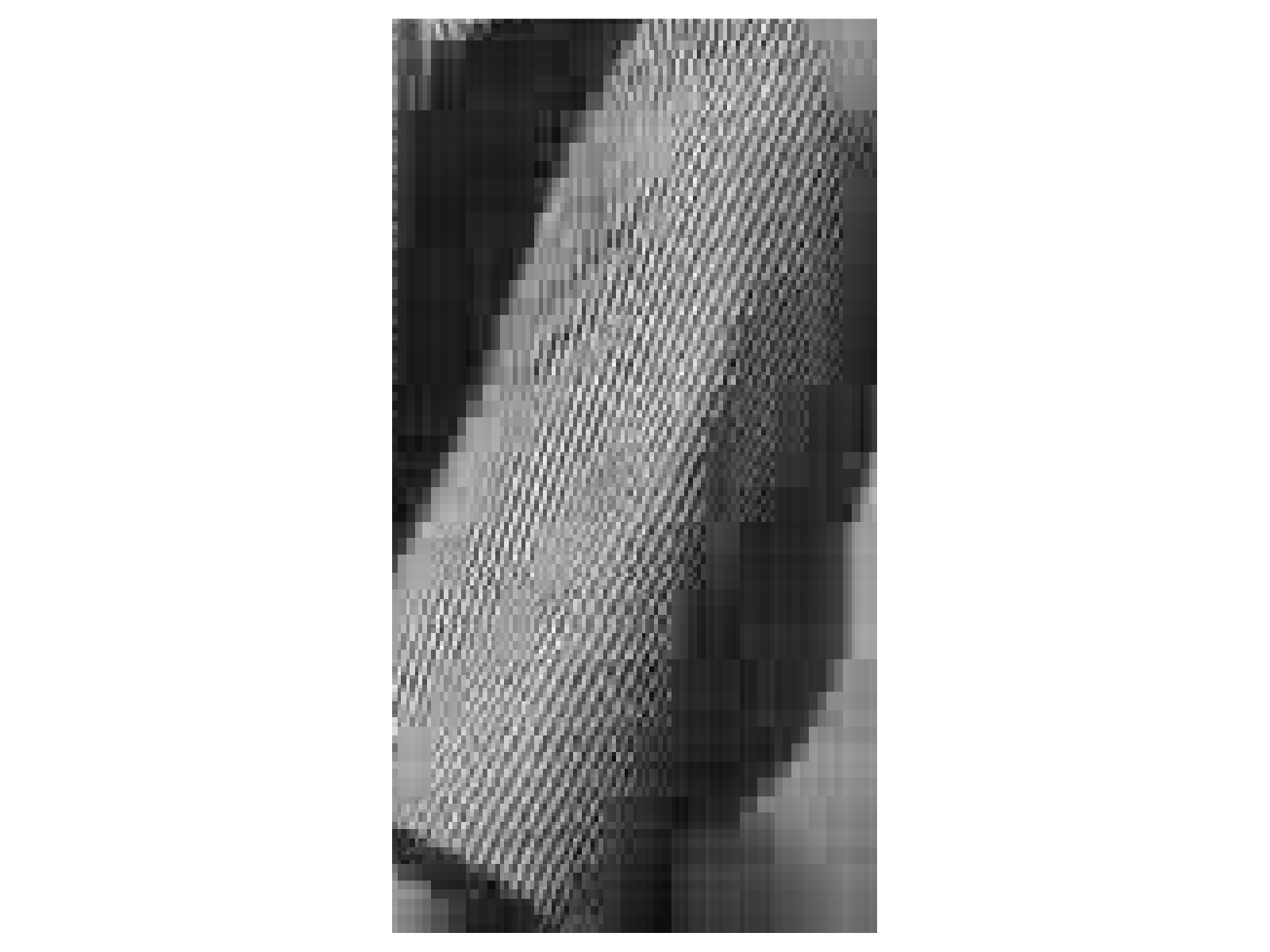}
    \centering\includegraphics[width=.475\textwidth]{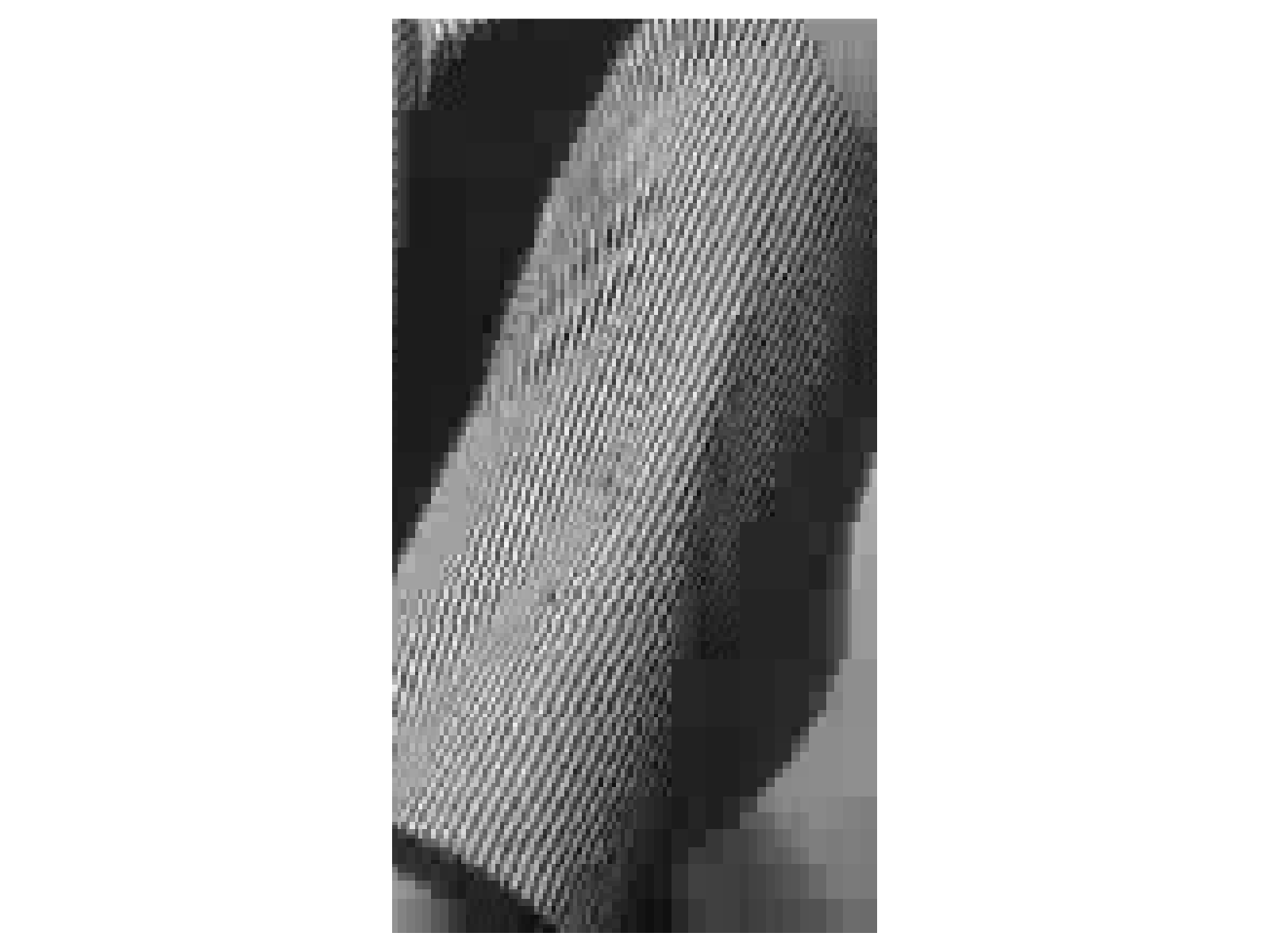}
    \caption{left leg}
  \end{subfigure}
  \caption{Zoomed up versions of Fig.~\ref{fig:barbara_full}.
    Methods used from top left to bottom right are: Haar; GHWT c2f; GHWT f2c;
    and eGHWT, respectively. Online viewing is recommended for the details.}
  \label{fig:barbara_part}
\end{figure}

In order to examine the quality of approximations visually,
Fig.~\ref{fig:barbara_full} displays those approximations using
only $1/32 = 3.125\%$ of the most significant expansion coefficients
for each basis.
The eGHWT best basis outperforms all the others with least blocky artifacts.
To examine the visual quality of these approximations further,
Fig.~\ref{fig:barbara_part} shows the zoomed-up face and left leg of those
approximations. Especially for the leg region that has some specific texture,
i.e., stripe patterns, the eGHWT outperformed the rest of the methods.
The performance is measured by PSNR (peak signal-to-noise ratio). 
\begin{hide}
Given an $m\times n$ monochrome image $I$ and its approximation $K$, the PSNR is defined as
\begin{align*}
    PSNR &= 10 \log_{10}\frac{(\max_{ij} I_{ij})^2}{MSE}\\
    MSE & = \frac{1}{mn}\sum^m_{i = 1}\sum^n_{j=1}(I_{ij} - K_{ij})^2
\end{align*}
\end{hide}

We note that both this approximation experiment on the Barbara image and
that on the vehicular volume count data on the Toronto street network discussed
in Sect.~\ref{sec:toronto}, the order of the performance was the same:
eGHWT > GHWT f2c > graph Haar > GHWT c2f $\geq$ graph Walsh.
This order in fact makes sense for those input datasets. The eGHWT could simply
search much larger ONBs than the GHWT best bases; the graph Haar basis is
a part of the GHWT f2c dictionary; and the graph Walsh basis is a part of
both the GHWT c2f and f2c dictionaries. The graph Walsh basis for these graph
signals was too global while the GHWT c2f best basis tended to select nonlocal
features compared to the eGHWT, GHWT f2c, and the graph Haar.

\subsubsection{The Haar Transform for Images with Non-Dyadic Size}
\label{sec:nondyadic}
For images of non-dyadic size, there is no straightforward way to obtain the
partition trees in a non-adaptive manner unlike the dyadic Barbara image of
$512 \times 512$ in the previous subsection.
This is a common problem faced by the classical Haar and wavelet transforms
as well: they were designed to work for images of dyadic size.
Non-dyadic images are often modified by zero padding, even extension at
the boundary pixels, or other methods before the Haar transform is applied.
We propose to apply the Haar transform on a non-dyadic image without
modifying the input image using the eGHWT dictionary.

To obtain the binary partition trees, we need to cut an input image
$\bI \in \Rf^{M \times N}$
horizontally or vertically into two parts recursively. Apart from using the
affinity matrices as we did for the vehicular volume count data on the Toronto
street network in Section~\ref{sec:toronto} and for the term-document matrix
analysis in \cite{IRION-SAITO-MLSP16}, we propose to use the
\emph{penalized total variation} (PTV) cost to partition a non-dyadic input
image.
Denote the two sub-parts of $\bI$ as $\bI_1$ and $\bI_2$.
We search for the optimal cut such that
\begin{displaymath}
\text{Penalized Total Variation Cost} \define \frac{\| \bI_1 \|_{\mathrm{TV}}}{|\bI_1|^p} + \frac{\| \bI_2 \|_{\mathrm{TV}}}{|\bI_2|^p}\ (p > 0)
\end{displaymath}
is minimized, where $\| \bI_k \|_{\mathrm{TV}} \define \sum_{i,j} ( |I_k[i+1,j] - I_k[i,j]| + |I_k[i,j+1] - I_k[i,j]|)$, and $|\bI_k|$ indicates the number of pixels
in $\bI_k$, $k=1, 2$. The denominator is used to make sure that the size of
$\bI_1$ and that of $\bI_2$ are close so that the tree becomes nearly balanced.
Recursively applying the horizontal cut on the rows of $\bI$ and the vertical cut
on the columns of $\bI$ will give us two binary partition trees. We can then
select the 2D Haar basis from the eGHWT dictionary or search for the best basis
with minimal cost (note that this cost function for the best-basis search is
the $\ell^1$-norm of the expansion coefficients, and is different from the
PTV cost above).

To demonstrate this strategy, we chose an image patch of size $100 \times 100$
around the face part from the original $512 \times 512$ Barbara image so that
it is non-dyadic. To determine the value of $p$, we need to balance between the
total variation and structure of the partition tree.
Larger $p$ means less total variation value after split but a more balanced
partition tree may be obtained. The value of $p$ can be fine-tuned based on
the evaluation of the final task, for example, the area under the curve of the
relative approximation error in the compression task~\cite{IRION-SAITO-TSIPN}.
In the numerical experiments below, after conducting preliminary partitioning
experiments using the PTV cost, we decided to choose $p=3$.

For comparison, we also used the \texttt{dwt} function supplied by
the \texttt{Wavelets.jl} package~\cite{JuliaWavelets} for the classical Haar
transform. We examined three different scenarios here: 1) directly input the
Barbara face image of size $100 \times 100$; 2) zero padding to four sides
of the original image to make it $128 \times 128$;
and 3) even reflection at the borders to make it 128 x 128.

\begin{figure}
  \centering
  \includegraphics[width=0.5\textwidth]{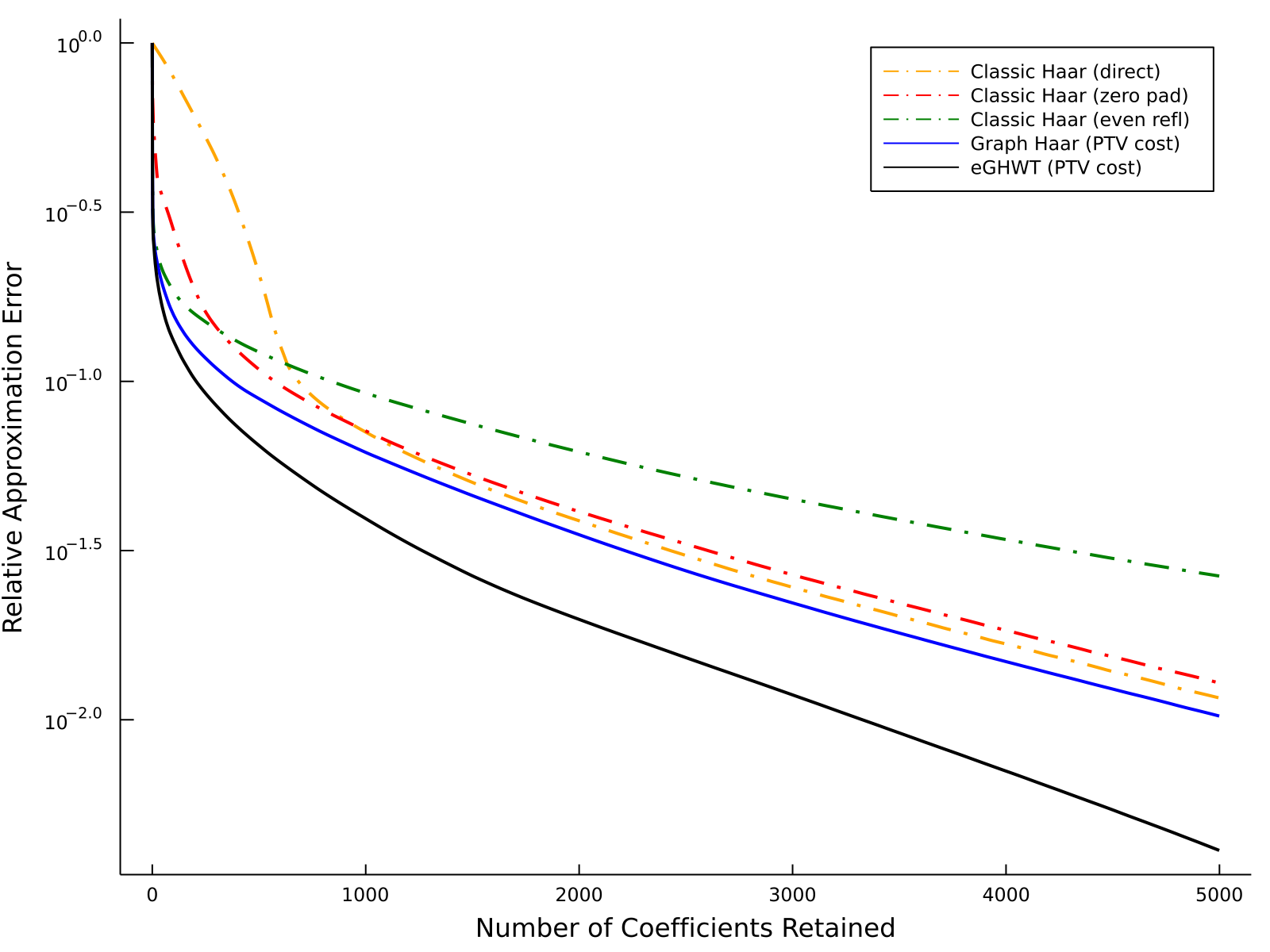}
  \caption{Comparison of approximation performance on the face part (of size
    $100 \times 100$ pixels) of the Barbara image: the classical Haar
    transform with various setups, the graph Haar basis, and the eGHWT best
    basis}
  \label{fig:unb_haar}
\end{figure}

Figure~\ref{fig:unb_haar} shows that the relative $\ell^2$-error curves of
these five methods. Note that we plotted these curves as a function of
the number of coefficients retained instead of the fraction of coefficients
retained that were used in Figs.~\ref{fig:toronto_l2} and \ref{fig:barbara_l2}.
This is because the zero-padded version and the even-reflection version have
$128 \time 128$ pixels although the degree of freedom is the same
$100 \times 100$ throughout the experiments.

Clearly, the graph-based methods, i.e., the graph Haar basis and the eGHWT best
basis outperformed the classical Haar transform applied to the three prepared
input images. 
The classical Haar transform applied to the original face image of size
$100 \times 100$ did perform poorly in the beginning because the \texttt{dwt}
function stops the decomposition when the sample (or pixel) size becomes
an odd integer. In this case, after two levels of decomposition,
it stopped ($100 \rightarrow 50 \rightarrow 25$).
Hence, it did not fully enjoy the usual advantage of deeper decomposition.
The classical Haar transform applied to the even-reflected image was the
worst performer among these five because the even-reflected image of
$128 \times 128$ is not periodic. The implementation of the \texttt{dwt}
assumes the periodic boundary condition by default, and if it does not
generate continuous periodic image, it would generate artificially-large
expansion coefficients due to the discontinuous periodic boundary condition;
see \cite{SAITO-REMY-ACHA, YAMATANI-SAITO, SAITO-PLASMA-ENGLISH} for further
information. We can summarize that our graph-based transforms can handle
images of non-dyadic size without any artificial preparations (zero padding
and even reflection) with some additional computational expense (the minimization
of the PTV cost for recursive partitioning).

\newpage
\subsection{Another Way to Construct a Graph from an Image for Efficient Approximation}
We can view a digital image of size $M \times N$ as a signal on a graph
consisting of $MN$ nodes by viewing each pixel as a node. Note that the
underlying graph is not a regular 2D lattice of size $M \times N$.
Rather it is a graph reflecting the relationship or affinity between pixels.
In other words, $w_{ij}$, the weight of the edge between $i$th and $j$th
pixels in that graph should reflect the affinity between local region around
these two pixels, and this weight may not be $0$ even if $i$th and $j$th pixels
are remotely located.
In the classical setting, this idea has been used in image denoising 
  (the so-called \emph{bilateral filtering})~\cite{TOMASI-MANDUCHI} and
  image segmentation~\cite{SHI-MALIK}. On the other hand, Szlam et al.\ 
  have proposed a more general approach for associating graphs and
  \emph{diffusion processes} to datasets and functions on such datasets,
  which includes the bilateral filtering of \cite{TOMASI-MANDUCHI} as a
  special case, and have applied to image denosing and transductive learning.
  See the review article of Milanfar~\cite{MILANFAR-IMG-FILTERING} for
  further connections between these classical and modern techniques and
  much more.
    
Here we define the edge weight $w_{ij}$ as Szlam et al.~\cite{SZLAM-MAGG-COIF}
did:
\begin{equation}
\label{eqn:edgeweight}
  w_{ij} = \exp\left(\frac{-\| \bF[i] - \bF[j] \|^2_2}{\sigma_F}\right) \cdot
  \begin{cases}
    \exp\left(\frac{-\| \bx[i] - \bx[j] \|^2_2}{\sigma_x}\right) & \text{if $\| \bx[i] - \bx[j] \|_2 < r$}\\
    0 & \text{otherwise}
  \end{cases}
\end{equation}
where $\bx[i] \in \Rf^2$ is the spatial location (i.e., coordinate) of node
(pixel) $i$, and $\bF[i]$ is a feature vector based on intensity, color, or
texture information of the local region centered at that node. As one can see
in Eq.~\eqref{eqn:edgeweight}, the pixels located within a disk with center
$\bx[i]$ and radius $r$ are considered to be the neighbors of the $i$th pixel.
The scale parameters, $\sigma_F$ and $\sigma_x$ must be chosen appropriately.
Once we construct this graph, we can apply the eGHWT in a straightforward manner.

\begin{figure}
  \begin{subfigure}{.475\textwidth}
    \centering\includegraphics[width=\textwidth]{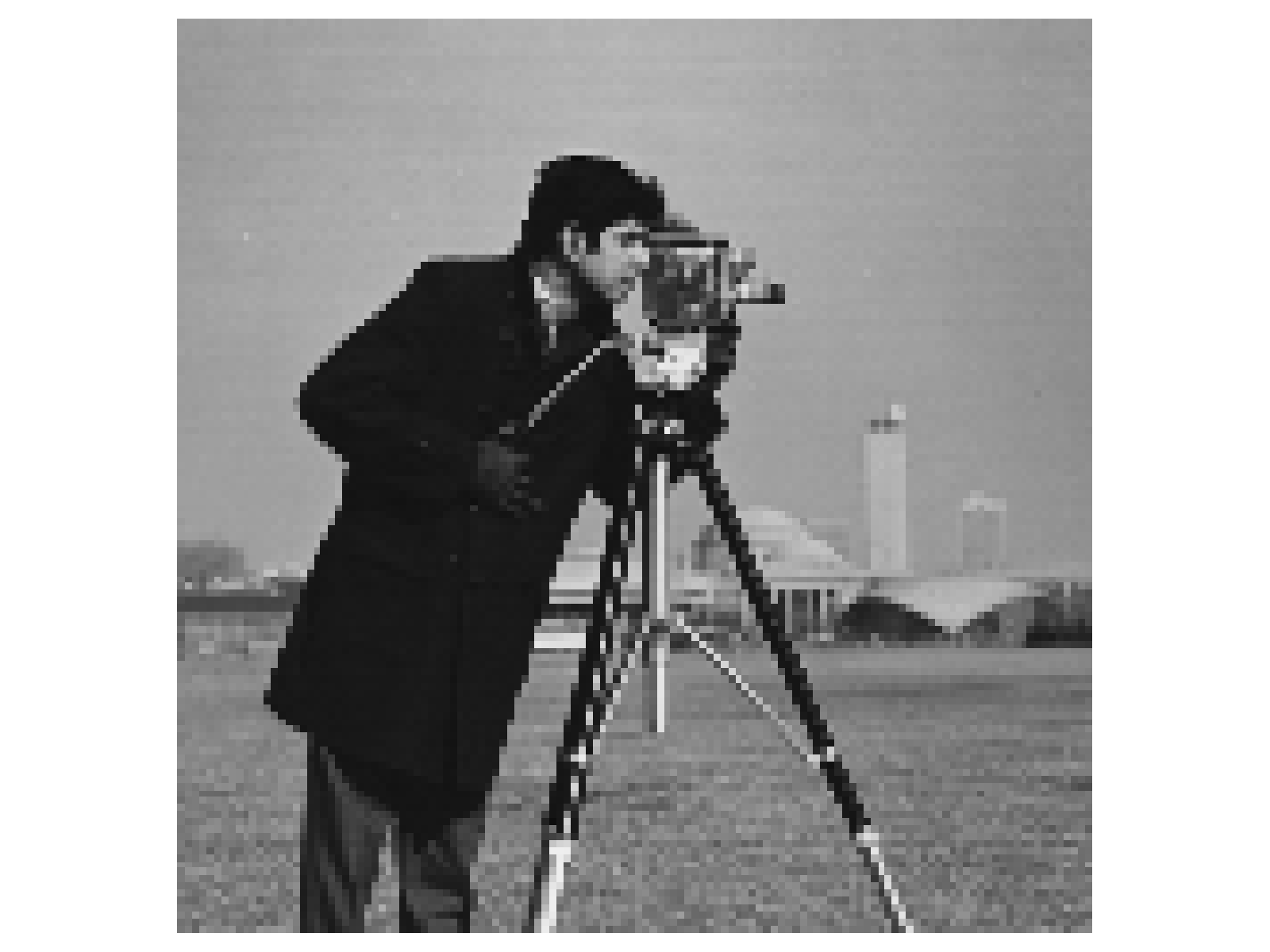}
    \caption{cameraman (subsampled)}
    \label{fig:cameraman}
  \end{subfigure}
  \quad
  \begin{subfigure}{.475\textwidth}
    \centering\includegraphics[width=\textwidth]{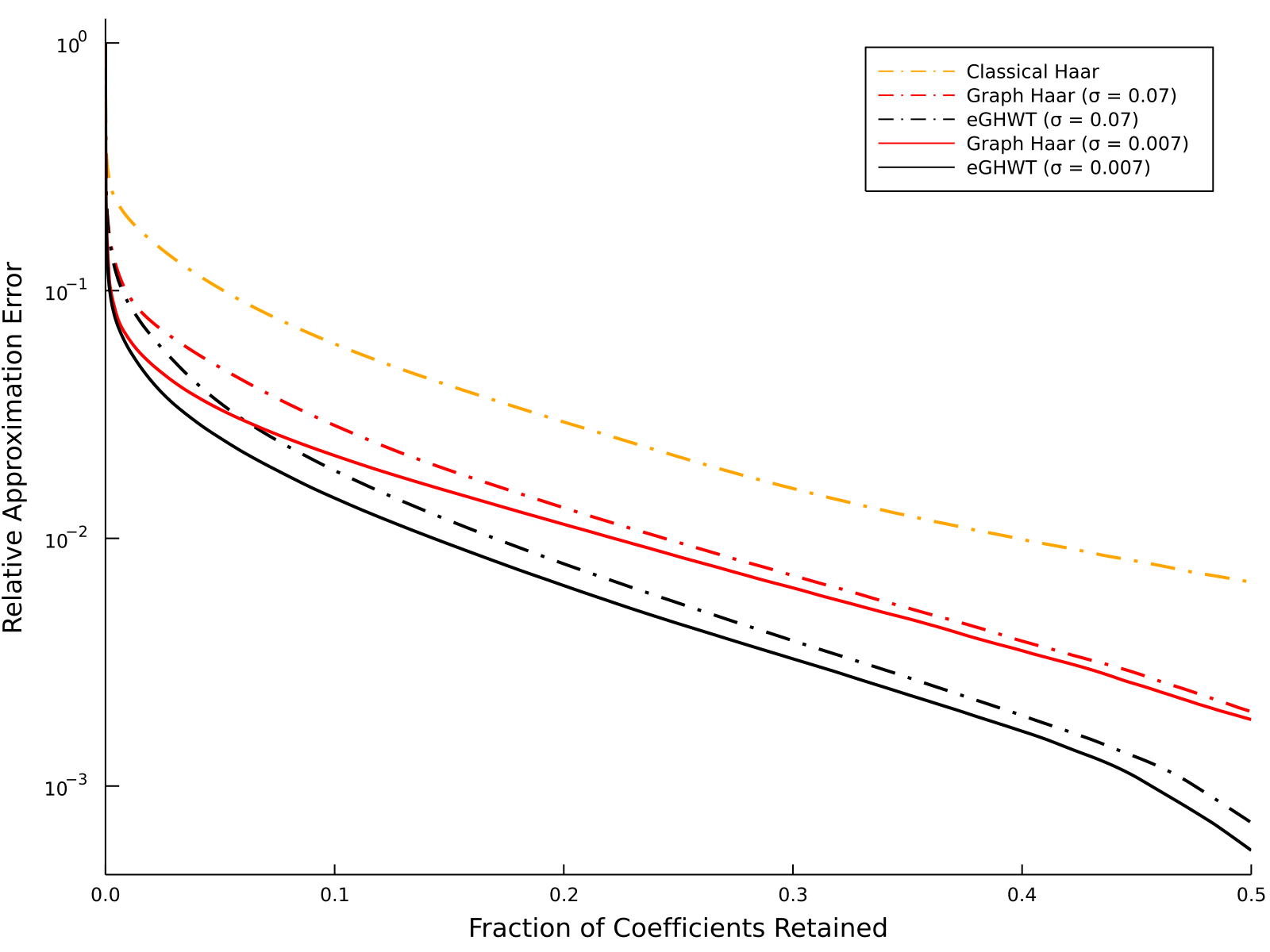}
    \caption{Relative $\ell^2$ approximation error}
    \label{fig:cameraman_l2}
  \end{subfigure}
  \caption{(a) The subsampled cameraman image of size $128 \times 128$;
    (b) Relative $\ell^2$ approximation error of (a) using five methods.}
\end{figure}

We examine two images here.
The first one (Fig.~\ref{fig:cameraman}) is the subsampled version of the
standard `cameraman' image; we subsampled the original cameraman image of size
$512 \times 512$ to $128 \times 128$ in order to reduce computational cost.
For the location parameters, we used $r=5$ and $\sigma_x=\infty$.
Note that $\sigma_x = \infty$ means that $w_{i \cdot}$ becomes an
indicator/characteristic function of a disk of radius $r$ with center $i$.
This setup certainly simplifies our experiments, and could sparsify the weight
matrix if $r$ is not too large.
On the other hand, the feature vector $\bF[i]$ of the $i$th pixel location,
we found that simply choosing raw pixel value as $\bF[i]$ can get good enough
results for relatively simple images (e.g., piecewise smooth without too much
high frequency textures) like the cameraman image.
However, the `Gaussian bandwidth' parameter for the features, $\sigma_F$, needs
to be tuned more carefully.
A possible tuning trick is to start from the median of all possible values of
the numerator in the exponential term in Eq.~\eqref{eqn:edgeweight}, i.e.,
$-\| \bF[i] - \bF[j] \|^2_2$~\cite{SZLAM-MAGG-COIF}, examine several
other values around that median, and then choose the value yielding the best
result. For more sophisticated approach to tune the Gaussian bandwidth,
see~\cite{LINDENBAUM-ETAL}.
In this example, we used $\sigma_F=0.007$ and $\sigma_F = 0.07$ after several
trials starting from the simple median approach in order to demonstrate
the effect of this parameter for the eGHWT best basis.

\begin{figure}
  \begin{subfigure}{.475\textwidth}
    \centering\includegraphics[width=\textwidth]{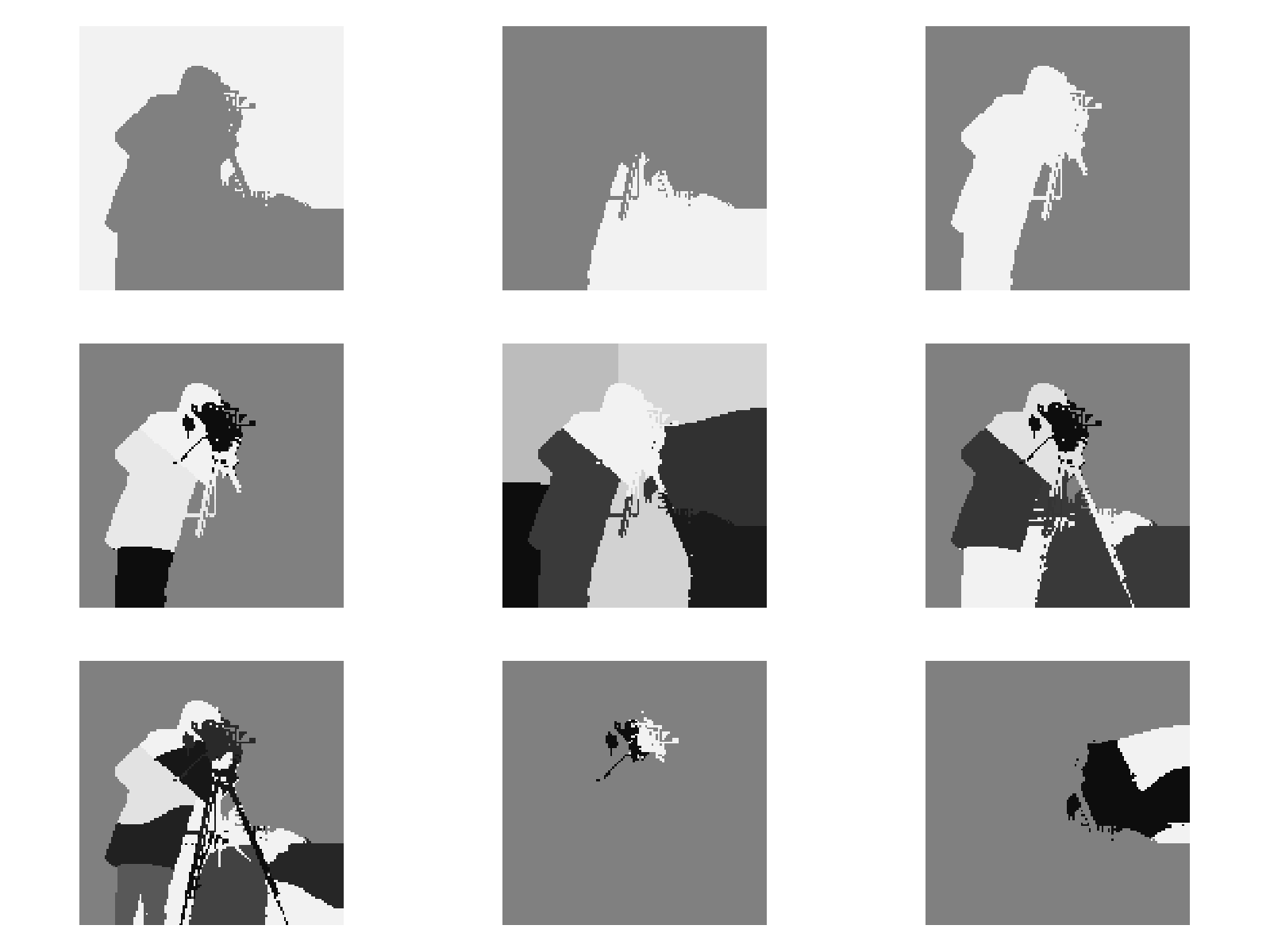}
    \caption{$\sigma_F = 0.007$}
    \label{fig:cameraman_top9_007}
  \end{subfigure}
  \quad
  \begin{subfigure}{.475\textwidth}
    \centering\includegraphics[width=\textwidth]{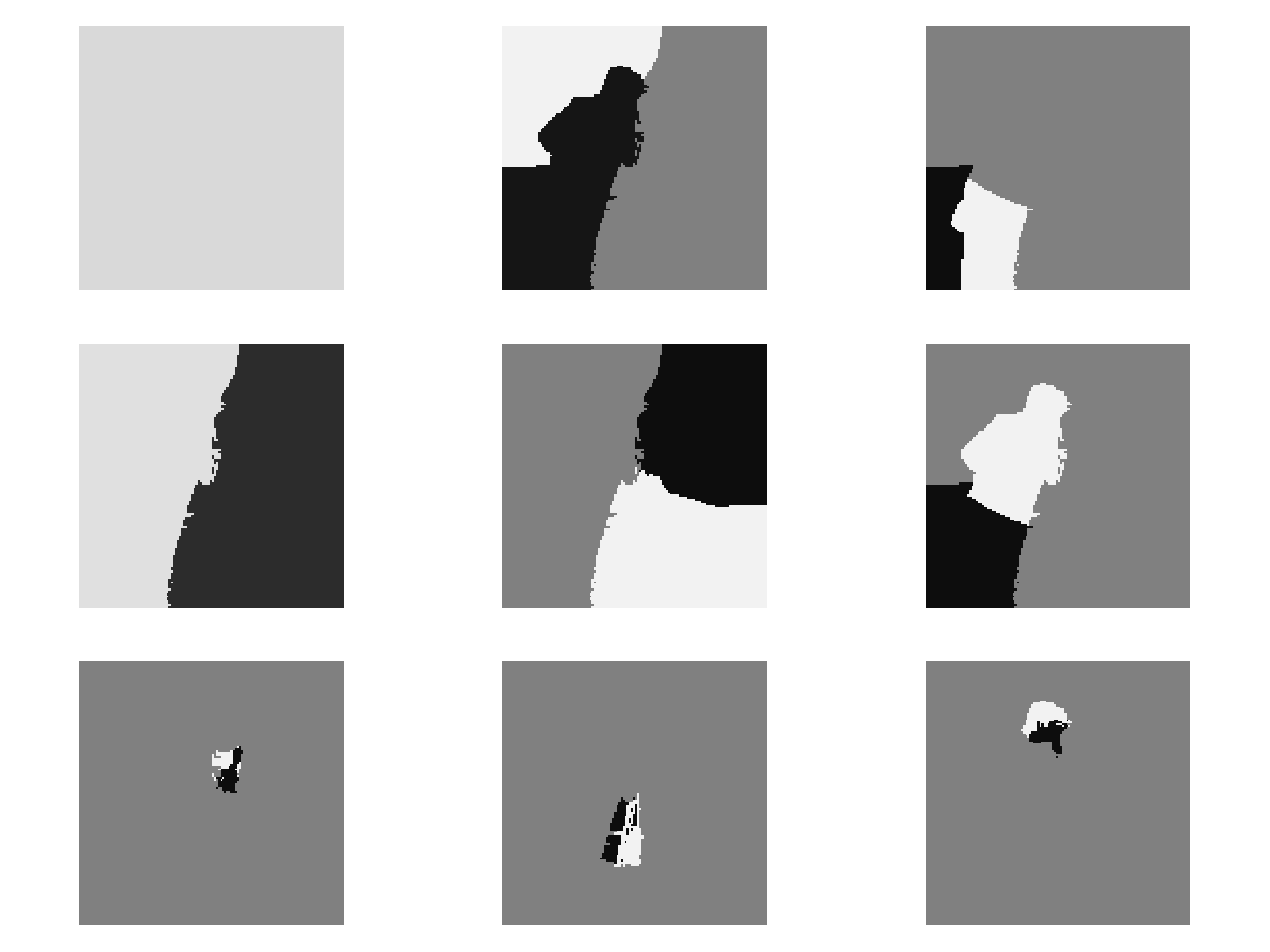}
    \caption{$\sigma_F = 0.07$}
    \label{fig:cameraman_top9_07}
  \end{subfigure}
  \caption{The top nine eGHWT best-basis vectors: (a) $\sigma_F = 0.007$;
    (b) $\sigma_F = 0.07$}
  \label{fig:cameraman_top9}
\end{figure}

Figure~\ref{fig:cameraman_l2} shows our results on the subsampled cameraman
image. Figure~\ref{fig:cameraman_l2} demonstrates that the decay rate of the
expansion coefficients w.r.t.\ the eGHWT best basis is much faster than that
of the classical Haar transform. Moreover, the eGHWT best-basis vectors extract
some meaningful features of the image. Figure~\ref{fig:cameraman_top9} shows the
eGHWT best-basis vectors corresponding to the largest nine expansion coefficients
in magnitude. We can see that the human part and the camera part are captured by
individual basis vectors.
Furthermore, in terms of capturing the objects and generating meaningful
segmentation, we observe that the results with $\sigma_F = 0.007$ shown in
Fig.~\ref{fig:cameraman_top9_007} are better than those with $\sigma_F = 0.07$
shown in Fig.~\ref{fig:cameraman_top9_07}. This observation coincides with
the better decay rate of the former than that of latter in
Fig.~\ref{fig:cameraman_l2}. Therefore, we can conclude that good segmentation
(or equivalently, a good affinity matrix setup) is quite important for
constructing a basis of good quality and consequently for approximating an input
image efficiently.
  
\begin{figure}
  \begin{subfigure}{.475\textwidth}
    \centering\includegraphics[width=\textwidth]{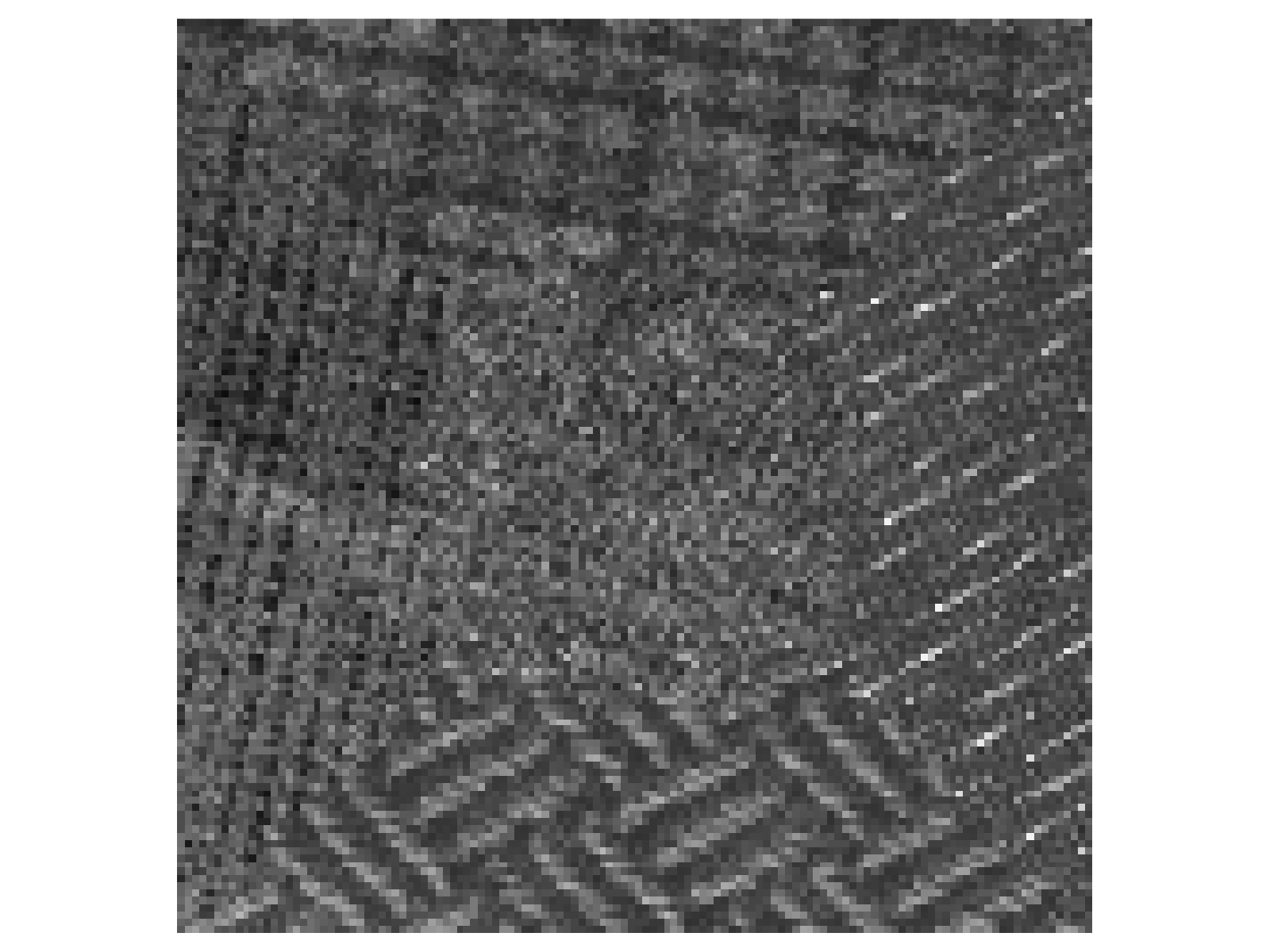}
    \caption{Subsampled Original}
    \label{fig:textures_orig}
  \end{subfigure}
  \quad
  \begin{subfigure}{.475\textwidth}
    \centering\includegraphics[width=\textwidth]{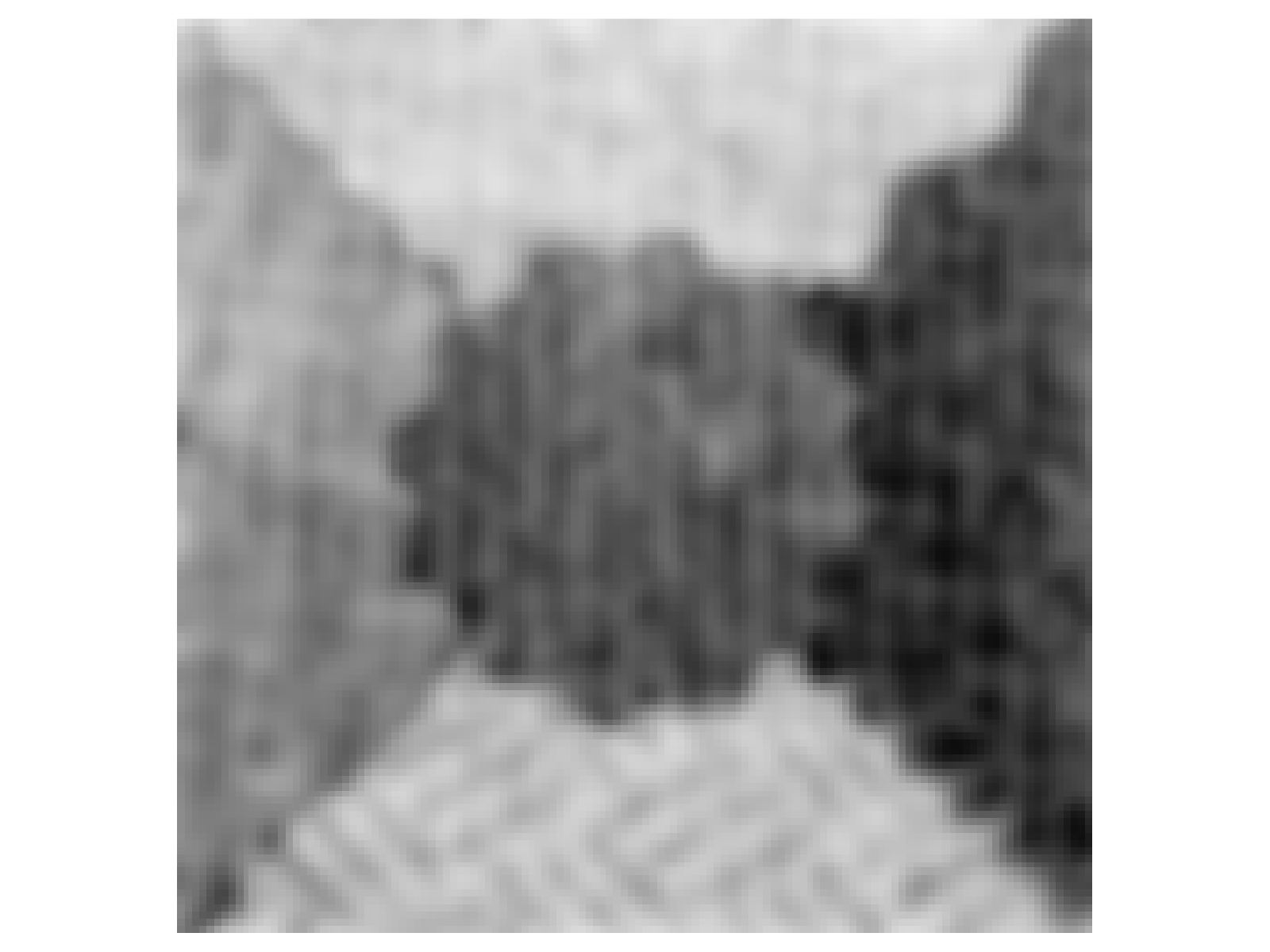}
    \caption{Mask from Gabor + PCA}
    \label{fig:textures_mask}
  \end{subfigure}
  \caption{(a) A composite texture image of size $128 \times 128$
    after subsampling the original image of size $512 \times 512$;
    (b) Mask image computed from PCA of 8 Gabor filtered images
    followed by subsampling to $128 \times 128$ pixels}
\end{figure}

\begin{figure}
  \begin{subfigure}{.475\textwidth}
    \centering\includegraphics[width=\textwidth]{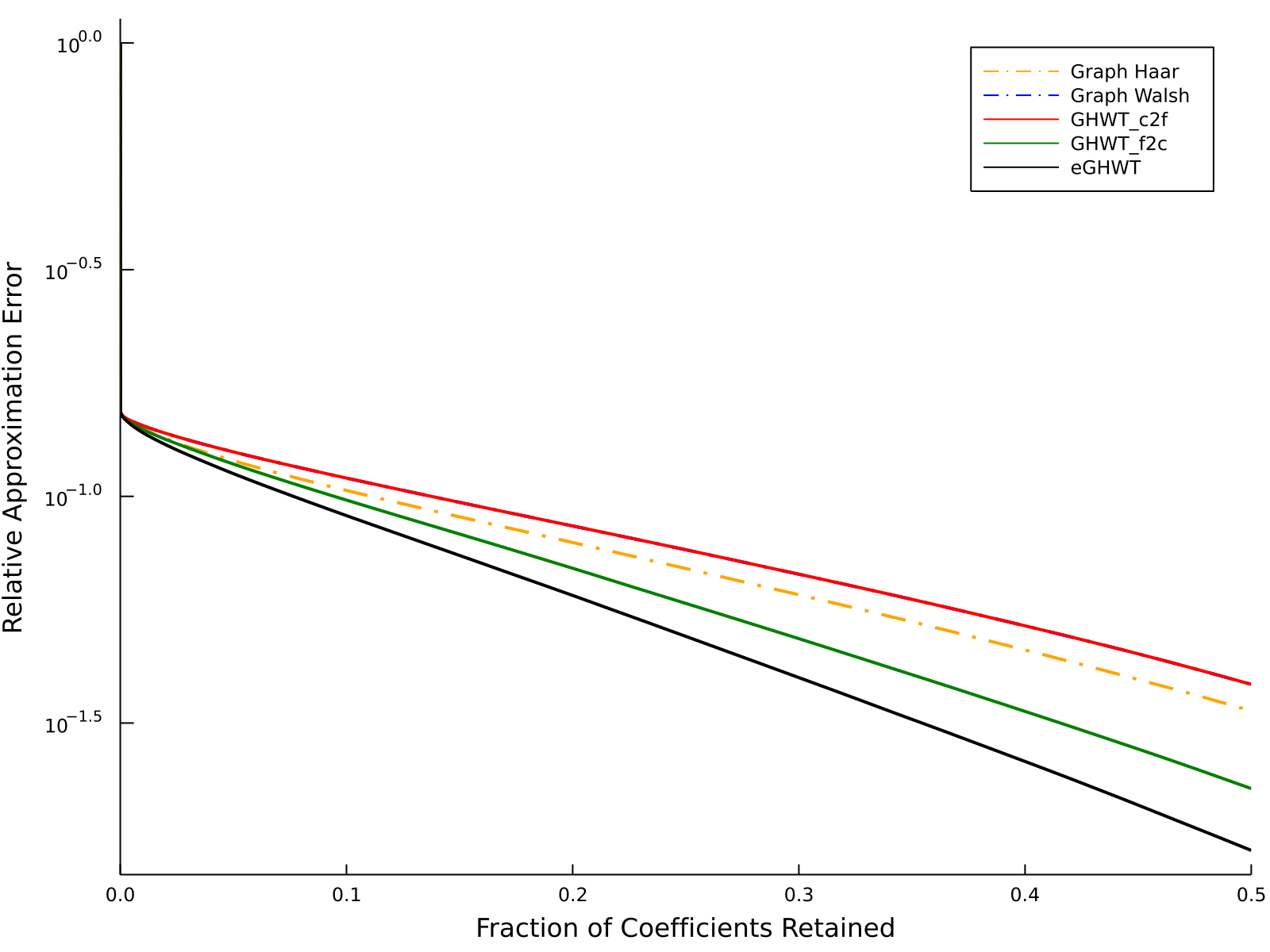}
    \caption{Relative $\ell^2$ approximation error}
    \label{fig:textures_l2}
  \end{subfigure}
\quad
  \begin{subfigure}{.475\textwidth}
    \centering\includegraphics[width=\textwidth]{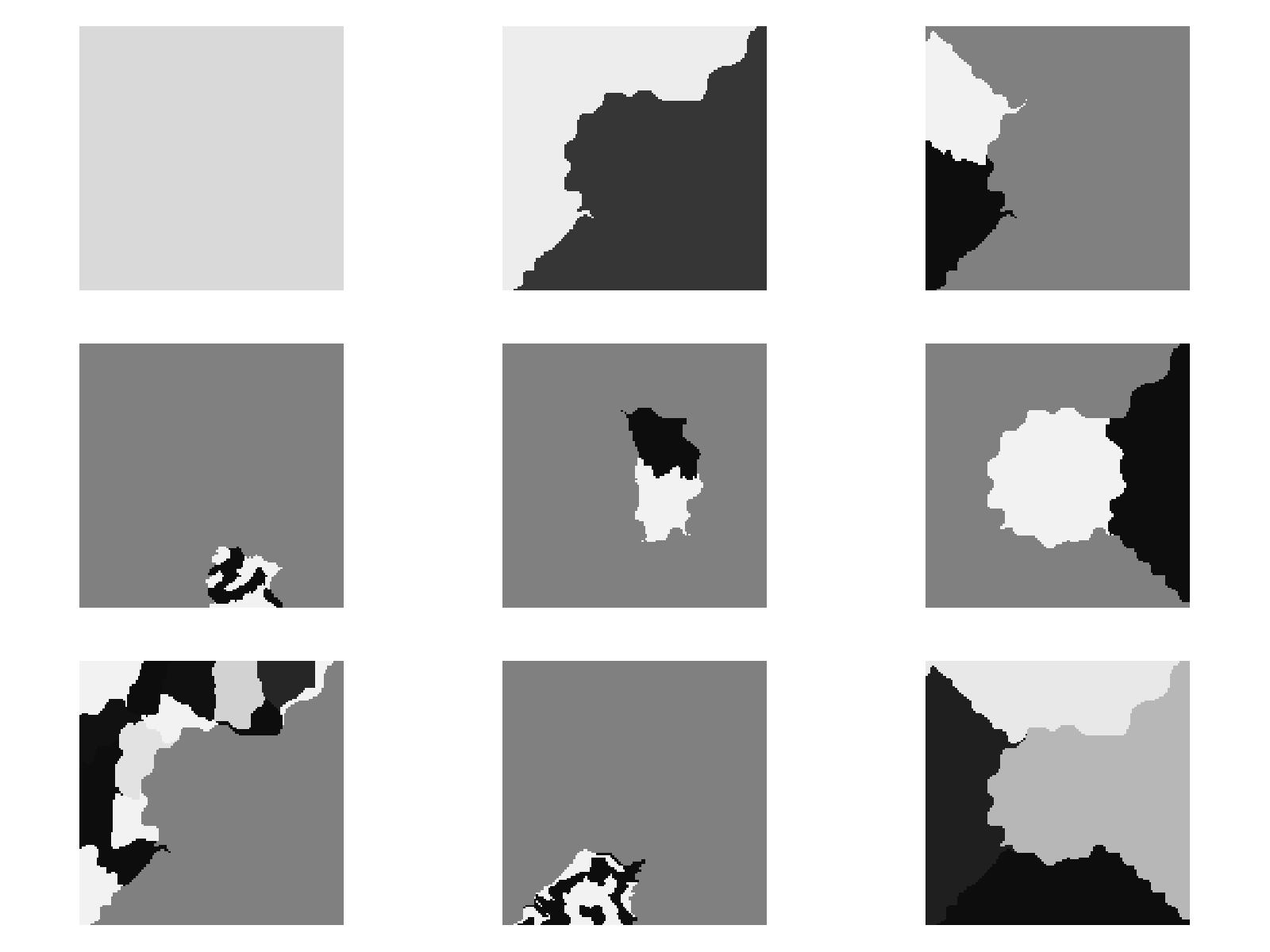}
    \caption{Top 9 eGHWT best-basis vectors}
    \label{fig:textures_top9}
  \end{subfigure}
  \caption{(a) Relative $\ell^2$ approximation error of Fig.~\ref{fig:textures_orig} using five methods; (b) Top 9 eGHWT best-basis vectors}
  \label{fig:texture_image_analysis}
\end{figure}

Our second example is a composite texture image shown in
Fig.~\ref{fig:textures_orig}, which was generated and analyzed by
Ojala et al.~\cite{OUTEX}.
Our method, i.e., applying the eGHWT on an image viewed as a graph,
allows us to generate basis vectors with irregular support that is adapted to
the structure of the input image as shown in the cameraman example above.
This strategy also works well on this composite texture image \emph{if we choose
the appropriate features in Eq.~\eqref{eqn:edgeweight} to generate a good weight
matrix}. The raw pixel values as the features, which we used for the cameraman
example, do not work in this composite texture image, which consists of highly
oscillatory anisotropic features. Ideally, we want to use a piecewise constant
image delineating each of the five textured regions for the weight matrix
computation. To generate such a piecewise constant image (or rather its
close approximation), we did the following: 
\begin{enumerate}
\item Apply a group of (say, $k$) 2D \emph{Gabor filters} of various frequencies
  and orientations on the original image of size $M \times N$;
\item Compute the absolute values of each Gabor filtered image to construct
  a nonnegative vector of length $k$ at each pixel location;
\item Standardize each of these $k$ components so that each component has mean
  $0$ and variance $1$;
\item Apply Principal Component Analysis (PCA)~\cite{JOLLIFFE2} on these
  standardized $MN$ vectors in $\Rf^k$;
\item Extract the first principal component (organized as a single
  image of the same size as the input image), normalize it to be in the range of
  $[0, 1]$, and use it as features in Eq.~\eqref{eqn:edgeweight}.
\end{enumerate}
In Step 1, the Gabor filters are used because they are quite efficient to
capture high frequency anisotropic features~\cite{GRIGORESCU-PETKOV-KRUIZINGA}.
In this particular image, we used four spatial frequencies: $0.2$, $0.3$, $0.4$,
and $0.5$ (unit: 1/pixel), and two different orientations: $\pi/3$ and
$5\pi/6$, for the Gabor filters. This generated a set of $k=8$ nonnegative
matrices in Step 2. We also note that the Gaussian bandwidth or the standard
deviation of the Gaussian envelop used in the Gabor filters needs to be
tuned appropriately. In our experiments, we used the wavelength-dependent
bandwidth, i.e., $\sigma = 3/\pi \cdot \sqrt{\ln2/2} \cdot \lambda$,
where $\lambda \in \{0.2, 0.3, 0.4, 0.5\}$, as \cite{PETKOV-KRUIZINGA} suggests.
The normalized first principal component
in Step 5 as an image is shown in Fig.~\ref{fig:textures_mask}.
The pixel values of this `mask' image are used as (scalar) features $\bF[i]$ in
Eq.~\eqref{eqn:edgeweight}.
Note that this mask computation was done on the original $512 \times 512$ image,
which was subsequently subsampled to have $128 \time 128$ pixels to match the
subsampled original image.
As for the Gaussian bandwidth, $\sigma_F$, in Eq.~\eqref{eqn:edgeweight},
we set $\sigma_F = 0.0005$ after several trials.
The other parameters, $(r, \sigma_x)$, were set as $(3, \infty)$.
Figure~\ref{fig:textures_l2} compares the performance of five different methods
in approximating the composite texture image shown in Fig.~\ref{fig:textures_orig}.
The order of approximation performance is the same as
the previous examples, i.e., the vehicular volume counts on the Toronto street
network in Fig.~\ref{fig:toronto_l2} and the Barbara image using the predefined
dyadic partitions in Fig.~\ref{fig:barbara_l2}: eGHWT > GHWT f2c > graph Haar
> GHWT c2f = graph Walsh. 
Figure~\ref{fig:textures_top9} displays the top nine eGHWT
best-basis vectors. We can see that the support of these basis vectors
approximately coincide with the five sections of the composite texture image,
and some of the basis vectors, particularly the fourth and eighth ones,
exhibit the oscillatory anisotropic features.

\section{Discussion}
\label{sec:disc}
In this article, we have introduced the \emph{extended Generalized Haar-Walsh
Transform} (eGHWT). After briefly reviewing the previous Generalized Haar-Walsh
Transform (GHWT), we have described how the GHWT can be improved with the new
best-basis algorithm, which is the generalization of the Thiele-Villemoes
algorithm~\cite{THIELE-VILLEMOES} for the graph setting. We refer to this whole
procedure of developing the extended Haar-Walsh wavelet packet dictionary on a
graph and selecting the best basis from it as the \emph{eGHWT}.
Moreover, we have developed the 2D eGHWT for matrix signals by viewing them as
tensor products of two graphs, which is a generalization of
the Lindberg-Villemoes algorithm~\cite{LINDBERG-VILLEMOES} for the graph setting.

We then showcased some applications of the eGHWT. When analyzing graph signals,
we demonstrated the improvement over the GHWT on synthetic and real data.
For the simple synthetic 6-node signal, we showed that the best basis from the
eGHWT can be selected by neither the GHWT c2f dictionary nor the GHWT f2c
dictionary and it had the smaller cost than the GHWT c2f/f2c best bases could
provide.
On the vehicular volume data on the Toronto street network, the eGHWT had the
best approximation performance among the methods we considered.
Then we proceeded to the applications to image approximation.
After demonstrated the superiority of the eGHWT combined with the predetermined
recursive partitioning on the original Barbara image of dyadic size, 
we proposed the use of the data adaptive recursive partitioning using the
penalized total variation for images of non-dyadic size, and again showed
the superiority of the eGHWT and the graph Haar transform over the classical
Haar transform applied to the preprocessed versions of a non-dyadic input image.
Finally, we demonstrated that the eGHWT could be applied to a graph generated
from an input image by carefully choosing the edge weights that encode the
similarity between pixels and their local neighbors, to get not only
superior approximation performance but also meaningful and interpretable
basis vectors that have non-rectangular supports and can extract certain
features and attributes of an input image.

The eGHWT basis dictionary is constructed upon the binary partition tree (or
a tensor product of binary partition trees in the case of 2D signals).
Currently, we use the Fiedler vectors of random-walk-normalized Laplacian
matrices to form the binary partition tree. However, as we have mentioned
earlier, our method is so flexible that any graph cut method or general
clustering method can be used, as long as the binary partition tree is formed.

If one is interested in \emph{compressing} a given graph signal instead of
  simply approximating it, where the encoding bitrate becomes an important issue,
  the eGHWT should be still useful and efficient.
  There is no need to transmit the eGHWT best-basis vectors.
  If both the sender and the receiver have the eGHWT
  algorithm, then the only information the sender needs to transmit is:
  1) the input graph structure via its adjacency matrix (which is often quite
  sparse and efficiently compressible); 2) the graph partitioning information to
  reproduce the hierarchical bipartition tree of the input graph; 3) the indices
  of the eGHWT best basis within this tree; and 4) the retained coefficients
  (after appropriate quantization) and their indices within the eGHWT best basis.

Another major contribution of our work is the software package we have developed.
Based on the MTSG toolbox written in
MATLAB\textregistered\, by Jeff Irion~\cite{MTSG},
we have developed the \texttt{MultiscaleGraphSignalTransforms.jl} package~\cite{MultiscaleGraphSignalTransforms}
written entirely in the \emph{Julia} programming language~\cite{JULIA},
which includes the new eGHWT implementation for 1D and 2D signals as well as
the natural graph wavelet packet dictionaries that our group has recently
developed~\cite{CLONINGER-LI-SAITO}.
We hope that interested readers will download the software themselves, and
conduct their own experiments with it: \\
\url{https://github.com/UCD4IDS/MultiscaleGraphSignalTransforms.jl}.

The readers might feel that the graphs we have used in this article are
  rather restrictive: we have only used 2D irregular grids (i.e., the Toronto
  street map) and 2D lattices (i.e., the standard digital images).
  However, we want to note that applying the eGHWT to more general graphs
  (e.g., social network graphs, etc.) is quite straightforward, and no
  algorithmic modification is necessary as long as an input graph is simple,
  connected, and undirected. As we have discussed earlier,
  the performance of the eGHWT for even such general graphs is always superior
  to that of the GHWT, the graph Haar basis, and the graph Walsh basis.
  We strongly encourage interested readers to try out our software package
  described above for such general graph signals.

There remains a lot of related projects to be done.
The most urgent one is to implement the version of the eGHWT that can handle
\emph{multiple} graph signals (on a given fixed graph).
In the classical setting, Wickerhauser proposed the so-called
\emph{Joint Best Basis} (JBB) algorithm~\cite[Sect.~11.2]{WICK-WPK} while Saito
proposed the so-called \emph{Least Statistically-Dependent Basis} (LSDB)
algorithm \cite{SAITO-LSDB3}, which correspond to the fast and approximate
version of the PCA and the ICA, respectively.
Moreover, our group has also developed the Local Discriminant Basis (LDB)
algorithm that can extract distinguishing local features for signal
classification problems~\cite{SAITO-COIF-JMIV, SAITO-COIF-GESHWIND-WARNER}.
We have already implemented the HGLET and the GHWT that can handle
multiple graph signals and that can compute the JBB, the LSDB, and the LDB.
Hence, making that version of the eGHWT is relatively straightforward, and
we plan to do so very soon.

Another important project, which we are currently pursuing, is to use an
appropriate subset of the scaling vectors in the eGHWT dictionary for
estimating good initialization to start the
\emph{Nonnegative Matrix Factorization} (NMF) algorithms that are of iterative
nature; see, e.g., \cite{NMF-BOOK}.

Finally, as a long-term research project, we plan to extend the GHWT and eGHWT
\emph{beyond} matrix-form data, i.e., to \emph{tensorial} data
(see, e.g., \cite{QI-LUO}), which seems quite promising and worth trying.

We plan to report our investigation and results on these projects at a later date.


\begin{acknowledgements}
This research was partially supported by the US National Science Foundation
grants DMS-1418779, DMS-1912747, CCF-1934568; the US Office of
Naval Research grant N00014-20-1-2381. In addition, Y.\ S.\ was supported by
2017--18 Summer Graduate Student Researcher Award by the UC Davis Office of
Graduate Studies.
The authors thank Haotian Li of UC Davis for constructing the graph of the
Toronto street network.
A preliminary version of this article was presented at the SPIE Conference
on Wavelets and Sparsity XVIII, August 2019, San Diego, CA~\cite{SHAO-SAITO-SPIE}.

\end{acknowledgements}


\begin{thebibliography}{10}
\providecommand{\url}[1]{{#1}}
\providecommand{\urlprefix}{URL }
\expandafter\ifx\csname urlstyle\endcsname\relax
  \providecommand{\doi}[1]{DOI~\discretionary{}{}{}#1}\else
  \providecommand{\doi}{DOI~\discretionary{}{}{}\begingroup
  \urlstyle{rm}\Url}\fi

\bibitem{JULIA}
Bezanson, J., Edelman, A., Karpinski, S., Shah, V.B.: Julia: {A} fresh approach
  to numerical computing.
\newblock SIAM Review \textbf{59}(1), 65--98 (2017).
\newblock \urlprefix\url{https://doi.org/10.1137/141000671}

\bibitem{BREMER-COIF-MAGG-SZLAM}
Bremer, J.C., Coifman, R.R., Maggioni, M., Szlam, A.: Diffusion wavelet
  packets.
\newblock Appl. Comput. Harmon. Anal. \textbf{21}(1), 95--112 (2006).
\newblock \urlprefix\url{https://doi.org/10.1016/j.acha.2006.04.005}

\bibitem{CHUNG-LU}
Chung, F., Lu, L.: Complex Graphs and Networks.
\newblock No. 107 in CBMS Regional Conference Series in Mathematics. Amer.
  Math. Soc., Providence, RI (2006).
\newblock \urlprefix\url{https://doi.org/10.1090/cbms/107}

\bibitem{CLONINGER-LI-SAITO}
Cloninger, A., Li, H., Saito, N.: Natural graph wavelet packet dictionaries.
\newblock J. Fourier Anal. Appl. \textbf{27}, Article \#41 (2021).
\newblock \urlprefix\url{https://doi.org/10.1007/s00041-021-09832-3}.
\newblock A part of ``Topical Collection: Harmonic Analysis on Combinatorial
  Graphs''

\bibitem{COIF-GAVISH}
Coifman, R.R., Gavish, M.: Harmonic analysis of digital data bases.
\newblock In: J.~Cohen, A.I. Zayed (eds.) Wavelets and Multiscale Analysis:
  Theory and Applications, Applied and Numerical Harmonic Analysis, pp.
  161--197. Birkh\"auser, Boston, MA (2011).
\newblock \urlprefix\url{https://doi.org/10.1007/978-0-8176-8095-4_9}

\bibitem{COIF-MAGG-DW}
Coifman, R.R., Maggioni, M.: Diffusion wavelets.
\newblock Appl. Comput. Harmon. Anal. \textbf{21}(1), 53--94 (2006).
\newblock \urlprefix\url{https://doi.org/10.1016/j.acha.2006.04.004}

\bibitem{COIF-WICK}
Coifman, R.R., Wickerhauser, M.V.: Entropy-based algorithms for best basis
  selection.
\newblock IEEE Trans. Inform. Theory \textbf{38}(2), 713--718 (1992).
\newblock \urlprefix\url{https://doi.org/10.1109/18.119732}

\bibitem{EASLEY-KLEINBERG}
Easley, D., Kleinberg, J.: Networks, Crowds, and Markets: Reasoning and a
  Highly Connected World.
\newblock Cambridge Univ. Press, New York (2010)

\bibitem{FIEDLER}
Fiedler, M.: A property of eigenvectors of nonnegative symmetric matrices and
  its application to graph theory.
\newblock Czechoslovak Math. J. \textbf{25}, 619--633 (1975).
\newblock \urlprefix\url{http://eudml.org/doc/12900}

\bibitem{GRIGORESCU-PETKOV-KRUIZINGA}
Grigorescu, S.E., Petkov, N., Kruizinga, P.: Comparison of texture features
  based on {G}abor filters.
\newblock IEEE Trans. Image Process. \textbf{11}(10), 1160--1167 (2002).
\newblock \urlprefix\url{https://doi.org/10.1109/TIP.2002.804262}

\bibitem{RATIOCUT}
Hagen, L., Kahng, A.B.: New spectral methods for ratio cut partitioning and
  clustering.
\newblock IEEE Trans. Comput.-Aided Des. \textbf{11}(9), 1074--1085 (1992).
\newblock \urlprefix\url{https://doi.org/10.1109/43.159993}

\bibitem{HAMMOND-VANDERGHEYNST-GRIBONVAL}
Hammond, D.K., Vandergheynst, P., Gribonval, R.: Wavelets on graphs via
  spectral graph theory.
\newblock Appl. Comput. Harmon. Anal. \textbf{30}(2), 129--150 (2011).
\newblock \urlprefix\url{https://doi.org/10.1016/j.acha.2010.04.005}

\bibitem{MultiscaleGraphSignalTransforms}
Irion, J., Li, H., Saito, N., Shao, Y.: Multiscalegraphsignaltransforms.jl.
\newblock \url{https://github.com/UCD4IDS/MultiscaleGraphSignalTransforms.jl}
  (2021)

\bibitem{IRION-SAITO-GHWT}
Irion, J., Saito, N.: The generalized {H}aar-{W}alsh transform.
\newblock In: Proc. 2014 IEEE Workshop on Statistical Signal Processing, pp.
  472--475 (2014).
\newblock \urlprefix\url{https://doi.org/10.1109/SSP.2014.6884678}

\bibitem{IRION-SAITO-SPIE}
Irion, J., Saito, N.: Applied and computational harmonic analysis on graphs and
  networks.
\newblock In: M.~Papadakis, V.K. Goyal, D.~{Van De Ville} (eds.) Wavelets and
  Sparsity XVI, Proc. SPIE 9597 (2015).
\newblock \urlprefix\url{https://doi.org/10.1117/12.2186921}.
\newblock Paper \# 95971F

\bibitem{MTSG}
Irion, J., Saito, N.: {MTSG}\_{T}oolbox.
\newblock \url{https://github.com/JeffLIrion//MTSG_Toolbox} (2015)

\bibitem{IRION-SAITO-MLSP16}
Irion, J., Saito, N.: Learning sparsity and structure of matrices with
  multiscale graph basis dictionaries.
\newblock In: A.~Uncini, K.~Diamantaras, F.A.N. Palmieri, J.~Larsen (eds.)
  Proc. 2016 IEEE 26th International Workshop on Machine Learning for Signal
  Processing (MLSP) (2016).
\newblock \urlprefix\url{https://doi.org/10.1109/MLSP.2016.7738892}

\bibitem{IRION-SAITO-TSIPN}
Irion, J., Saito, N.: Efficient approximation and denoising of graph signals
  using the multiscale basis dictionaries.
\newblock IEEE Trans. Signal Inform. Process. Netw. \textbf{3}(3), 607--616
  (2017).
\newblock \urlprefix\url{https://doi.org/10.1109/TSIPN.2016.2632039}

\bibitem{JANSEN-NASON-SILVERMAN}
Jansen, M., Nason, G.P., Silverman, B.W.: Multiscale methods for data on graphs
  and irregular multidimensional situations.
\newblock J. R. Stat. Soc. Ser. B, Stat. Methodol. \textbf{71}(1), 97--125
  (2008).
\newblock \urlprefix\url{https://doi.org/10.1111/j.1467-9868.2008.00672.x}

\bibitem{JOLLIFFE2}
Jolliffe, I.T.: Principal Component Analysis and Factor Analysis, 2nd edn.,
  chap.~7.
\newblock Springer New York, New York, NY (2002).
\newblock \urlprefix\url{https://doi.org/10.1007/b98835}

\bibitem{KALOFOLIAS-ETAL}
Kalofolias, V., Bresson, X., Bronstein, M., Vandergheynst, P.: Matrix
  completion on graphs.
\newblock In: Neural Information Processing Systems workshop ``Out of the Box:
  Robustness in High Dimension'' (2014).
\newblock \urlprefix\url{https://arxiv.org/abs/1408.1717}

\bibitem{LEE-NADLER-WASSERMAN}
Lee, A., Nadler, B., Wasserman, L.: Treelets---an adaptive multi-scale basis
  for sparse unordered data.
\newblock Ann. Appl. Stat. \textbf{2}, 435--471 (2008).
\newblock \urlprefix\url{https://doi.org/10.1214/07-AOAS137}

\bibitem{LINDBERG-VILLEMOES}
Lindberg, M., Villemoes, L.F.: Image compression with adaptive {H}aar-{W}alsh
  tilings.
\newblock In: A.~Aldroubi, A.F. Laine, M.A. Unser (eds.) Wavelet Applications
  in Signal and Image Processing VIII, Proc. SPIE 4119, pp. 911--921 (2000).
\newblock \urlprefix\url{https://doi.org/10.1117/12.408575}

\bibitem{LINDENBAUM-ETAL}
Lindenbaum, O., Salhov, M., Yeredor, A., Averbuch, A.: Gaussian bandwidth
  selection for manifold learning and classification.
\newblock Data Min. Knowl. Discov. \textbf{34}, 1676--1712 (2020).
\newblock \doi{10.1007/s10618-020-00692-x}.
\newblock \urlprefix\url{https://doi.org/10.1007/s10618-020-00692-x}

\bibitem{LOVASZ-BOOK}
Lov\'{a}sz, L.: Large Networks and Graph Limits, \emph{Colloquium
  Publications}, vol.~60.
\newblock Amer. Math. Soc., Providence, RI (2012)

\bibitem{VONLUX}
von Luxburg, U.: A tutorial on spectral clustering.
\newblock Stat. Comput. \textbf{17}(4), 395--416 (2007).
\newblock \urlprefix\url{https://doi.org/10.1007/s11222-007-9033-z}

\bibitem{MILANFAR-IMG-FILTERING}
Milanfar, P.: A tour of modern image filtering: {N}ew insights and methods,
  both practical and theoretical.
\newblock IEEE Signal Processing Magazine \textbf{30}(1), 106--128 (2013).
\newblock \urlprefix\url{https://doi.org/10.1109/MSP.2011.2179329}

\bibitem{MURTAGH-Haar}
Murtagh, F.: The {H}aar wavelet transform of a dendrogram.
\newblock J. Classification \textbf{24}(1), 3--32 (2007).
\newblock \urlprefix\url{https://doi.org/10.1007/s00357-007-0007-9}

\bibitem{NMF-BOOK}
Naik, G.R. (ed.): Non-negative Matrix Factorization Techniques: Advances in
  Theory and Applications.
\newblock Springer (2016).
\newblock \urlprefix\url{https://doi.org/10.1007/978-3-662-48331-2}

\bibitem{NEWMAN2}
Newman, M.: Networks, 2nd edn.
\newblock Oxford Univ. Press (2018)

\bibitem{OUTEX}
Ojala, T., Maenp\"a\"a, T., Pietik\"ainen, M., Viertola, J., Kyl\"onen, J.,
  Huovinen, S.: Outex - new framework for empirical evaluation of texture
  analysis algorithms.
\newblock In: Proceedings of 16th International Conf. on Pattern Recognition,
  vol.~1, pp. 701--706 (2002).
\newblock \urlprefix\url{https://doi.org/10.1109/ICPR.2002.1044854}

\bibitem{ORTEGA-ETAL}
Ortega, A., Frossard, P., J.Kova{\v{c}}evi{\'c}, Moura, J.M.F., Vandergheynst,
  P.: Graph signal processing: {O}verview, challenges, and applications.
\newblock Proc. IEEE \textbf{106}(5), 808--828 (2018).
\newblock \urlprefix\url{https://doi.org/10.1109/JPROC.2018.2820126}

\bibitem{PETKOV-KRUIZINGA}
Petkov, N., Kruizinga, P.: Computational models of visual neurons specialised
  in the detection of periodic and aperiodic oriented visual stimuli: bar and
  grating cells.
\newblock Biological Cybernetics  (1997).
\newblock \urlprefix\url{https://doi.org/10.1007/s004220050323}

\bibitem{QI-LUO}
Qi, L., Luo, Z.: Tensor Analysis: Spectral Theory and Special Tensors.
\newblock SIAM, Philadelphia, PA (2019).
\newblock \urlprefix\url{https://doi.org/10.1137/1.9781611974751}

\bibitem{VIRIDIS}
Rudis, B., Ross, N., Garnier, S.: The viridis color palettes.
\newblock
  \url{https://cran.r-project.org/web/packages/viridis/vignettes/intro-to-viridis.html}
  (2018)

\bibitem{RUSTAMOV}
Rustamov, R.M.: Average interpolating wavelets on point clouds and graphs.
\newblock arXiv:1110.2227 [math.FA] (2011)

\bibitem{SAITO-LSDB3}
Saito, N.: Image approximation and modeling via least statistically dependent
  bases.
\newblock Pattern Recognition \textbf{34}, 1765--1784 (2001)

\bibitem{SAITO-PLASMA-ENGLISH}
Saito, N.: Laplacian eigenfunctions and their application to image data
  analysis.
\newblock J. Plasma Fusion Res. \textbf{92}(12), 904--911 (2016).
\newblock
  \urlprefix\url{http://www.jspf.or.jp/Journal/PDF_JSPF/jspf2016_12/jspf2016_12-904.pdf}.
\newblock In Japanese

\bibitem{SAITO-COIF-JMIV}
Saito, N., Coifman, R.R.: Local discriminant bases and their applications.
\newblock J. Math. Imaging Vis. \textbf{5}(4), 337--358 (1995).
\newblock Invited paper

\bibitem{SAITO-COIF-SONIC}
Saito, N., Coifman, R.R.: Extraction of geological information from acoustic
  well-logging waveforms using time-frequency wavelets.
\newblock Geophysics \textbf{62}(6), 1921--1930 (1997).
\newblock \urlprefix\url{https://doi.org/10.1190/1.1444292}

\bibitem{SAITO-COIF-GESHWIND-WARNER}
Saito, N., Coifman, R.R., Geshwind, F.B., Warner, F.: Discriminant feature
  extraction using empirical probability density estimation and a local basis
  library.
\newblock Pattern Recognition \textbf{35}(12), 2841--2852 (2002)

\bibitem{SAITO-REMY-ACHA}
Saito, N., Remy, J.F.: The polyharmonic local sine transform: {A} new tool for
  local image analysis and synthesis without edge effect.
\newblock Appl. Comput. Harmon. Anal. \textbf{20}(1), 41--73 (2006).
\newblock \urlprefix\url{https://doi.org/10.1016/j.acha.2005.01.005}

\bibitem{SHAO-SAITO-SPIE}
Shao, Y., Saito, N.: The extended generalized {H}aar-{W}alsh transform and
  applications.
\newblock In: D.~{Van De Ville}, M.~Papadakis, Y.M. Lu (eds.) Wavelets and
  Sparsity XVIII, Proc. SPIE 11138 (2019).
\newblock \urlprefix\url{https://doi.org/10.1117/12.2528923}.
\newblock Paper \#111380C

\bibitem{SHI-MALIK}
Shi, J., Malik, J.: Normalized cuts and image segmentation.
\newblock IEEE Trans. Pattern Anal. Machine Intell. \textbf{22}(8), 888--905
  (2000).
\newblock \urlprefix\url{https://doi.org/10.1109/34.868688}

\bibitem{SHUMAN-ETAL}
Shuman, D.I., Narang, S.K., Frossard, P., Ortega, A., Vandergheynst, P.: The
  emerging field of signal processing on graphs.
\newblock IEEE Signal Processing Magazine \textbf{30}(3), 83--98 (2013).
\newblock \urlprefix\url{https://doi.org/10.1109/MSP.2012.2235192}

\bibitem{SZLAM-MAGG-COIF}
Szlam, A.D., Maggioni, M., Coifman, R.R.: Regularization on graphs with
  function-adapted diffusion processes.
\newblock Journal of Machine Learning Research \textbf{9}(55), 1711--1739
  (2008).
\newblock \urlprefix\url{http://jmlr.org/papers/v9/szlam08a.html}

\bibitem{SZLAM-HAAR-GRAPH}
Szlam, A.D., Maggioni, M., Coifman, R.R., Bremer Jr., J.C.: Diffusion-driven
  multiscale analysis on manifolds and graphs: top-down and bottom-up
  constructions.
\newblock In: M.~Papadakis, A.F. Laine, M.A. Unser (eds.) Wavelets XI, Proc.
  SPIE 5914 (2005).
\newblock \urlprefix\url{https://doi.org/10.1117/12.616931}.
\newblock Paper \# 59141D

\bibitem{JuliaWavelets}
{The Julia DSP Team}: Wavelets.jl.
\newblock \url{https://github.com/JuliaDSP/Wavelets.jl} (2021)

\bibitem{THIELE-VILLEMOES}
Thiele, C.M., Villemoes, L.F.: A fast algorithm for adapted time-frequency
  tilings.
\newblock Appl. Comput. Harmon. Anal. \textbf{3}(2), 91--99 (1996).
\newblock \urlprefix\url{https://doi.org/10.1006/acha.1996.0009}

\bibitem{TOMASI-MANDUCHI}
Tomasi, C., Manduchi, R.: Bilateral filtering for gray and color images.
\newblock In: Sixth International Conference on Computer Vision (IEEE Cat.
  No.98CH36271), pp. 839--846 (1998).
\newblock \urlprefix\url{https://doi.org/10.1109/ICCV.1998.710815}

\bibitem{TREMBLAY-BORGNAT}
Tremblay, N., Borgnat, P.: Subgraph-based filterbanks for graph signals.
\newblock IEEE Trans. Signal Process. \textbf{64}(15), 3827--3840 (2016).
\newblock \urlprefix\url{https://doi.org/10.1109/TSP.2016.2544747}

\bibitem{WICK-WPK}
Wickerhauser, M.V.: {Adapted Wavelet Analysis from Theory to Software}.
\newblock A~K~Peters, Ltd., Wellesley, MA (1994)

\bibitem{YAMATANI-SAITO}
Yamatani, K., Saito, N.: Improvement of {DCT}-based compression algorithms
  using {Poisson's} equation.
\newblock IEEE Trans. Image Process. \textbf{15}(12), 3672--3689 (2006).
\newblock \urlprefix\url{https://doi.org/10.1109/TIP.2006.882005}

\end{thebibliography}

\end{document}